\documentclass{bmcart}
\usepackage{amsthm,amsmath}
\usepackage[utf8]{inputenc} 

\usepackage{amsfonts,amssymb,graphics,graphicx,epsfig,bbm}
\usepackage[colorlinks=true,citecolor=blue,linkcolor=blue,urlcolor=blue]{hyperref}
\usepackage{graphicx}
\usepackage{subfigure}
\usepackage{amsmath}
\usepackage{epsfig}
\usepackage{dcolumn}
\usepackage{bm}
\usepackage{color}
\usepackage{times}
\usepackage{epstopdf}
\usepackage{amssymb}
\usepackage{amstext}
\usepackage{latexsym}
\usepackage{hyperref}
\usepackage{amsfonts}
\usepackage{psfrag}
\usepackage{soul,xcolor}
\usepackage[normalem]{ulem}
\usepackage{dsfont}
\usepackage{txfonts}
\usepackage{float}
\newcommand{\tabincell}[2]{\begin{tabular}{@{}#1@{}}#2\end{tabular}}

\newcommand{\ie}{{\it{i.e.~}}}

\newcommand{\ket}[1]{\vert #1 \rangle}
\newcommand{\bra}[1]{\langle #1 \vert}

\usepackage[numbers,sort&compress]{natbib}
\startlocaldefs
\endlocaldefs

\begin{document}

\begin{frontmatter}

\begin{fmbox}
\dochead{Research}


\title{Superconducting Circuit Architecture for Digital-Analog Quantum Computing}


\author[addressref={aff1}, ]{J. Yu}
\author[addressref={aff2,aff3}, ]{J. C. Retamal}
\author[addressref={aff4,aff5}, ]{M. Sanz}
\author[addressref={aff1,aff4,aff5,aff6},corref={aff1},email={enr.solano@gmail.com} ]{E. Solano}
\author[addressref={aff2}, corref={aff2}, email={pancho.albarran@gmail.com}]{F. Albarr\'an-Arriagada}


\address[id=aff1]{International Center of Quantum Artificial Intelligence for Science and Technology (QuArtist) and Physics Department, Shanghai University, 200444 Shanghai, China}
\address[id=aff2]{Departamento de F\'isica, Universidad de Santiago de Chile (USACH), Avenida Ecuador 3493, 9170124, Santiago, Chile}
\address[id=aff3]{Center for the Development of Nanoscience and Nanotechnology 9170124, Estaci\'on Central, Santiago, Chile}
\address[id=aff4]{Department of Physical Chemistry, University of the Basque Country UPV/EHU, Apartado 644, 48080 Bilbao, Spain}
\address[id=aff5]{IKERBASQUE, Basque Foundation for Science, Plaza Euskadi 5, 48009 Bilbao, Spain}
\address[id=aff6]{Kipu Quantum, Kurwenalstrasse 1, 80804 Munich, Germany}


\end{fmbox}


\begin{abstractbox}
\begin{abstract} 
We propose a superconducting circuit architecture suitable for digital-analog quantum computing (DAQC) based on an enhanced NISQ family of nearest-neighbor interactions. DAQC makes a smart use of digital steps (single qubit rotations) and analog blocks (parametrized multiqubit operations) to outperform digital quantum computing algorithms. Our design comprises a chain of superconducting charge qubits coupled by superconducting quantum interference devices (SQUIDs). Using magnetic flux control, we can activate/deactivate exchange interactions, double excitation/de-excitations, and others. As a paradigmatic example, we present an efficient simulation of an $\ell\times h$ fermion lattice (with $2<\ell \leq h$), using only $2(2\ell+1)^2+24$ analog blocks. The proposed architecture design is feasible in current experimental setups for quantum computing with superconducting circuits, opening the door to useful quantum advantage with fewer resources.
\end{abstract}
\end{abstractbox}
%

\end{frontmatter}



\color{black}\section{Introduction}
It is known that the calculation of the exact dynamics of a quantum many-body system is in general a challenging task. When the system is complex enough, analytical and numerical solutions are not possible. However, quantum simulation (QS) allows us to overcome this difficulty by using controllable and manipulable quantum systems, known as quantum simulators, to study another non-controllable one~\cite{Georgescu2014RMP, Cirac2012NatPhys}. QS can be classified into three different groups: analog quantum simulations (AQS), digital quantum simulations (DQS), and digital-analog quantum simulations (DAQS). Formally, DQS and Digital-Quantum Computing (DQC) are equivalent, and we will claim the same for DAQS and Digital-Analog Quantum Computing (DAQC).

In AQS, we may reproduce a given target Hamiltonian for some parameter regimes of the simulated model and of the quantum simulator, which is not universal~\cite{Aedo2018PRA, Braumuller2017NatCommun}. In DQS, we can perform a sequence of quantum gates in the quantum simulator, which is similar to what happens in DQC. In this case, we can approximate any unitary evolution, specifically, the unitary evolution of the target Hamiltonian model or the given quantum algorithm~\cite{Lloyd1996Science}. Even if DQC is universal, it is less accurate than AQS and Analog Quantum Computing, requiring quantum error correction to scale up and being impractical for current platforms. On the other hand, the recently proposed DAQS aims at getting the best of analog and digital paradigms, with more versatility in the target models or algorithms, higher accuracy, efficient administration of coherence time, and more suitable for current noise intermediate scale quantum (NISQ) architectures \cite{Preskill2017Quantum}. In DAQS, we use a continuous set of complex many-body interactions as a resource (analog blocks), offered naturally by the architecture possibilities of the quantum platform used as quantum computer. We complement it with accessible continuous sets of single-qubit operations (digital steps), providing versatility of target Hamiltonian models or algorithms~\cite{ParraRodriguez2020PRA,Celeri2021arXiv,GarciaMolina2021arXiv,GonzalezRaya2021PRXQuantum}.

Several physical platforms have been used as quantum simulators, such as optical lattices~\cite{QiuNPJQI}, trapped ions~\cite{Blatt2012NatPhys} and superconducting circuits~\cite{Paraoanu2014JLTP,Schmidt2013AnnPhys,Devoret2013Science}. The latter has particular features allowing for current scalability and design flexibility, and produced the recent claim of quantum supremacy in quantum computing~\cite{Arute2019Nature,Wu2021PRL}. At the same time, diverse target models have been proposed and implemented, principally for DQS and AQS, such as quantum chemistry systems~\cite{Lanyon2010Nat,ArgueloLuengo2019Nature,Babbush2014SciRep,MacDonell2021ChemSci}, high-energy physics~\cite{Lamata2007PRL,Gerritsma2011PRL,Gerritsma2010Nature,Bauer2019arXiv}, and condensed matter physics~\cite{Casanova2012PRL,Hensgens2017Nature,Kim2010Nature}.  In the long run to reach fault-tolerant quantum computers in the future, proposals for DAQS appear promising for this and next generation of co-design quantum computers~\cite{Arrazola2016SciRep,AlbarranArriagada2018PRA,Martin2020PRR,Babukhin2020PRA}. To achieve that, it would be desirable to widely enhance the DAQC algorithmic mappings as well as the variety of accessible quantum computer geometries and topologies.

In this work, we propose a superconducting circuit design for DAQC. It is composed of a chain of charge qubits coupled through grounded superconducting quantum interference devices (SQUIDs), where the nearest-neighbor qubits are off resonance (different energy gap), similar ideas have been proposed using variable electrostatic fields~\cite{Averin2003PRL, Hutter2006EPL}. In this proposal, the SQUIDs modify the resonant condition among nearest-neighbor qubits, which allows us to produce different and independent interactions like an exchange or double excitation/de-excitation term. Also, as the SQUIDs are physically distant, we could manipulate them individually, activating/deactivating several interactions for obtaining parametrized multiqubit gates, also proving a large family of analog multibody Hamiltonians, being suitable for efficient implementations of DAQC protocols. Finally, we test our architecture performing the simulation of the Fermi-Hubbard model, where we need only $2(2\ell + 1)^2 + 24$ analog blocks for the simulation of a $\ell\times h$ ($\ell\le h$) Fermi-Hubbard lattice.
\section{The model}
\subsection{Two-qubit model}

\begin{figure}[t]
\includegraphics[width=0.5\linewidth]{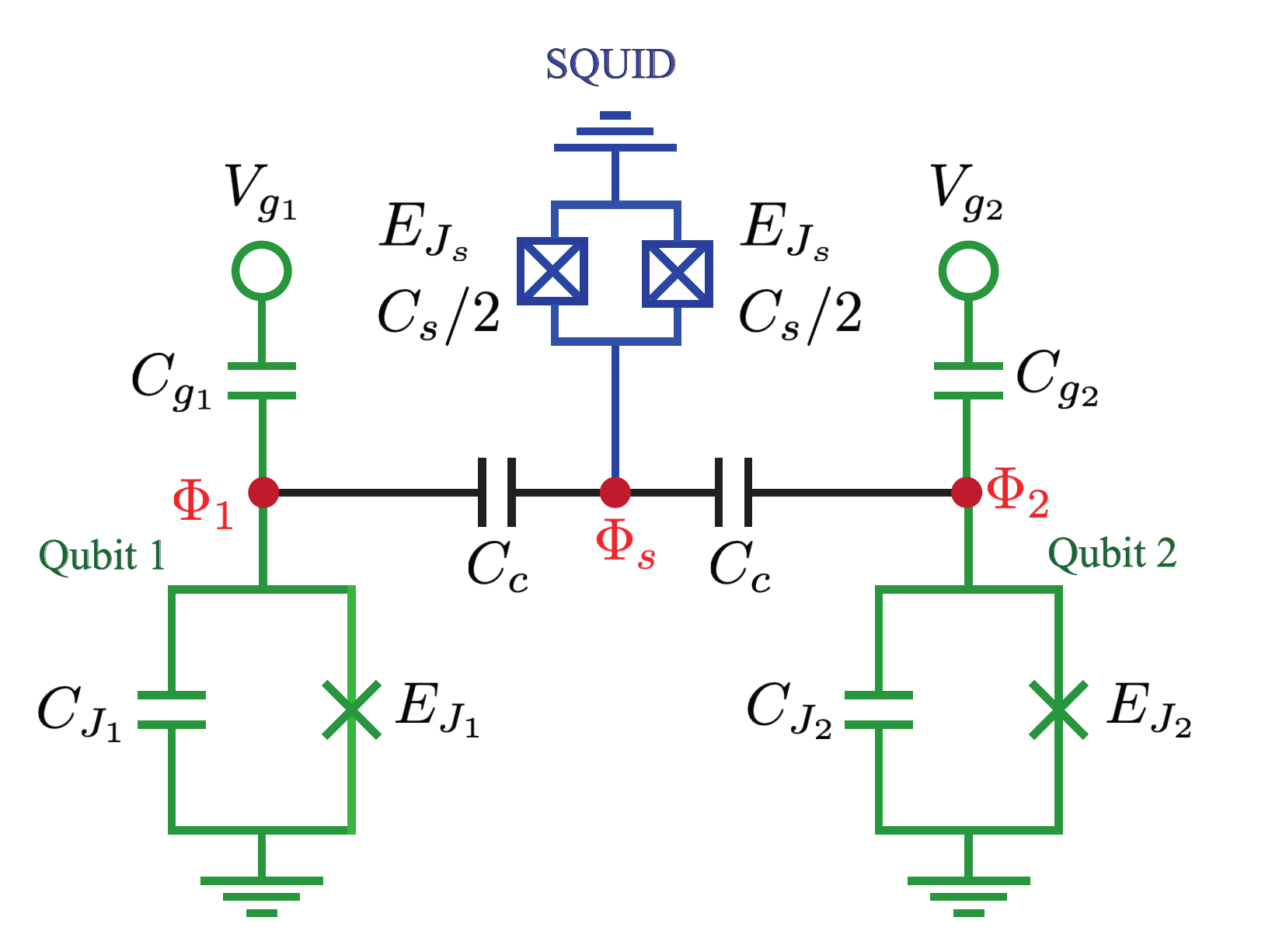}
\caption{Circuit diagram of two charged qubits (green) coupled through a grounded SQUID (blue). 
$E_{J_{1(2)}}$, $C_{J_{1(2)}}$, $C_{g_{1(2)}}$, and $V_{g_{1(2)}}$ are the Josephson energy, Josephson capacitance, gate capacitance, and gate voltage of the qubit 1(2) respectively. $E_{J_{s}}$ and $C_{s}$ are the Josephson energy and effective capacitance of the SQUID. Moreover, $C_c$ is the coupling capacitance, and $\Phi_1$, $\Phi_2$, and $\Phi_s$ are node fluxes that define the degrees of freedom of the circuit.}
\label{Fig01}
\end{figure}

First, let us consider the two-qubit system showed in Fig.~\ref{Fig01}. It consists of two charge qubits coupled through a grounded symmetric SQUID. The Lagrangian of the circuit is given by
\begin{align}\nonumber
\mathcal{L} &= \sum_{j=1}^2\left[\frac{C_{g_j}}{2}(\Phi^{\prime}_j-V_{g_j})^2+\frac{C_{J_j}}{2}\Phi^{\prime 2}_j+E_{J_j}\cos{(\varphi_j)}\right]  \\
 & + \frac{C_{s}}{2}{\Phi_s}^{\prime 2}+E_{J_s}^{\textrm{eff}}\cos{(\varphi_s)}+\frac{C_{c}}{2}(\Phi^{\prime}_1-\Phi^{\prime}_s)^2+\frac{C_{c}}{2}(\Phi^{\prime}_s-\Phi^{\prime}_2)^2,  
\label{Eq01}
\end{align}
where $f^{\prime}=(d/dt)f(t)$, represent the time derivate of a function $f$, $E_{J_s}^{\textrm{eff}}=2E_{J_s}\cos{(\varphi_{ext})}$ is the effective Josephson energy of the SQUID, and $\varphi_j=2\pi\Phi_j/\Phi_0$ is the the superconducting phase, with the superconducting flux quantum $\Phi_0=h/2e$, and $2e$ is the electrical charge of a Cooper pair. Moreover $\Phi_1$, $\Phi_2$, and $\Phi_s$ are node fluxes defined in Fig.~\ref{Fig01}. The effective inductance of the SQUID $L_{J_s}({\varphi_{ext}})=({\Phi_0}/{2\pi})^2/E_{J_s}^{\textrm{eff}}$ can be tuned by the external magnetic flux $\varphi_{ext}(t)$, providing a tunable boundary condition~\cite{Johansson2010PRA,Wilson2011Nature}.

By applying the Legendre transformation, we obtain the Hamiltonian (see appendix \ref{AppA})
\begin{align}
\mathcal{H}&=\sum_{j=1}^2\mathcal{H}_{\textrm{qubit}}^j+\mathcal{H}_{\textrm{SQUID}}+\mathcal{H}_{\textrm{coupling}},
\label{Eq02}
\end{align}
with
\begin{align}\nonumber
&\mathcal{H}_{\textrm{qubit}}^j=\frac{1}{2\tilde{C}_{J_j}}(Q_j-2e\tilde{n}_{g_j})^2-E_{J_j}\cos{\varphi_j},
\\\nonumber
&\mathcal{H}_{\textrm{SQUID}}=\frac{1}{2\tilde{C}_{J_s}}(Q_s-2e\tilde{n}_{g_s})^2- E_{J_s}^{\textrm{eff}}\cos{(\varphi_s)},
\\
&\mathcal{H}_{\textrm{coupling}}=g_{12}Q_1 Q_2+g_{1s}Q_1 Q_s+g_{2s}Q_2 Q_s,
\label{Eq03}
\end{align}
where $Q_j={\partial L}/{\partial \Phi^{\prime}_j}$ is the charge (conjugate momenta) of the j$th$~node given by
\begin{eqnarray}
Q_{1(2)}=(C_{1(2)}+C_c)\Phi^{\prime}_{1(2)}-C_{g_{1(2)}}V_{g_{1(2)}}-C_c\Phi^{\prime}_{s},~ Q_{s}=(C_s+2C_c)\Phi^{\prime}_{s}-C_c(\Phi^{\prime}_{1}+\Phi^{\prime}_{2}),
\end{eqnarray}
and the effective Josephson capacitances are defined as
\begin{align}
\tilde{C}_{J_{1(2)}}=\frac{C_\star^3}{C_{2(1)}(2C_c+C_s)+C_c(C_c+C_s )},\quad\tilde{C}_{J_s}=\frac{C_cC_\star^3}{(C_c+C_1)(C_c+C_2)},
\label{Eq05}
\end{align}
with $C_\star^3=C_c(C_1+C_2)(C_s+C_c)+C_c^2C_s+C_1C_2(2C_c+C_s)$, and $C_{j}=C_{g_j}+C_{J_j}$ ($j=\{1,2\}$). Moreover, the gate-charge numbers read
\begin{align}\nonumber
&\tilde{n}_{g_{1(2)}}=-\frac{C_{g_{1(2)}}}{2e}V_{g_{1(2)}}-\frac{\tilde{C}_{J_{1(2)}}C_c^2C_{g_{2(1)}}}{2eC_\star^3}V_{g_{2(1)}},\\
&\tilde{n}_{g_s}=-\frac{\tilde{C}_{J_s}C_c}{2eC_\star^3}\bigg(C_{g_1}(C_{2} + C_c )V_{g_1}+C_{g_2}(C_{1} + C_c )V_{g_2}\bigg) \, ,
\label{Eq05}
\end{align}
and couplings strengths are given by
\begin{eqnarray}
g_{12}=\frac{C^2_c}{C^3_\star}, \quad g_{1s(2s)}=\frac{C_c(C_{2(1)} + C_c )}{C_\star^3} \, .
\label{Eq06}
\end{eqnarray}

\begin{figure*}
\centering
\includegraphics[width=0.95\textwidth]{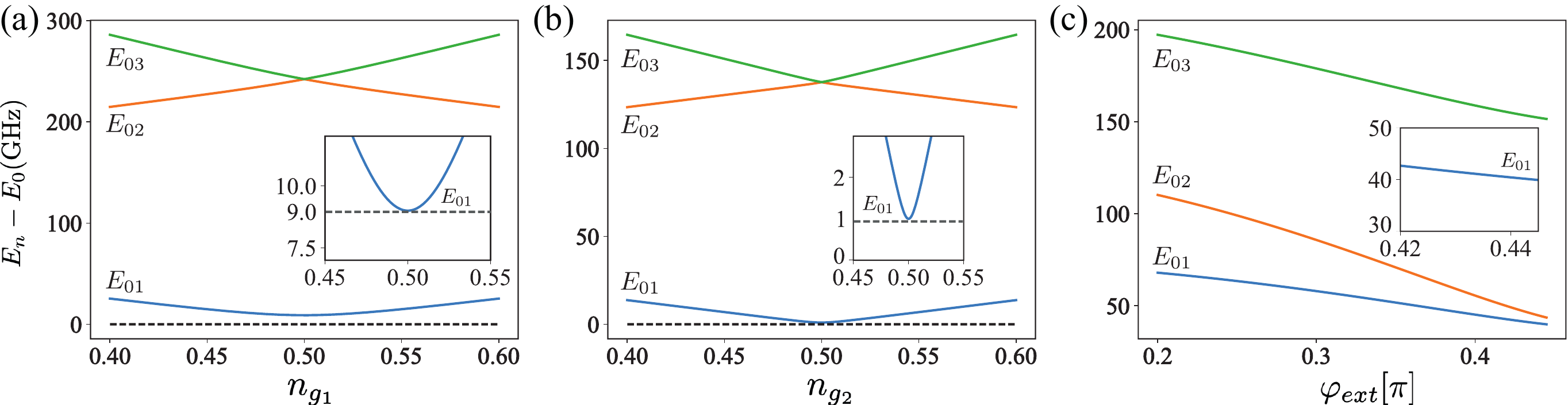}
\caption{Transitions $E_n-E_0$ (n = 1, 2, 3) of the qubits and SQUID Hamiltonians given in Eq.~(\ref{Eq03}). (a) Qubit 1: $E_{J_1} /E_{C_1}=0.303$. (b) Qubit 2: $E_{J_2} /E_{C_2}=0.058$. (c) SQUID: $E_{J_s}$= 50($\rm{GHz}$) and $C_s$=12[$fF$].}
\label{Fig02}
\end{figure*}

\begin{figure}[h]
\includegraphics[width=0.5\linewidth]{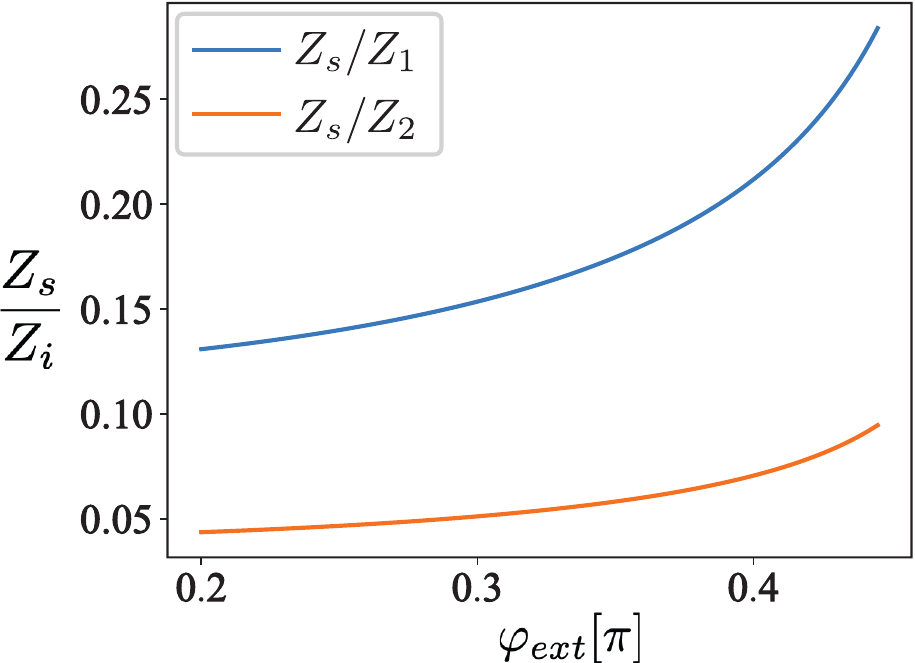}
\caption{Ratio between the SQUID impedance $Z_s$ and the qubit 1(2) impedance $Z_{1(2)}$ as a function of the external magnetic flux $\varphi_{ext}$.}
\label{fig:impedance}
\end{figure}
\begin{equation}
Q_s=-C_{c}\left(\frac{Q_{1}+C_{g_1}V_{g_1}}{C_{1}+C_c}+\frac{Q_{2}+C_{g_2}V_{g_2}}{C_{2}+C_c}\right).
\label{Eq08}
\end{equation}

Here, we consider the regime of high plasma frequency for the SQUID, where the charge energy is small compared to the Josephson energy, and the plasma frequency of the SQUID is far exceeding the frequency of the qubits (see Fig.~\ref{Fig02}), then we can consider $\Phi^{\prime}_s\ll\Phi^{\prime}_1(\Phi^{\prime}_2)$ and $\Phi^{\prime\prime}_s\ll\Phi^{\prime\prime}_1(\Phi^{\prime\prime}_2)$~\cite{Johansson2010PRA}. In addition, we also consider the low impedance for the SQUID  (see Fig.~\ref{fig:impedance}), which allow us consider $\Phi_s\ll\Phi_1(\Phi_2)$. Based on the above conditions, we obtain the next relation for $Q_s$ (see appendix~\ref{AppA})

Now, using the Euler-Lagrange equations we obtain (see appendix~\ref{AppA})
\begin{align}\nonumber
&\left(C_{1(2)}+C_c\right)\Phi^{\prime\prime}_{1(2)}-C_c\Phi^{\prime\prime}_s+\frac{2\pi}{\Phi_0}E_{J_{1(2)}}\sin{(\varphi_{1(2)})}=0,\\
&-C_{c}\Phi^{\prime\prime}_1-C_{c}\Phi^{\prime\prime}_2+2C_s\Phi^{\prime\prime}_s+\frac{2\pi E_{J_s}^{\textrm{eff}}}{\Phi_0}\sin{(\varphi_s)}=0,
\end{align}
using the same above conditions we get the relation for ${\varphi_s}$ as
\begin{align}
&{\varphi_s}=\frac{-C_c}{E_{J_s}^{\textrm{eff}}}\left(\frac{E_{J_1}\sin{(\varphi_1)}}{C_1+C_c}+\frac{E_{J_2}\sin{(\varphi_2)}}{C_2+C_c}\right),
\label{Eq10}
\end{align}
where we approximate $\sin{(\varphi_s)}\approx\varphi_s$. 

We note that, we can write the charge in the  node $j$ as $Q_j=2en_j$. Promoting the classical variables $\{n_j,\varphi_j\}$ to quantum operators $\{\hat{n}_j,\hat{\varphi}_j\}$ with the commutation relation $[e^{i\hat{\varphi}_{j}},\hat{n}_{j}]=e^{i\hat{\varphi}_{j}}$\cite{Vool2017arXiv}, and applying Eqs.~({\ref{Eq08}}) and (\ref{Eq10}) to Eq.~(\ref{Eq02}), we obtain the quantum mechanical Hamiltonian describing our circuit as
\begin{align}
\hat{\mathcal{H}}=\sum_{j=1}^2\hat{\mathcal{H}}_{sub}^j+\gamma_{12}(\varphi_{ext})\sin{(\hat{\varphi}_1)}\sin{(\hat{\varphi}_2)} \, ,
\label{Eq11}
\end{align}
where the effective coupling strength
\begin{align}
\gamma_{12}(\varphi_{ext})=\frac{C_{c}^2E_{J_1}E_{J_2}}{(C_{1}+C_c)(C_{2}+C_c)E_{J_s}^{\textrm{eff}}} \, ,
\label{Eq12}
\end{align}
and $\hat{\mathcal{H}}_{sub}^j$ is the Hamiltonian of the $j${th} subsystem, given by 
\begin{align}
\hat{\mathcal{H}}_{sub}^j=4E_{C_j}(\hat{n}_{j}-\bar{n}_{g_j})^2-E_{J_j}\cos{(\hat{\varphi}_j)}+\gamma_j(\varphi_{ext})\sin{(\hat{\varphi}_j)}^2 ,
\label{Eq13}
\end{align}
with $E_{C_j}=e^2/2(C_j+C_c)$, $\bar{n}_{g_j}=-C_{g_j}V_{g_j}/2e$, and
\begin{align}
\gamma_j(\varphi_{ext})=\frac{C_{c}^2E_{J_j}^2}{2E_{J_s}^{\textrm{eff}}(C_{j}+C_c)^2} \, .
\label{Eq14}
\end{align}

 In the following discussion, we consider $\bar{n}_{g_1}=\bar{n}_{g_2}=0.5$, and $\hbar=1$. It is convenient to write the circuit Hamiltonian in the charge basis, it means $\hat{n}_i=\sum_{n_j}n_j|n_j\rangle\langle n_j|$ and $\cos{(\hat{\varphi}_j)}=1/2(\sum_{n_j}|n_j\rangle\langle n_j+1|+\sum_{n_j}|n_j+1\rangle \langle n_j|)$~\cite{Vool2017arXiv}.  Due to the anharmonicity of $\hat{\mathcal{H}}_{sub}^j$ (see appendix~\ref{AppA}), we can perform the two-level approximation in order to obtain the effective Hamiltonian
\begin{align}
\hat{\mathcal{H}}=\frac{\omega_1}{2}\sigma_1^z+\frac{\omega_2}{2}\sigma_2^z+\frac{\gamma_{12}(\varphi_{ext})}{4}\sigma_1^y\sigma_2^y \, ,
\label{Eq15}
\end{align}
 where $\omega_1=E_{J_1}$, $\omega_2=E_{J_2}$ and $\sigma^{\alpha}_j$ is Pauli matrix element of the $j$th charge qubit and $\mathbb{I}$ is the identity operator.

Now, we will consider the external flux $\varphi_{ext}$ to be composed by a DC signal and a small AC signal as $\varphi_{ext}=\varphi_{ext}(t)=\varphi_{DC}+\varphi_{AC}(t)$, where 
\begin{align}
\varphi_{AC}(t)=A_1\cos{(\nu_1 t +\tilde{\varphi}_1)}+A_2\cos{(\nu_2 t +\tilde{\varphi}_2)} \, ,
\label{Eq16}
\end{align}
with $|A_{1}|, |A_{2}|\ll |\varphi_{DC}|$, with which we can approximate
\begin{align}
\frac{1}{E_{J_s}^{\textrm{eff}}}\approx\frac{1}{\bar{E}_{J_s}}\left[1+\frac{\sin{({\varphi}_{DC})}}{\cos{({\varphi}_{DC}})}{\varphi}_{AC}(t)\right] \, ,
\label{Eq17}
\end{align}
where $\bar{E}_{J_s}=2E_{J_s}\cos{({\varphi}_{DC})}$. Then, we can rewrite the Hamiltonian in Eq. (\ref{Eq15}) as
\begin{align}
\hat{\mathcal{H}}&=\frac{\omega_1}{2}\sigma_1^z+\frac{\omega_2}{2}\sigma_2^z+\left[g_0+g_1\varphi_{AC}(t)\right]\sigma_1^y\sigma_2^y \, ,
\label{Eq18}
\end{align}
where
\begin{align}
g_0&=\frac{C_{c}^2E_{J_1}E_{J_2}}{4(C_{1}+C_c)(C_{2}+C_c)\bar{E}_{J_s}}, ~g_1=\frac{C_{c}^2E_{J_1}E_{J_2}}{4(C_{1}+C_c)(C_{2}+C_c)\bar{E}_{J_s}}\frac{\sin{({\varphi}_{DC})}}{\cos{({\varphi}_{DC})}} \, .
\label{Eq19}
\end{align}

Now, we write the Hamiltonian of  Eq. (\ref{Eq18}) in the interaction picture with respect to $\hat{\mathcal{H}}_0=\sum^2_{i=j}\omega_j\sigma_j^z/2$ and perform the rotating wave approximation (RWA), obtaining
\begin{align}\nonumber
\hat{\mathcal{H}}_I&\approx-\frac{g_1}{2}\sigma_1^-\sigma_2^-\bigg(A_1e^{i\tilde{\varphi}_1}e^{i(\nu_1-\mu_{12})t}+A_2e^{i\tilde{\varphi}_2}e^{i(\nu_2-\mu_{12})t}\bigg)\\\nonumber
&+\frac{g_1}{2}\sigma_1^-\sigma_2^+\bigg(A_1e^{i\tilde{\varphi}_1}e^{i(\nu_1-\Delta_{12})t}+A_2e^{i\tilde{\varphi}_2}e^{i(\nu_2-\Delta_{12})t}\bigg)\\\nonumber
&+\frac{g_1}{2}\sigma_1^+\sigma_2^-\bigg(A_1e^{-i\tilde{\varphi}_1}e^{-i(\nu_1-\Delta_{12})t}+A_2e^{-i\tilde{\varphi}_2}e^{-i(\nu_2-\Delta_{12})t}\bigg)\\
&-\frac{g_1}{2}\sigma_1^+\sigma_2^+\bigg(A_1e^{-i\tilde{\varphi}_1}e^{-i(\nu_1-\mu_{12})t}+A_2e^{-i\tilde{\varphi}_2}e^{-i(\nu_2-\mu_{12})t}\bigg). 
\label{Eq20}
\end{align}
Here, we make use of $\Delta_{12}=\omega_1-\omega_2$, $\mu_{12}=\omega_1+\omega_2$ and we neglect the fast oscillating terms proportional to exp$({\pm i(\Delta_{12}+\nu_{1(2)})t})$, exp$({\pm i(\mu_{12}+\nu_{1(2)})t})$, exp$({\pm i\Delta_{12}t})$, and exp$({\pm i\mu_{12}t})$. As the qubits are far from resonance and considering $\{g_0, A_{1} g_1/2, A_{2} g_1/2\}\ll\{\Delta_{12}, \mu_{12}, \nu_1, \nu_2\}$, the RWA is justified (for more details see appendix~\ref{AppA}). Considering $\nu_1=\Delta_{12}$ and $\nu_2=\mu_{12}$, the Hamiltonian in Eq.~(\ref{Eq20}) turns
\begin{align}\nonumber
\hat{\mathcal{H}}_I&=\frac{g_1}{4}\bigg((A_1\cos{\tilde{\varphi}_1}-A_2\cos{\tilde{\varphi}_2})\sigma_1^x\sigma_2^x-(A_1\sin{\tilde{\varphi}_1}+A_2\sin{\tilde{\varphi}_2})\sigma_1^x\sigma_2^y\\
&+(A_1\sin{\tilde{\varphi}_1}-A_2\sin{\tilde{\varphi}_2})\sigma_1^y\sigma_2^x+(A_1\cos{\tilde{\varphi}_1}+A_2\cos{\tilde{\varphi}_2})\sigma_1^y\sigma_2^y\bigg) \, ,
\label{Eq21}
\end{align}
where we neglect the fast oscillating terms proportional to exp(${\pm i(\Delta_{12}-\nu_{2})t}$) and exp(${\pm i(\mu_{12}-\nu_{1})t}$). We recall that, for a proper choice of the phases $\tilde{\varphi}_1$ and $\tilde{\varphi}_2$ in Eq.~(\ref{Eq21}), we can engineer different interactions as those in Tab.~\ref{Tab01}.

\begin{table}[t]
\centering
\begin{tabular*} {3.4cm}{llllllllllll}
\hline  
\hline  
Operator \ \ \ &$\tilde{\varphi}_1$ \ \ \ \ \ \  & $\tilde{\varphi}_2$\\ 
\hline
$\sigma_y^1 \sigma_y^2$ & 2$\pi$     &2$\pi$     \\  
$-\sigma_y^1 \sigma_y^2$ & $\pi$     &$\pi$     \\  
$\sigma_x^1 \sigma_x^2$       &$2\pi$& $\pi$    \\   
$-\sigma_x^1 \sigma_x^2$       &$\pi$& $2\pi$    \\   
$\sigma_y^1 \sigma_x^2$       &${1}/{2}\pi$ &  ${3}/{2}\pi$   \\  
$-\sigma_y^1 \sigma_x^2$       &${3}/{2}\pi$ &  ${1}/{2}\pi$   \\  
$\sigma_x^1 \sigma_y^2$       &    ${3}/{2}\pi$ &    ${3}/{2}\pi$ \\ 
$-\sigma_x^1 \sigma_y^2$       &    ${1}/{2}\pi$ &    ${1}/{2}\pi$ \\ 
\hline 
\hline 
\end{tabular*}  
\caption{Interactions produced by different choosing of the phases $\tilde{\varphi}_1$ and $\tilde{\varphi}_1$ in Eq. (\ref{Eq21}).}
\label{Tab01}
\end{table}

\subsection{Three-qubit model}
\begin{figure}[h!]
\centering
\includegraphics[width=0.8\linewidth]{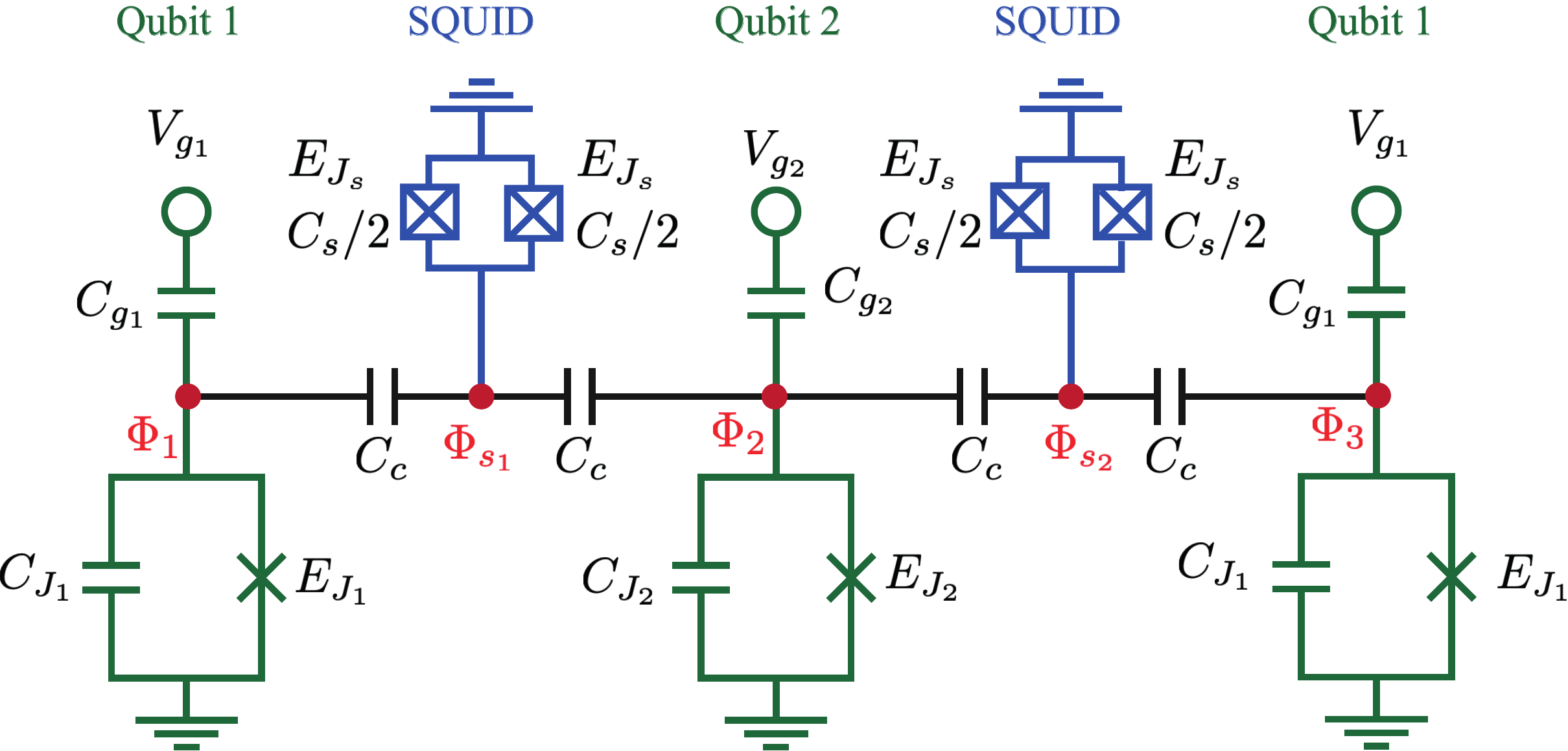}
\caption{Effective circuit diagram of three charge qubits (green) couple through grounded SQUIDs (blue) with Josephson energy $E_{J_s}$ and effective capacitor $C_{s}/2$. Moreover $C_c$ is the coupling capacitance, and $\Phi_1$, $\Phi_2$, $\Phi_3$, $\Phi_{s_1}$ and $\Phi_{s_2}$ are node fluxes that define the degrees of freedom of the qubits and SQUIDs.}
\label{Fig04}
\end{figure}
In the three-qubit model, we consider the circuit given by Fig.~\ref{Fig04}. It is composed of a chain of three charge qubits coupled through grounded SQUIDs. As in the previous case, we consider far off-resonance nearest-neighbor qubits. Following the same procedure of the two-qubits model, we get the next effective Hamiltonian (see appendix~\ref{AppB})
\begin{align}
\hat{\mathcal{H}}=\sum_{\ell=1}^3\frac{\omega_\ell}{2}\sigma_\ell^z+\sum_{j=1}^2\left[g^{(j)}_0+g^{(j)}_1\varphi^{(j)}_{AC}(t)\right]\sigma_j^y\sigma_{j+1}^y,
\label{Eq22}
\end{align}
where $\omega_3=\omega_1=E_{J_1}$, $\omega_2=E_{J_2}$, and the time-dependent signal reads
\begin{eqnarray}
\varphi^{(j)}_{AC}(t)=A^{(j)}_1\cos{(\nu^{(j)}_1 t +\tilde{\varphi}^{(j)}_1)}+A^{(j)}_2\cos{(\nu^{(j)}_2 t +\tilde{\varphi}^{(j)}_2)}.
\end{eqnarray}
Moreover the coupling strength $g^{(j)}_0$ and $g^{(j)}_1$ $(j=\{1,2\})$ are given by
\begin{align}
&g^{(j)}_0=\frac{C_{c}^2E_{J_1}E_{J_2}}{4(C_{1}+C_c)(C_{2}+2C_c)\bar{E}^{(j)}_{J_s}},~g^{(j)}_1=\frac{C_{c}^2E_{J_1}E_{J_2}}{4(C_{1}+C_c)(C_{2}+2C_c)\bar{E}^{(j)}_{J_s}}\frac{\sin{\left({\varphi}^{(j)}_{DC}\right)}}{\cos{\left({\varphi}^{(j)}_{DC}\right)}},~~~~~~~~~
\end{align}
with $\bar{E}^{(j)}_{J_s}=2E_{J_s}\cos{\left({\varphi}^{(j)}_{DC}\right)}$. To visualize the dynamics of the system, we write the Hamiltonian in the interaction picture. After we consider the resonant conditions $\nu^{(1)}_1=\nu^{(2)}_1=\Delta_{12}$ and $\nu^{(1)}_2=\nu^{(2)}_2=\mu_{12}$ and neglect the fast oscillating terms, the Hamiltonian in the interaction picture reads
\begin{align}
\hat{\mathcal{H}}_I&=\hat{\mathcal{H}}^{1,2}_I+\hat{\mathcal{H}}^{2,3}_I
\end{align}
where 
\begin{align}\nonumber
\hat{\mathcal{H}}^{j,j+1}_I&=\frac{g^{(j)}_1}{4}\bigg[\left(A^{(j)}_1\cos{\tilde{\varphi}^{(j)}_1}-A^{(j)}_2\cos{\tilde{\varphi}^{(j)}_2}\right)\sigma_j^x\sigma_{j+1}^x\\\nonumber
&+\left((-1)^jA^{(j)}_1\sin{\tilde{\varphi}^{(j)}_1}-A_2\sin{\tilde{\varphi}^{(j)}_2}\right)\sigma_j^x\sigma_{j+1}^y\\\nonumber
&+\left((-1)^{j+1}A^{(1)}_1\sin{\tilde{\varphi}^{(j)}_1}-A^{(j)}_2\sin{\tilde{\varphi}^{(j)}_2}\right)\sigma_j^y\sigma_{j+1}^x\\
&+\left(A^{(j)}_1\cos{\tilde{\varphi}^{(j)}_1}+A^{(j)}_2\cos{\tilde{\varphi}^{(j)}_2}\right)\sigma_j^y\sigma_{j+1}^y\bigg],
\label{Eq26}
\end{align}
is the interaction Hamiltonian between $j$th and $(j+1)$th qubit. By choosing proper phase parameters, we can engineer different interaction operators between adjacent qubits, like in the previous case.

\begin{center}
\begin{figure*}
\includegraphics[width=0.9\textwidth]{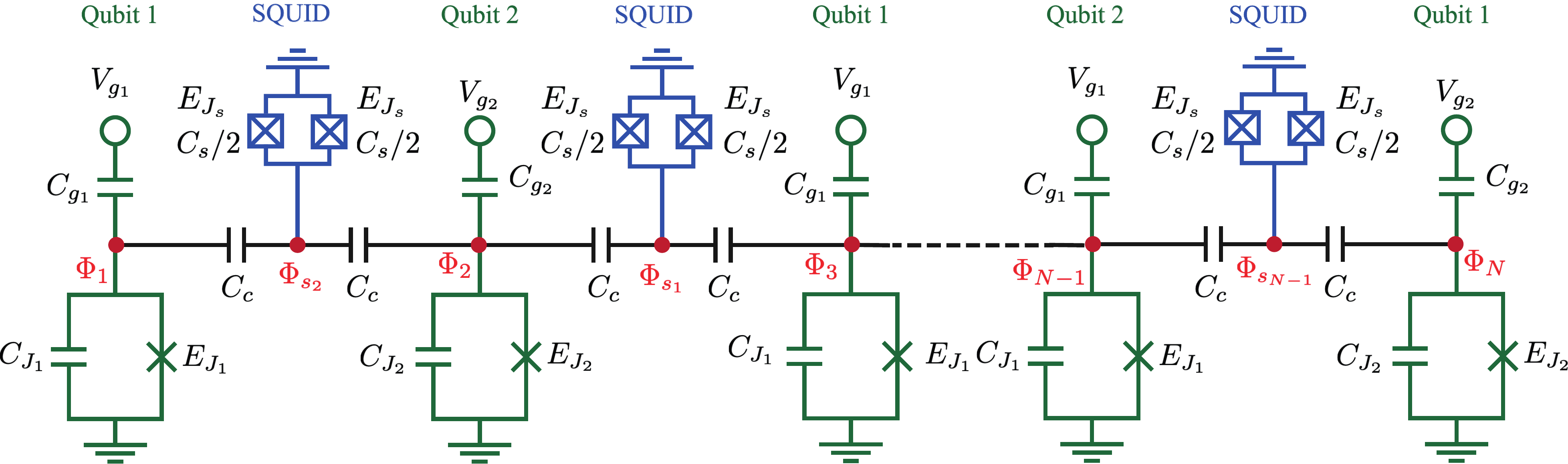}
\caption{General circuit design of a chain of charge qubits (green part) coupling through grounded SQUIDs (blue part) with Josephson energy $E_{J_s}$ and effective capacitor $C_{s}$. $\{\Phi_1,\,\Phi_{s_{1}},\dots,\Phi_{s_{N-1}},\Phi_N\}$ are node fluxes that define the degrees of freedom of the circuit.}
\label{Fig05}
\end{figure*}
\end{center}

It is possible to generalize this expression for a chain of $\ell$ qubits coupled through grounded SQUIDs (see Fig. \ref{Fig05}), where we define the qubits in odd positions as qubit 1 with frequency $\omega_1$ and the qubits in even positions as qubit 2 with frequency $\omega_2$. In the following discussion, we consider the amplitude of the two harmonic signals to be the same, that is $A^{(j)}_1=A^{(j)}_2=A$ and the coupling strength $g^{(j)}_0=g_0$ and $g^{(j)}_1=g_1$.

By considering the resonant conditions $\nu^{(j)}_1=\Delta_{12}$, $\nu^{(j)}_2=\mu_{12}$ and choosing proper phase parameters $\tilde{\varphi}^{(j)}_1$ and $\tilde{\varphi}^{(j)}_2$, we can engineer again a family of interactions between nearest-neighbor qubits as is shown in Tab. \ref{Tab02}. Note that the phase $\tilde{\varphi}^{j}_1$ required to achieve $\pm\sigma_j^x\sigma_{j+1}^y$ and $\pm\sigma_j^y\sigma_{j+1}^x$ are different for odd and even $j$.

\begin{table}[h]
\centering
\begin{tabular*} {5cm}{llllllllllll}
\hline  
\hline  
Operator \ \ \ &$\tilde{\varphi}^{(j)}_1$ \ \ \ \ \ \  & $\tilde{\varphi}^{(j)}_2$\\ 
\hline
$\sigma^y_j \sigma^y_{j+1}$ & 2$\pi$     &2$\pi$     \\  
$-\sigma^y_j \sigma^y_{j+1}$ & $\pi$     &$\pi$     \\  
$\sigma^x_j \sigma^x_{j+1}$       &$2\pi$& $\pi$    \\   
$-\sigma^x_j \sigma^x_{j+1}$       &$\pi$& $2\pi$    \\   
$\sigma^y_j \sigma^x_{j+1}$       &$(1+(-1)^j/2)\pi$ &  ${3}/{2}\pi$   \\  
$-\sigma^y_j \sigma^x_{j+1}$       &$(1+(-1)^{j+1}/2)\pi$&  ${1}/{2}\pi$   \\  
$\sigma^x_j \sigma^y_{j+1}$       &$(1+(-1)^{j+1}/2)\pi$&    ${3}/{2}\pi$ \\ 
$-\sigma^x_j \sigma^y_ {j+1}$       & $(1+(-1)^{j}/2)\pi$ &    ${1}/{2}\pi$ \\ 
\hline 
\hline 
\end{tabular*}  
\caption{Interactions between $j$th qubit and $(j+1)$th qubit produced by different choice of the phases $\tilde{\varphi}^{(j)}_1$ and $\tilde{\varphi}^{(j)}_1$, where we take odd $j$ and even $j$ both into account.}
\label{Tab02}
\end{table}

The controllability and flexibility of the interactions that our proposal offers, give us the possibility to the implement of a large variety of Hamiltonians in an analog way, such as Dzyaloshinskii-Moriya, $XY$, homogeneous and inhomogeneous spin chains. Such analog Hamiltonians could be very useful for DAQS and DAQC, where we can use such analog Hamiltonians like a resource (complex multibody gate)  for simulating more complex systems, like quantum chemistry physics, condensed matter phenomena in spin lattices~\cite{Molnar2019ADP,Rota2017PRB}, and shortcuts to adiabaticity in digitezed adiabatic quantum computing~\cite{Hegade2021PRApplied}. 

The approach we presented in this work is intimately linked to the nature of the Jordan-Wigner mapping that requires the quantum simulation algorithm follows a linear sorting of the lattice sites to reproduce the fermionic anti-commutation relation, being a qubit-chain a natural simulator of the fermion models. Naturally, the search for an experimentally feasible fermion to qubit mapping approaching a two-dimensional lattice beyond the Jordan-Wigner transformation is an open question that deserves further investigation. There is a recent work that proposed a novel mapping in this direction, but the experimental realization is still an open question~\cite{Derby2021PRB}. Also, the use of qubit lattices for the Jordan-Wigner transformation has been proposed; nevertheless, the gates number scaling is the same as that in the qubit-chain case~\cite{Celeri2021arXiv}.

The current proposal could be extended to a two-dimensional (2D) array of charged qubits coupled through grounded SQUIDs generating a more complex family of Hamiltonians, opening the door to more efficient simulations. Nevertheless, for 2D structures, we can have the non-trivial problem of cross-talk between the different SQUID and loops in the circuit, requiring a deep feasibility study which is not the scope of this article. In the next section, we show a particular example about the efficient DAQC of a complex system, the Fermi-Hubbard model. This example will illustrate the versatility of our design.

\section{Digital-Analog Quantum Computation}
\begin{figure}[h!]
	\centering
	\includegraphics[width=0.6\linewidth]{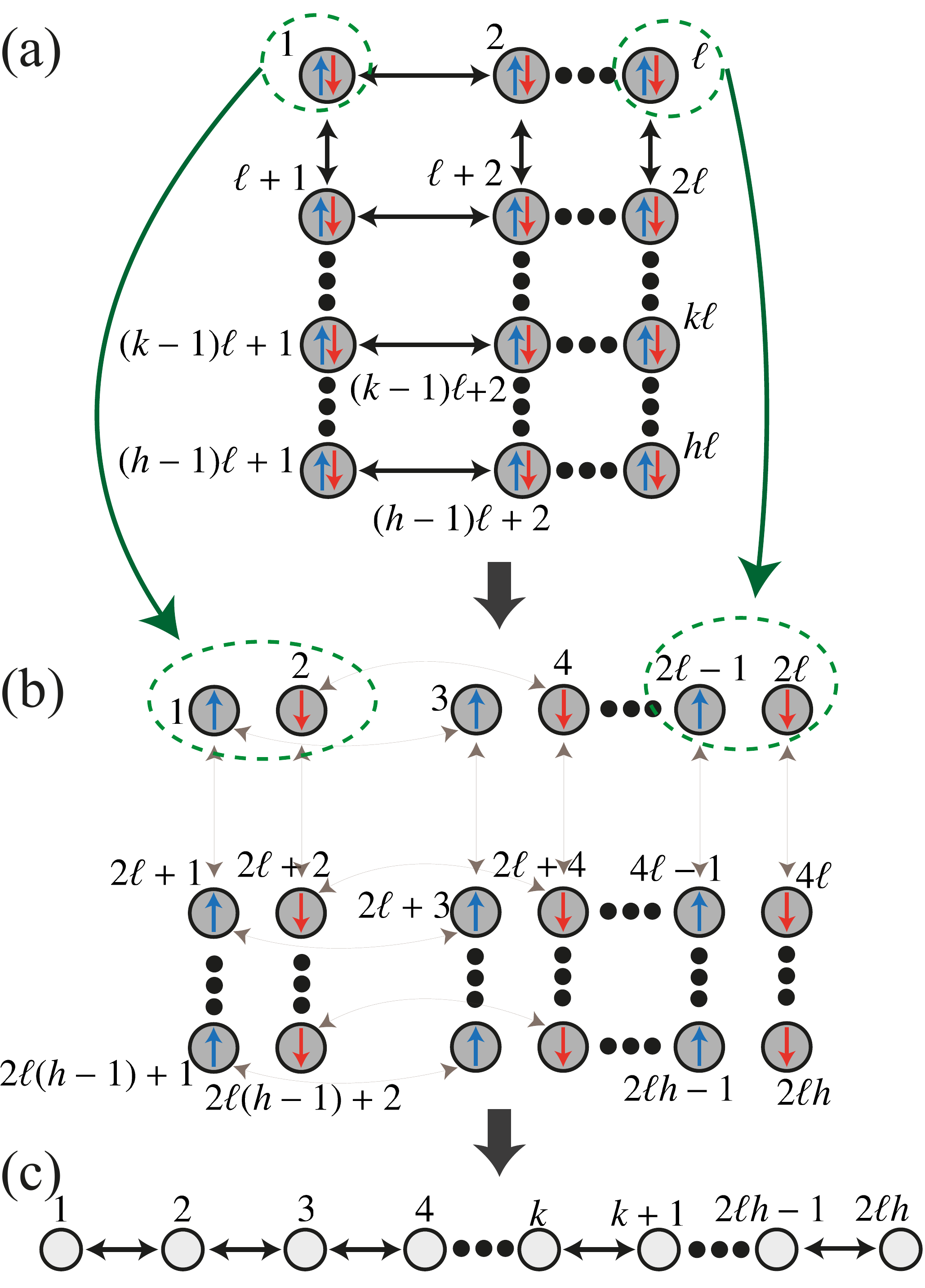}
	\caption{Mapping of the fermionic lattice into a spin chain. (a) $h\times\ell$ lattice where we have and spin-up or spin-down fermion in each site. (b) $h\times2\ell$ lattice where we in each odd site we have and spin-up fermion and in each even site we have a spin-down fermion, it represents the first mapping. (c) Spin chain with $2\ell h$ sites resulting after the Wigner-Jordan Mapping.}
	\label{Fig06}
\end{figure}

Hubbard model represents the interaction between the neighboring sites, which is defined by hopping element and Coulombic interaction on the same site, called on-site interaction~\cite{Tasaki1998JPCM}. In this section, we are interested in the simulation of the hopping terms of a $\ell\times h$ fermion-lattice (with $\ell \leq h$). The Hamiltonian of a $\ell \times h$ fermion-lattice (see Fig.~\ref{Fig06}~(a)) reads
\begin{align}
\mathcal{H}_{\textrm{Hubb}}=\mathcal{A}\sum_{\alpha=\{\uparrow,\downarrow\}}\sum_{\langle j,k\rangle}\big(c^{\dagger}_{j,\alpha}c_{k,\alpha}+c^{\dagger}_{k,\alpha}c_{j,\alpha}\big)+\mathcal{B}\sum_{j}n_{j,\uparrow}n_{j,\downarrow} \, , 
\label{Eq27}
\end{align}
where $c_{j,\alpha}^{\dagger}$ ($c_{j,\alpha}$) are the creation (annihilation) operators of the $j$th site, with the number operator $n_{j,\uparrow(\downarrow)}=c^{\dagger}_{j,\uparrow(\downarrow)}c_{j,\uparrow(\downarrow)}$, and spin-$\alpha$, with $\alpha=\{\uparrow,\downarrow\}$. To suppress the index $\alpha$, we map the $\ell\times h$ lattice to a equivalent $2\ell\times h$ spin-less lattice by
\begin{align}
c_{j,\uparrow}^{\dagger}\rightarrow b_{2j-1}^{\dagger},\quad c_{j,\downarrow}^{\dagger}\rightarrow b_{2j}^{\dagger} \, ,
\label{Eq28}
\end{align}
where $b_{k}^{\dagger}$ are the creation operation over the site $k$ for the lattice given by Fig.~\ref{Fig06}~(b). Using these operators the Hamiltonian Eq.~(\ref{Eq27}) can be rewrite as
\begin{align}
\mathcal{H}_{\textrm{Hubb}}&=\mathcal{A}\Bigg(\sum_{k=0}^{h-1}\Bigg[\sum_{j=1}^{\ell-1}\bigg(b_{2k\ell+2j-1}^{\dagger}b_{2k\ell+2j+1}+b_{2k\ell+2j+1}^{\dagger}b_{2k\ell+2j-1}\bigg)\nonumber\\
&+\sum_{j=1}^{\ell-1}\bigg(b_{2k\ell+2j}^{\dagger}b_{2k\ell+2(j+1)}+b_{2k\ell+2(j+1)}^{\dagger}b_{2k\ell+2j}\bigg)\Bigg]\nonumber\\
&+\sum_{k=0}^{h-2}\sum_{j=1}^{2\ell}\bigg[b_{2k\ell+j}^{\dagger}b_{2(k+1)\ell+j}+b_{2(k+1)\ell+j}^{\dagger}b_{2k\ell+j}\bigg]\Bigg)\nonumber\\
&+\mathcal{B}\Bigg(\sum_{j=1}^{h\ell}b^{\dagger}_{2j-1}b_{2j-1}b^{\dagger}_{2j}b_{2j}\Bigg) \, .
\label{Eq29}
\end{align}
Finally, we map the $2\ell\times h$ fermion lattice to a spin$-1/2$ chain using the Wigner-Jordan transformation (see Fig.~\ref{Fig06}~(c)), where we represent $b_j$ and $b_j^{\dagger}$ as a combination of Pauli matrices
\begin{eqnarray}
b_j=&&\bigg[\prod_{l=1}^{j-1}(-\sigma_{l}^z)\bigg]\sigma_j=(-1)^{j-1}\frac{1}{2}\bigg[\prod_{l=1}^{j-1}\sigma_{l}^z\bigg](\sigma_j^x-i\sigma_j^y),\nonumber\\
b_j^{\dagger}=&&\bigg[\prod_{l=1}^{j-1}(-\sigma_{l}^z)\bigg]\sigma_j^{\dagger}=(-1)^{j-1}\frac{1}{2}\bigg[\prod_{l=1}^{j-1}\sigma_{\ell}^z\bigg](\sigma_j^x+i\sigma_j^y) \, .
\label{Eq30}
\end{eqnarray}
Before to write the equivalent spin chain Hamiltonian, we  define the operator
\begin{align}
U_{(j,k)}^{\alpha,\beta}=e^{-i\frac{\pi}{4}(\sigma_{j-1}^{\alpha}\sigma_j^{\alpha}+\sigma_k^{\beta}\sigma_{k+1}^{\beta})}={U}_{j-1}^{\alpha}{U}_{k}^{\beta},
\label{U}
\end{align}
where $j\ne k$, and ${U}_{j}^{\alpha}=e^{-i\frac{\pi}{4}\sigma_{j}^{\alpha}\sigma_{j+1}^{\alpha}}$.  After some algebraic manipulation, we obtain (for details see appendix~\ref{AppC})
\begin{align}
\mathcal{H}_{\textrm{Hubb}}=\mathcal{H}_{\textrm{hori}}+\mathcal{H}_{\textrm{verti}}+\mathcal{H}_{\textrm{coul}},
\label{Eq32}
\end{align}
where $\mathcal{H}_{\textrm{hori}}$, $\mathcal{H}_{\textrm{verti}}$ and $\mathcal{H}_{\textrm{coul}}$ are given by
\begin{align}\nonumber
\mathcal{H}_{\textrm{hori}}&=\frac{\mathcal{A}}{2}\bigg(\bigg[U_{(1,2)}^{x^\dagger} H_{1,2}^{(x,y)}U_{(1,2)}^{x} \bigg]+\bigg[U_{(1,2)}^{y} H_{1,2}^{(y,x)}U_{(1,2)}^{y^\dagger} \bigg]+\bigg[U_{(1,3)}^{x^\dagger} H_{1,3}^{(x,y)}U_{(1,3)}^{x} \bigg]\\\nonumber
&+\bigg[U_{(1,3)}^{y} H_{1,3}^{(y,x)}U_{(1,3)}^{y^\dagger} \bigg]+\bigg[U_{(2,4)}^{x^\dagger} H_{2,4}^{(x,y)}U_{(2,4)}^{x} \bigg]+\bigg[U_{(2,4)}^{y} H_{2,4}^{(y,x)}U_{(2,4)}^{y^\dagger} \bigg]\\\nonumber
&+\bigg[U_{(2,5)}^{x^\dagger} H_{2,5}^{(x,y)}U_{(2,5)}^{x} \bigg]+\bigg[U_{(2,5)}^{y} H_{2,5}^{(y,x)}U_{(2,5)}^{y^\dagger} \bigg]\bigg),\\\nonumber
\mathcal{H}_{\textrm{verti}}&=\frac{\mathcal{A}}{2}\sum_{j=1}^{2\ell}\bigg[\bigg(\tilde{U}_{j,1}^{(x,x)}\tilde{U}_{j,2}^{(y,y)}...\tilde{U}_{j,\ell-2}^{(x,x)}\tilde{U}_{j,\ell-1}^{(y,y)}\tilde{U}_{j}^{x^\dagger}\bigg) \cdot\Theta_j^{x,y} \bigg(\tilde{U}_{j}^{x}U_{j,\ell-1}^{(y,y)^\dagger}  U_{j,\ell-2}^{(x,x)^\dagger}...U_{j,2}^{(y,y)^\dagger}U_{j,1}^{(x,x)^\dagger}\bigg) \nonumber\\
&+\bigg(U_{j,1}^{(y,y)}U_{j,2}^{(x,x)}...U_{j,\ell-2}^{(y,y)}U_{j,\ell-1}^{(x,x)}U_{j}^{y}\bigg)\cdot\Theta_j^{y,x} \bigg(\tilde{U}_{j}^{y^\dagger} \tilde{U}_{j,\ell-1}^{(x,x)^\dagger}\tilde{U}_{j,\ell-2}^{(y,y)^\dagger}...U_{j,2}^{(x,x)^\dagger}U_{j,1}^{(y,y)^\dagger}\bigg)\bigg], \nonumber\\ 
\mathcal{H}_{\textrm{coul}}&=\frac{\mathcal{B}}{4}\sum_{j=1}^{h\ell}(\sigma_{2j-1}^z+\mathbb{I})(\sigma_{2j}^z+\mathbb{I}),
\label{Eq33}
\end{align}
which correspond to the Hamiltonian of the horizontal hopping, vertical hopping and coulomb interaction respectively, with
\begin{align}\nonumber
U_{(n,i)}^{a}&=\prod_{k=0}^{h-1} \prod_{j=1}^{m_n} U^a_{2k\ell+4(j-1)+i},~H_{n,i}^{(a,b)}=\sum_{k=0}^{h-1}\sum_{j=1}^{m_n}\sigma_{2k\ell+4j-5+i}^a \sigma_{2k\ell+4(j-1)+i}^b,\\\nonumber
\Theta_j^{a,b}&=\sum_{k=0}^{h-2}\sigma_{2k\ell+j+\ell-1}^a\sigma_{2k\ell+j+\ell}^b,~\tilde{U}_{j,i}^{(a,a)}=\prod_{k=0}^{h-2}U_{2k\ell+j+i,2k\ell+j+2\ell-i}^{(a,a)},\\
\tilde{U}_{j}^{a}&=\prod_{k=0}^{h-2}U_{2k\ell+j+\ell}^{a} \, .
\label{Eq34}
\end{align}
and
\begin{align}
m_1=\frac{2\ell-1-(-1)^{\ell+1}}{4},\quad
m_2=\frac{2\ell-3+(-1)^{\ell+1}}{4} \, .
\label{Eq35}
\end{align}

\begin{figure}[t!]
\centering
\includegraphics[width=0.5\linewidth]{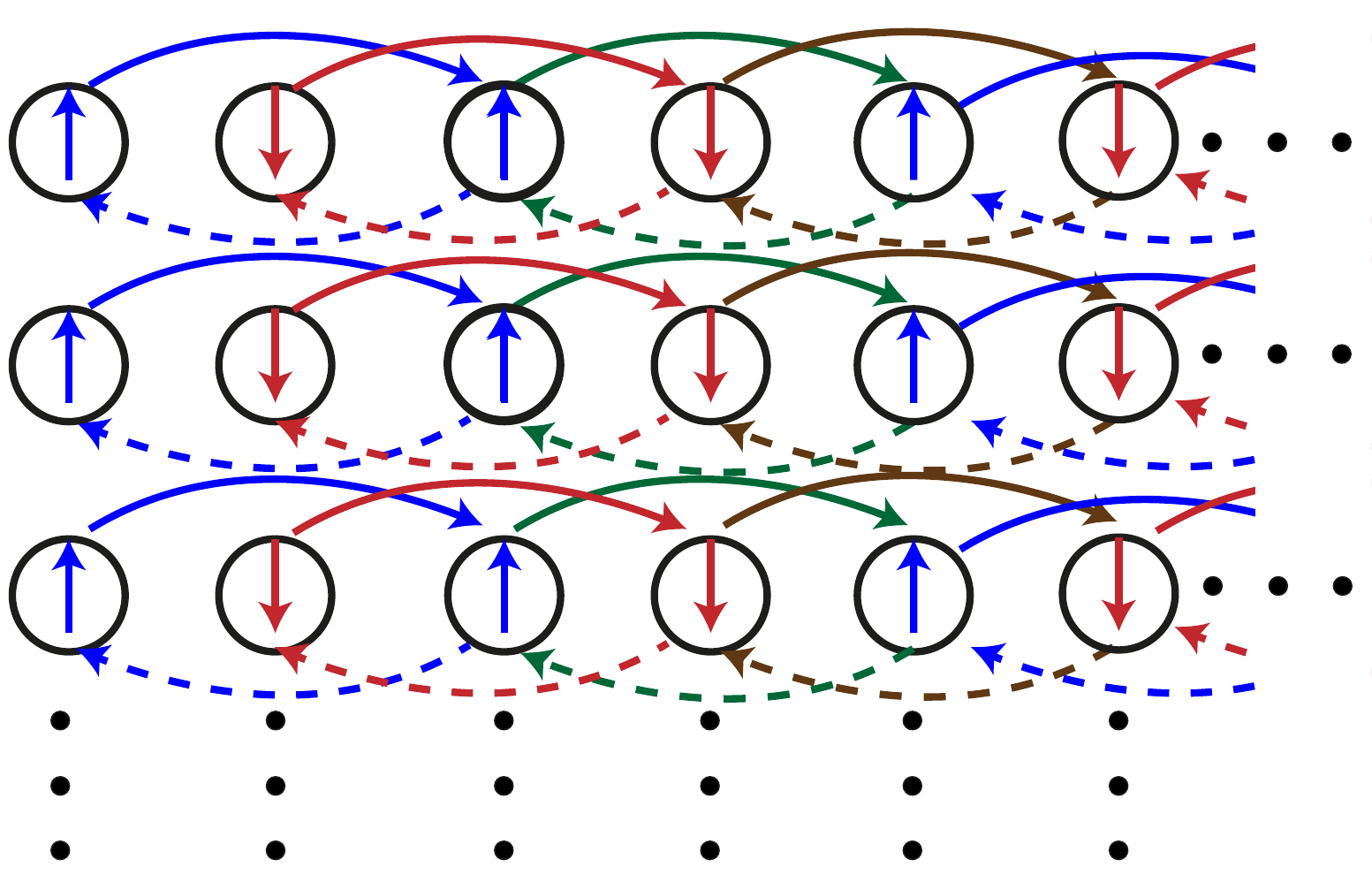}
\caption{Diagram for the different horizontal hopping interactions. Different colors show the set of interactions that can be performed at the same time in an analog way. Solid arrows are for forward hopping and dashed arrows for backward hopping.}
\label{Fig07}
\end{figure}

We notice that each term of $\mathcal{H}_{\textrm{hori}}$ in Eq.~(\ref{Eq33}) correspond to the set of horizontal hopping interactions which do not share sites in the lattice, these eight interactions are represented by arrows with different colors in Fig.~$\ref{Fig07}$ (blue, red, green and brown), and different textures (solid and dashed). Also, we highlight that each interactions given by Eq.~(\ref{Eq34}) can be performed in an analog way since the sub-gates involved are applied in different qubits and can be done together without interfering with each other (see appendix~\ref{AppC}), {giving us the analog resource for the digital-analog simulation}. The number $m_1$ is the number of the hopping terms corresponding to the blue(red) solid/dash arrows, and $m_2$ is the number of the hopping terms corresponding to the green(brown) solid/dash arrows (see Fig. ~\ref{Fig07}), where $m_1+m_2=\ell-1$.

Now, we approximate the time evolution of our system using the Trotter expansion~\cite{Suzuki1990JP} as follows
\begin{align}
e^{-i\mathcal{H}_{\textrm{Hubb}}t}&\approx \left[e^{-i\mathcal{H}_{\textrm{hori}}t/n}e^{-i\mathcal{H}_{\textrm{verti}}t/n}e^{-i\mathcal{H}_{\textrm{coul}}t/n}\right]^n\nonumber\\
&= \left[U_{\textrm{hori}}(t/n)U_{\textrm{verti}}(t/n)U_{\textrm{coul}}(t/n)\right]^n
\label{Eq36}
\end{align}
where $U_{\textrm{word}}(t)=e^{-i\mathcal{H}_{\textrm{word}}t}$. Using Eq.~(\ref{Eq33}) we have
\begin{align}
U_{\textrm{hori}}(t/n)&\approx U_{(1,2)}^{x^\dagger}e^{-i\frac{\mathcal{A}t}{2n}H_{1,2}^{(x,y)}}U_{(1,2)}^{x}\cdot U_{(1,2)}^{y} e^{-i\frac{\mathcal{A}t}{2n}H_{1,2}^{(y,x)}}U_{(1,2)}^{y^\dagger}\nonumber\\
&\cdot U_{(1,3)}^{x^\dagger}e^{-i\frac{\mathcal{A}t}{2n}H_{1,3}^{(x,y)}}U_{(1,3)}^{x}\cdot U_{(1,3)}^{y} e^{-i\frac{\mathcal{A}t}{2n}H_{1,3}^{(y,x)}}U_{(1,3)}^{y^\dagger}\nonumber\\
&\cdot U_{(2,4)}^{x^\dagger}e^{-i\frac{\mathcal{A}t}{2n}H_{2,4}^{(x,y)}}U_{(2,4)}^{x}\cdot U_{(2,4)}^{y} e^{-i\frac{\mathcal{A}t}{2n}H_{2,4}^{(y,x)}}U_{(2,4)}^{y^\dagger}\nonumber\\
&\cdot U_{(2,5)}^{x^\dagger}e^{-i\frac{\mathcal{A}t}{2n}H_{2,5}^{(x,y)}}U_{(2,5)}^{x}\cdot U_{(2,5)}^{y} e^{-i\frac{\mathcal{A}t}{2n}H_{2,5}^{(y,x)}}U_{(2,5)}^{y^\dagger}\nonumber\\
U_{\textrm{verti}}(t/n)&\approx \prod_{j=1}^{2\ell}\bigg[\bigg(\tilde{U}_{j,1}^{(x,x)}\tilde{U}_{j,2}^{(y,y)}...\tilde{U}_{j,\ell-2}^{(x,x)}\tilde{U}_{j,\ell-1}^{(y,y)}\tilde{U}_{j}^{x^\dagger}\bigg) \nonumber \\
&\cdot e^{-i\frac{\mathcal{A}t}{2n}\Theta_j^{x,y}} \bigg(\tilde{U}_{j}^{x}U_{j,\ell-1}^{(y,y)^\dagger}  U_{j,\ell-2}^{(x,x)^\dagger}...U_{j,2}^{(y,y)^\dagger}U_{j,1}^{(x,x)^\dagger}\bigg) \nonumber\\
&\cdot \bigg(U_{j,1}^{(y,y)}U_{j,2}^{(x,x)}...U_{j,\ell-2}^{(y,y)}U_{j,\ell-1}^{(x,x)}U_{j}^{y}\bigg)\nonumber\\
&\cdot e^{-i\frac{\mathcal{A}t}{2n}\Theta_j^{y,x}} \bigg(\tilde{U}_{j}^{y^\dagger} \tilde{U}_{j,\ell-1}^{(x,x)^\dagger}\tilde{U}_{j,\ell-2}^{(y,y)^\dagger}...U_{j,2}^{(x,x)^\dagger}U_{j,1}^{(y,y)^\dagger}\bigg)\bigg], \nonumber\\ 
U_{\textrm{coul}}(t/n)&\approx \prod_{j=1}^{h\ell}e^{-i\frac{\mathcal{B}t}{4n}(\sigma_{2j-1}^z+\mathbb{I})(\sigma_{2j}^z+\mathbb{I})}\nonumber\\
&=\prod_{j=1}^{h\ell}e^{-i\frac{\mathcal{B}t}{4n}(\sigma_{2j-1}^z\sigma_{2j}^z)}e^{-i\frac{\mathcal{B}t}{4n}(\sigma_{2j-1}^z+\sigma_{2j}^z)} \, .
\label{Eq37}
\end{align}

First, for $U_{\textrm{hori}}$ we simulate eight types of interactions (see Fig.~\ref{Fig07} and Eq.~(\ref{Eq33})), each of these interactions need three gates, as we mention above each of this gate can be simulated in an analog way. Therefore, to simulate $U_{\textrm{hori}}$ (the horizontal hopping) we need $24$ gates, as we can see in Eq.~(\ref{Eq34}).  
\begin{figure}[t!]
\centering
\includegraphics[width=0.5\linewidth]{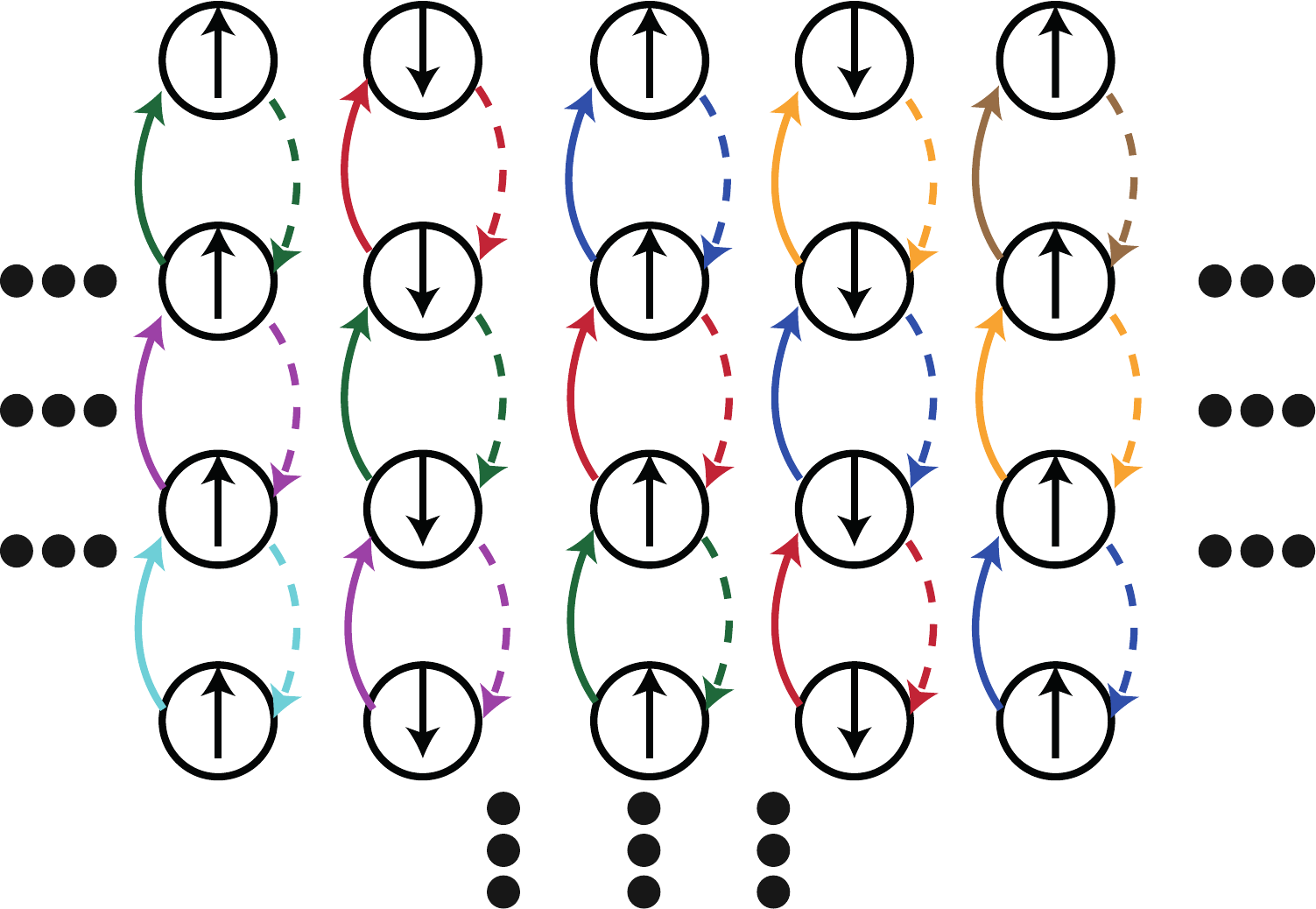}
\caption{Diagram for the different vertical hopping interactions. Different colors show the set of interactions that can be performed at the same time in an analog way. Solid arrows are for forward hopping and dashed arrows for backward hopping.}
\label{Fig08}
\end{figure}

Second, for $U_{\textrm{verti}}$, \ie all the vertical hopping terms, we need $2(2\ell+1)$ types of interactions, they are shown in Fig.~\ref{Fig08} with different colors and different textures. We note that all the interactions of the same type can be performed at the same time, due to they do not share sites during its implementation, \ie all the interactions with the same color and same texture can be simulated in parallel. Now, to simulate each type of interaction we need $2\ell+1$ analog gates (see Eqs.~(\ref{Eq33}) and (\ref{Eq34})). Then, to simulate all the vertical hopping terms, we need $2(2\ell+1)^2$ analog gates. Therefore to simulate all the hopping terms in a $\ell\times h$ fermion lattice with $\ell\le h$ we need only $2(2\ell+1)^2+24=8\ell^2+8\ell+26$ gates.

 {It is necessary to highlight that to our knowledge, the more efficient proposal for the quantum simulation of the Hubbard model is a trapped ion one~\cite{Lamata2014EPJ}, which uses a multi-body entangling gate, needing $8(2h\ell-\ell-h)+20$ gates for a $\ell\times h$ fermion lattice. It means that for a square lattice $(\ell=h)$, the trapped ion proposal needs $2(8\ell^2-8\ell+10)$ gates, which for large $\ell$ need almost two times more gates than our proposal. It means that even if superconducting circuits cannot use multi-body entangling gates like trapped ions, it still is useful by a suitable design as in this work.}

On the other hand, $U_{\textrm{coul}}(t/n)$, correspond to the free energy of the original Hubbard model (Eq.~(\ref{Eq27})) and we will not consider for the Hopping dynamics simulation. Nevertheless, it can be simulated using three gates, \ie the analog interaction $\sum_j\sigma_{2j-1}^x\sigma_{2j}^x$ plus two rotations in the $y$-axis (the local terms $\sigma_z^{j}$ can be mapped correspond to the free energy of our simulator).

The simulation time can be easily derivate as follow. Each type of interaction involve unitary gate of the form $\mathcal{U}_a=\textrm{exp}(-\frac{i\mathcal{A}}{2}\frac{t}{n}\hat{O})$ and of the form $\mathcal{U}_b=\textrm{exp}(-i\frac{\pi}{4}\hat{O})$. From Eq.~(\ref{Eq37}) we obtain that the time for each kind of gates ($a$ and $b$) is
\begin{align}
 \tau_a=\frac{\mathcal{A}t}{Ag_1n},\quad  \tau_b=\frac{\pi}{2Ag_1} \, ,
 \label{Eq38}
\end{align}
respectively. From the simulation of $U_{\textrm{hori}}$ we have $8$ gates of the class $a$, and from $U_{\textrm{verti}}$ we have $2(2\ell+1)$ gates of the class $a$. Then the time necessary to perform all these gates is $(4\ell+10)\tau_a$. As we have a total of $2(2\ell+1)^2+24$ gates, the number of type $b$ gates is $8\ell^2+4\ell+16$. Therefore, the total time for the simulation is
\begin{align}
\tau_{\textrm{sim}}&=(4\ell+10)\tau_a + (8\ell^2+4\ell+16)\tau_b=(4\ell+10)\frac{\mathcal{A}t}{Ag_1n}+(4\ell^2+2\ell+8)\frac{\pi}{Ag_1} \, .
\label{Eq39}
\end{align}

We note that for the case $\ell=2$, we need fewer gates, in particular, to simulate $U_{\textrm{hori}}$ we need half of the gates, it means $12$, for this case, the simulation time also decreases and is given by
\begin{align}
\tau_{\textrm{sim}}^{*}=(4\ell+6)\tau_a + (8\ell^2+4\ell+8)\tau_b=14\frac{\mathcal{A}t}{Ag_1n}+24\frac{\pi}{Ag_1} \, .
\label{Eq40}
\end{align}
{Finally, the character digital-analog of our simulation is given by the use of analog gates in each digital step, it means gates that act over several qubits simultaneously.} In the next section, we present the numerical results of a quantum simulation of the hopping interaction of a $2\times 3$ fermion-lattice. 

\section{Numerical results: $2\times3$ fermion lattice}
As we mention above, for the case of $\ell=2$ we only need $12$ gates to simulate $U_{\textrm{hori}}$, then for $2\times h$ lattice, we need $62$ gates per Trotter step

\begin{figure}[h]
\centering
\includegraphics[width=0.9\linewidth]{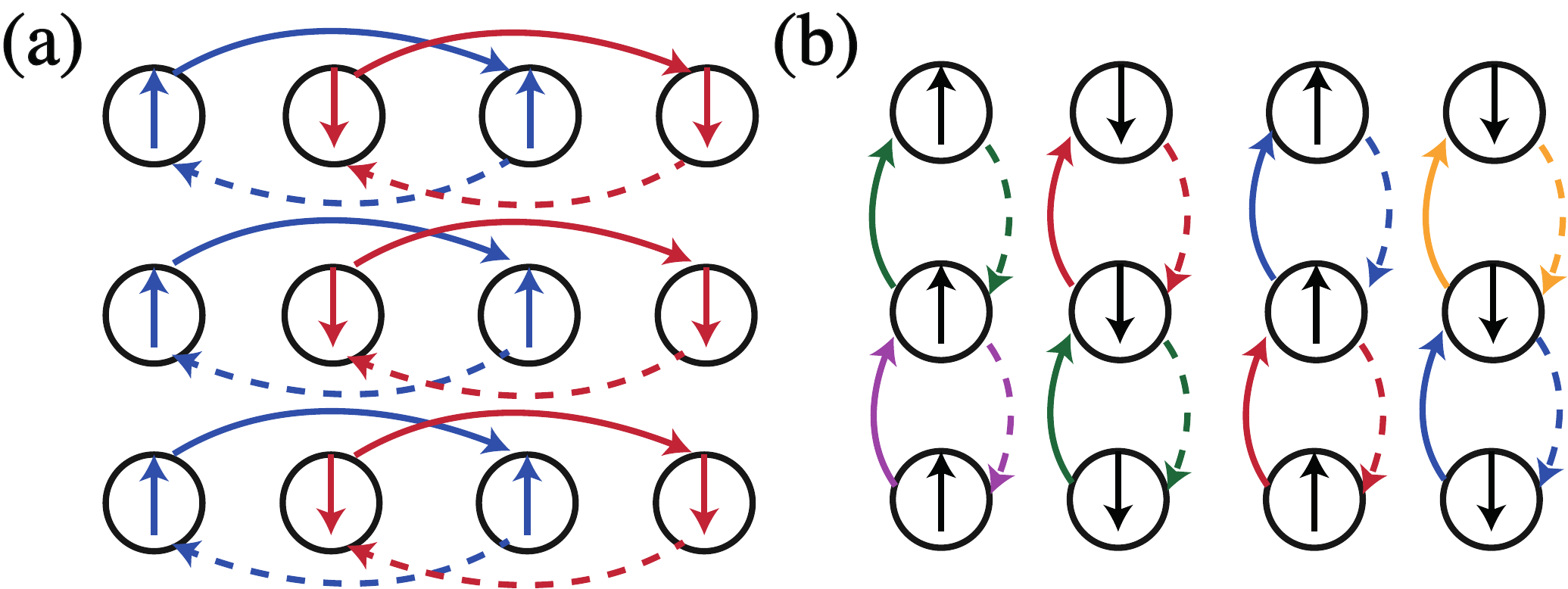}
\caption{Diagram for the different hopping interactions in a $2\times3$ fermion lattice, with the solid arrows to be forward hopping and dashed arrows to be backward hopping (a) Horizontal hopping. (b) Vertical hopping.}
\label{Fig09}
\end{figure}

Figure~\ref{Fig09} shows the types of hopping interactions to simulate for a $2\times 3$ fermion-lattice. In Fig.~\ref{Fig09} (a), we can see the four interactions to describe the horizontal hopping, where the solid arrows correspond to the forward hopping, and the dashed arrows correspond to the backward hopping. For vertical hopping, it requires ten types of interactions, as is shown in Fig.~\ref{Fig09} (b). The sequence of the gates for different hopping interactions is shown in appendix~\ref{AppD} and appendix~\ref{AppE}. 

\begin{figure}[h]
\centering
\includegraphics[width=0.7\linewidth]{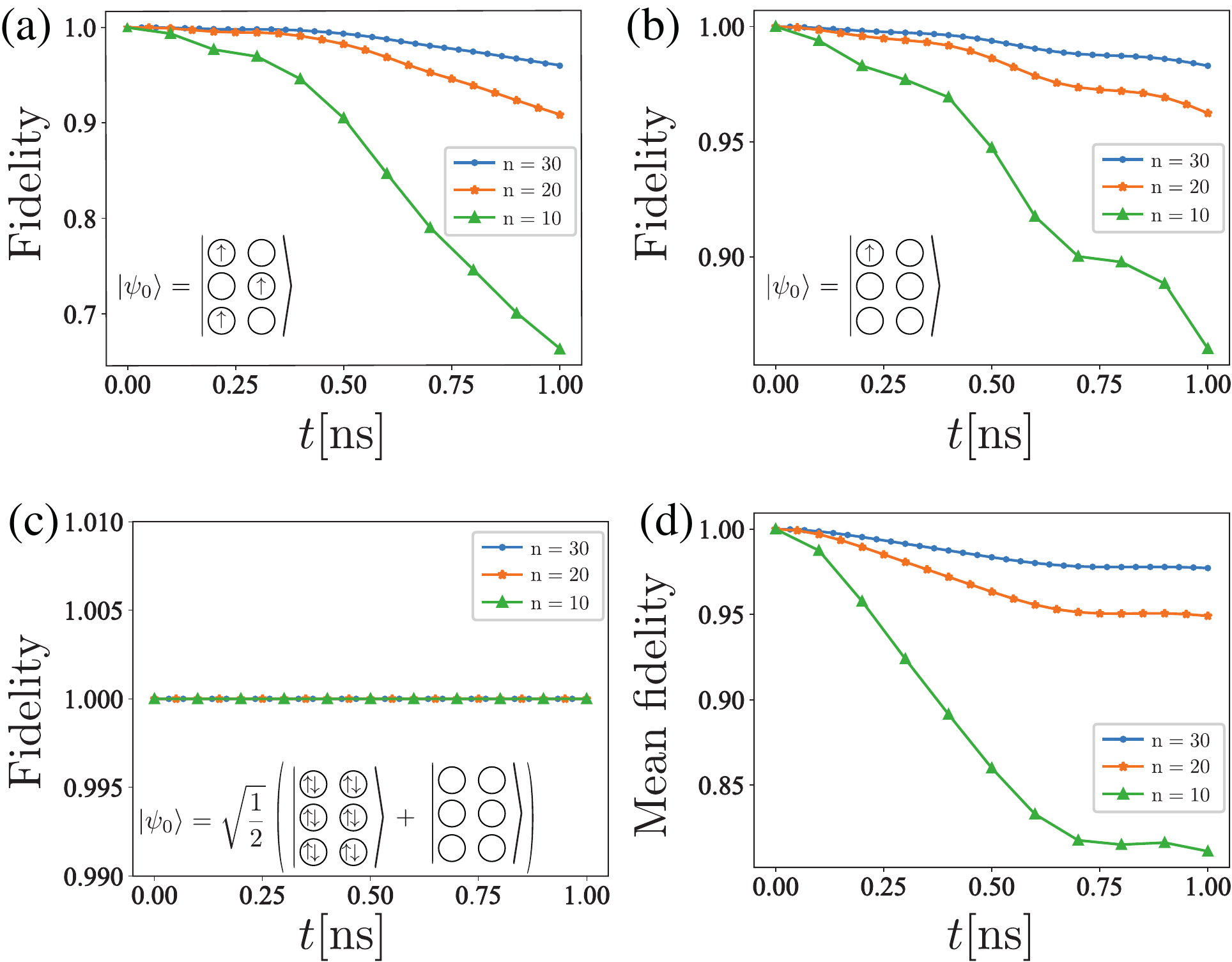}
\caption{(a, b, c) Fidelity between the perfect evolution and DAQC for different numbers of Trotter steps and different initial states$|\psi_0\rangle$ (shown in each subfigure). (d) Mean fidelity for 1000 random initial states for different Trotter steps. The physical parameters of the fermion model are shown Tab. \ref{Tab03}.}
\label{Fig10}
\end{figure}

For the simulation, we map the $2\times 3$ fermion lattice into a $12$ qubit chain described by the Hamiltonian Eq.~(\ref{Eq26}).  The parameters that we consider for the simulation are summarized in the table \ref{Tab03}, where $t$ represent the simulated time (evolution time of the system to be simulated).  Figure~\ref{Fig10}  show the fidelity $|\langle \psi_{\textrm{sim}}(t)|\psi(t)\rangle|^2$ of our simulation for different initial states and $10$, $20$ and $30$ Trotter steps, where $\ket{\psi}(t)$ is the state at time $t$ of the real model, and $\ket{\psi_{\textrm{sim}}(t)}$ is the state given by the simulation, which simulate the evolution at a time $t$. If we think the fermion lattice as a $3\times2$  matrix, where each element can be $\uparrow$, $\downarrow$ or vacuum, the initials states are: Fig.~\ref{Fig10}~(a), $\uparrow$ for the sites $(1,1)$, $(2,2)$ and $(3,1)$, and the rest in vacuum; Fig.~\ref{Fig10}~(b), $\uparrow$ for the sites $(1,1)$ and the rest in vacuum; Fig.~\ref{Fig10}~(c), quantum superposition between the state $\uparrow\downarrow$ for all sites and vacuum for all sites. Finally Fig.~\ref{Fig10}~(d), shows the mean fidelity over $1000$ random states. 

\begin{table}[h!]
\centering
\begin{tabular*} {3.8cm}{llll}
\hline  
\hline  
Fermion model &\\ 
\hline
$\mathcal{A}t$ &$4$  \\  
\hline
cQED model &\\ 
\hline
$\omega_1/2\pi$ &$9(\textrm{GHz})$  \\  
$\omega_2/2\pi$ &$1(\textrm{GHz})$  \\  
$g_0/2\pi$ &$0.05(\textrm{GHz})$  \\ 
$Ag_1/2\pi$ &$0.08(\textrm{GHz})$  \\    
\hline 
\hline 
\end{tabular*}  
\caption{Parameters for Fermion model of Eq. (\ref{Eq15}).}
\label{Tab03}
\end{table}

Table \ref{Tab04} summarize the time $\tau_a$, $\tau_b$ and $\tau_{\textrm{sim}}^*$ defining in Eqs.~(\ref{Eq34}) and (\ref{Eq36}), for the different number of Trotter steps. We can note that all the simulation times are below the $0.2$ [$\mu s$], which means that the simulation can be implemented with the current technology, where the coherent times for superconducting qubits are in the order of $100~[\mu s]$~\cite{Place2021}.

 \begin{table}[h!]
\centering
\begin{tabular*} {4.8cm}{llll}
\hline  
\hline  
n \ \ \ &$\tau_a$[ns] \ \ \ &$\tau_b$[ns] \ \ \ &$\tau_{\textrm{sim}}^*$[$\mu$s]\\ 
\hline
$10$ &0.796  &3.125  &0.161  \\  
$20$ &0.398  &3.125 &0.156  \\  
$30$ &0.265 & 3.125 &0.153   \\   
\hline 
\hline 
\end{tabular*}  
\caption{Times involved in the fermion lattice simulations for different number of Trotter steps. The corresponding parameters of the fermion model are shown in Tab. \ref{Tab03}.}
\label{Tab04}
\end{table}

\section{Conclusion}
We have designed a superconducting circuit architecture suitable for DAQC, in the sense of providing a wide family of analog Hamiltonians as source of analog blocks and more flexibility for this pragmatic quantum computing paradigm. We test our design by the numerical calculation of the quantum simulation of a $3\times 2$ fermion lattice described by the Hubbard model. We find that for a $\ell\times h$ lattice ($\ell\le h$), we need $2(2\ell+1)^2+24$ analog blocks, which depend only on one of the dimensions of the lattice and improves. To our knowledge, this result would improve previous achievements for simulating the Hubbard model by a factor $2$ for large square lattice ($\ell = h$) \cite{Lamata2014EPJ}. Moreover, the total simulation time for $30$ Trotter steps is less than $0.2\,\mu s$ with an ideal fidelity around $0.97$ (only digital error), which makes our proposal experimentally feasible. Finally, we consider this work provides an important boost to the DAQC paradigm, paving the way for computing and simulating complex systems in quantum platforms, while approaching us to useful quantum advantage with fewer algorithmic and hardware resources.

\section*{Appendix}
\appendix
\section{Simplified two-qubit Hamiltonian}
\label{AppA}

\begin{figure}[t!]
\centering
\includegraphics[width=0.9\linewidth]{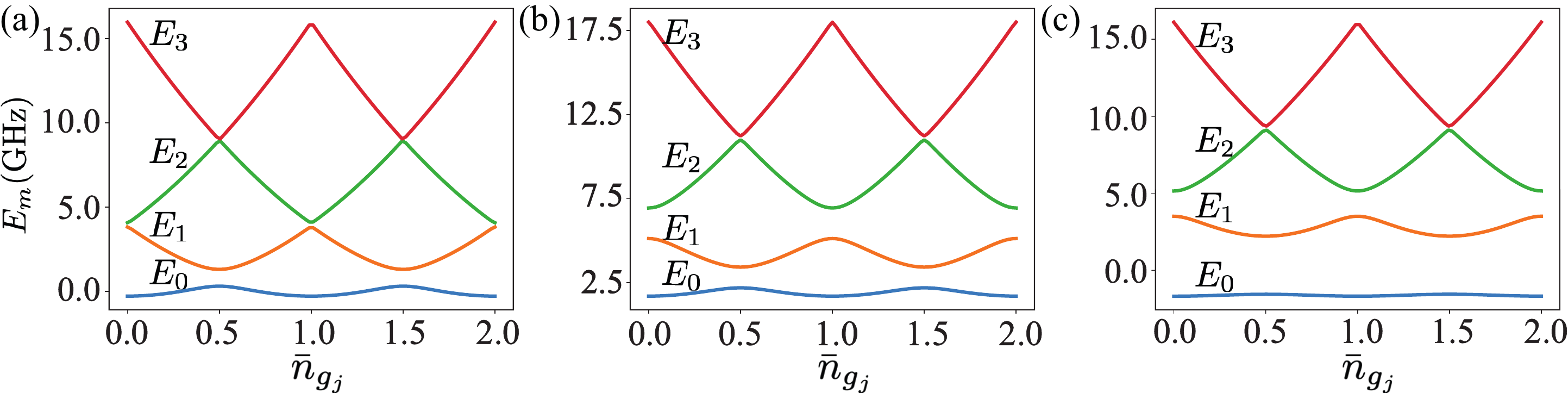}
\caption{Eigenvalue $E_m$(m=0, 1, 2, 3) of $\hat{\mathcal{H}}_{sub}^j$ in Eq.~(\ref{EqA12}) as a function of offset charge $\bar{n}_{g_j}$ for different ratio $E_{J_j}/E_{C_j}$. (a) $E_{J_j}/E_{C_j}=1$, and $\gamma_j(\varphi_{ext})/E_{C_j}=1$. (b) $E_{J_j}/E_{C_j}=1$, and $\gamma_j(\varphi_{ext})/E_{C_j}=4$. (c) $E_{J_j}/E_{C_j}=4$, and $\gamma_j(\varphi_{ext})/E_{C_j}=1$.}
\label{Fig11}
\end{figure}

For completeness, we describe the derivation of the simplified Hamiltonian of the two-qubit system, which consists of two charge qubits coupled through a grounded SQUID as shown in Fig.~\ref{Fig01} of the main text. The Lagrangian of the circuit reads
\begin{align}\nonumber
\mathcal{L}&=\sum_{j=1}^2\bigg[\frac{C_{g_j}}{2}(\Phi^{\prime}_j-V_{g_j})^2+\frac{C_{J_j}}{2}\Phi_1^{\prime 2}+E_{J_j}\cos{(\varphi_j)}\bigg]+E_{J_s}^{\textrm{eff}}\cos{(\varphi_s)}+\frac{C_{s}}{2}{\Phi}_s^{\prime 2}\\
&+\frac{C_{c}}{2}(\Phi^{\prime}_1-\Phi^{\prime}_s)^2+\frac{C_{c}}{2}(\Phi^{\prime}_s-\Phi^{\prime}_2)^2,
\label{EqA1}
\end{align}
where $\varphi_j=2\pi\Phi_j/\Phi_0$, with $\Phi_0=h/2e$ is the superconducting flux quantum and $2e$ is the electrical charge of a Cooper pair. Moreover $E_{J_s}^{\textrm{eff}}=2E_{J_s}\cos{(\varphi_{ext})}$ is the effective Josephson energy of the SQUID. Now, we calculate the conjugate momenta (node charge) $Q_{j}=\partial L/\partial  \Phi^{\prime}_{j}$
\begin{align}\nonumber
Q_{1(2)}&=C_{g_{1(2)}}(\Phi^{\prime}_{1(2)}-V_{g_{1(2)}})+C_{J_{1(2)}}\Phi^{\prime}_{1(2)}+C_c(\Phi^{\prime}_{1(2)}-\Phi^{\prime}_s),\\
Q_s&=C_s\Phi^{\prime}_s-C_{c}(\Phi^{\prime}_1-\Phi^{\prime}_s)-C_c(\Phi^{\prime}_2-\Phi^{\prime}_s).
\label{EqA2}
\end{align}
By applying the Legendre transformation $\mathcal{H}(\Phi_k,Q_k)=\sum_k Q_k\Phi^{\prime}_k-\mathcal{L}$, we obtain the Hamiltonian
\begin{align}\nonumber
\mathcal{H}&=\sum_{j=1}^2\bigg[\frac{1}{2\tilde{C}_{J_j}}(Q_j-2e\tilde{n}_{g_j})^2-E_{J_j}\cos{(\varphi_j)}\bigg]+\frac{1}{2\tilde{C}_{J_s}}(Q_s-2e\tilde{n}_{g_2})^2-E_{J_s}^{\textrm{eff}}\cos{(\varphi_2)}\\
&+g_{12}Q_1 Q_2+g_{1s}Q_1 Q_s+g_{2s}Q_2 Q_s,
\label{EqA3}
\end{align}
where the effective Josephson capacitances, the gate-charge numbers and the coupling strength are defined as follows
\begin{align}\nonumber
&\tilde{C}_{J_{1(2)}}=\frac{C_\star^3}{C_{2(1)}(2C_c+C_s)+C_c(C_c+C_s )},\quad \tilde{C}_{J_s}=\frac{C_\star^3}{(C_c+C_1)(C_c+C_2)},\\\nonumber
&\tilde{n}_{g_{1(2)}}=-\frac{C_{g_{1(2)}}}{2e}V_{g_{1(2)}}-\frac{\tilde{C}_{J_{1(2)}}C_c^2C_{g_{2(1)}}}{2eC_\star^3}V_{g_{2(1)}},\\\nonumber
& \tilde{n}_{g_s}=\frac{\tilde{C}_{J_s}C_c}{2eC_\star^3}\bigg(C_{g_1}(C_{2} + C_c )V_{g_1}+C_{g_2}(C_{1} + C_c )V_{g_2}\bigg),\\ &g_{12}=\frac{C^2_c}{C^3_\star}, \quad g_{1s(2s)}=\frac{C_c(C_{2(1)} + C_c )}{C_\star^3},
\end{align}
with $C_j=C_{g_j}+C_{J_j}$ ($j=\{1,2\}$), and $C_\star^3=C_c(C_1+C_2)(C_s+C_c)+C_c^2C_s+C_1C_2(2C_c+C_s)$.
In the following discussion, we calculate the simplified Hamiltonian between the two CPBs by applying the two approximations $\Phi^{\prime}_s\ll\Phi^{\prime}_{1(2)}$ ($\Phi^{\prime\prime}_s\ll\Phi^{\prime\prime}_{1(2)}$), and $\Phi_s\ll\Phi_{1(2)}$ \ie we consider the SQUID in phase regime with high plasma frequency, and meanwhile the low impedance. With the first approximation $\Phi^{\prime}_s\ll\Phi^{\prime}_{1(2)}$, we can neglect the terms proportional to $\Phi^{\prime}_s$ in Eq.~(\ref{EqA2}) obtaining the relation between nodes charge as follows
\begin{equation}
Q_s=-C_{c}\left(\frac{Q_{1}+C_{g_1}V_{g_1}}{C_{1}+C_c}+\frac{Q_{2}+C_{g_2}V_{g_2}}{C_{2}+C_c}\right).
\label{EqA5}
\end{equation}
Next, we derive the relation between the nodes flux, by calculating the Euler-Lagrange (E-L) equations ${\partial}L/{\partial\Phi_j}-{d({\partial L}/{\partial \Phi^{\prime}_j})}/{dt}=0$, which govern the dynamics of our system
\begin{align}\nonumber
&\left(C_{1(2)}+C_c\right)\Phi^{\prime\prime}_{1(2)}-C_c\Phi^{\prime\prime}_s+\frac{2\pi}{\Phi_0}E_{J_{1(2)}}\sin{(\varphi_{1(2)})}=0,\\
&-C_{c}\Phi^{\prime\prime}_1-C_{c}\Phi^{\prime\prime}_2+2C_s\Phi^{\prime\prime}_s+\frac{2\pi E_{J_s}^{\textrm{eff}}}{\Phi_0}\sin{(\varphi_s)}=0,
\label{EqA6}
\end{align}
and by applying the two approximations, we can neglect the terms proportional to $\Phi^{\prime\prime}_s$ in Eq.~(\ref{EqA6}), and meanwhile approximate $\sin{\varphi_s}=\varphi_s$ obtaining
\begin{eqnarray}
{\varphi_s}=-\frac{C_cE_{J_1}\sin{(\varphi_1)}}{E_{J_s}^{\textrm{eff}}(C_1+C_c)}-\frac{C_cE_{J_2}\sin{(\varphi_2)}}{E_{J_s}^{\textrm{eff}}(C_2+C_c)}.
\label{EqA7}
\end{eqnarray}
Moreover, with the approximation $\Phi_s\ll \Phi_{1(2)}$, we approximate $\cos{\varphi_s}\approx(1-{\varphi_s^2/2})$, with which we can keep the potential energy of the SQUID up to the second-order, and by replacing Eq.~(\ref{EqA5}), and Eq.~(\ref{EqA7}) in the Hamiltonian in Eq.~(\ref{EqA3}), we obtain
\begin{align}\nonumber
\mathcal{H}&=\sum_{j=1}^2\bigg(\frac{1}{2\bar{C}_{J_j}}(Q_j-2e\bar{n}_{g_j})^2-E_{J_j}\cos{(\varphi_j)}\\
&+\gamma_{j}(\varphi_{ext})\sin{(\varphi_j)}^2\bigg)+\gamma_{12}(\varphi_{ext})\sin{(\varphi_1)}\sin{(\varphi_2)},
\label{EqA8}
\end{align}
where the effective Josephson capacitance and gate-charge number
\begin{align}\nonumber
&\bar{C}_{J_{1(2)}}=\tilde{C}_{J_{1(2)}}+\frac{Cc^2}{C_s+C_c+\frac{C_cC_{2(1)}}{C_c+C_{2(1)}}}=C_{1(2)}+C_c,\\
& \bar{n}_{g_{1(2)}}=\tilde{n}_{g_{1(2)}}+\frac{C_{g_{2(1)}}V_{g_{2(1)}}}{2e\bigg(\frac{2C_{2(1)}}{C_c}+\frac{C_{2(1)}C_s}{C_c^2}+\frac{C_s}{C_c}+1\bigg)}=-\frac{{C}_{g_{1(2)}}V_{g_{1(2)}}}{2e},
\end{align}
which shows that the replacement of the $Q_s$ in terms of $Q_{1(2)}$ in the Hamiltonian in Eq.~(\ref{EqA3}) corrects the effective Josephson capacitances and gate-charge numbers of the simplified model. Moreover, we define
\begin{align}
\gamma_{j}(\varphi_{ext})=\frac{C^2_cE^2_{J_j}}{2E_{J_s}^{\textrm{eff}}(C_{j}+C_c)^2},\quad \gamma_{12}(\varphi_{ext})=\frac{C^2_cE_{J_1}E_{J_2}}{E_{J_s}^{\textrm{eff}}(C_1+C_c)(C_2+C_c)},
\label{EqA10}
\end{align}
which depend on the external flux $\varphi_{ext}$. Now by promoting the classical variables to quantum operators \ie $Q_j\rightarrow \hat{Q}_j=2e\hat{n}_j$ and $\varphi_j\rightarrow \hat{\varphi}_j$ with the commutation relation $[e^{i\hat{\varphi}_{j}},\hat{n}_{j}]=e^{i\hat{\varphi}_{j}}$, we obtain the following quantum Hamiltonian 
\begin{eqnarray}
\hat{\mathcal{H}}&=&\sum_{j=1}^2\hat{\mathcal{H}}_{sub}^j+\gamma_{12}(\varphi_{ext})\sin{(\hat{\varphi}_1)}\sin{(\hat{\varphi}_2)},
\label{EqA11}
\end{eqnarray}
where
\begin{eqnarray}
\hat{\mathcal{H}}_{sub}^j=4E_{C_j}(\hat{n}_{j}-\bar{n}_{g_j})^2-E_{J_j}\cos{(\hat{\varphi}_j)}+\gamma_j(\varphi_{ext})\sin{(\hat{\varphi}_j)}^2 ,
\label{EqA12}
\end{eqnarray}
is the Hamiltonian of the $j$th subsystem with the charge energy $E_{C_j}={e^2}/{2\bar{C}_{J_i}}$. In the following discussion, we consider $\bar{n}_{g_1}=\bar{n}_{g_2}=0.5$ and $\hbar=1$. Note that the free Hamiltonian of the subsystem $\mathcal{H}^j_{sub}$ in Eq.~(\ref{EqA12}) includes both the bare CPB Hamiltonian and the nonlinear term proportional to $\sin{(\hat{\varphi}_j)}^2$. To study how this extra term affects the anharmonicity of the subsystem, we plot the energy spectrum of the free Hamiltonian $\hat{\mathcal{H}}_{sub}^j$ as a function of offset charge $\bar{n}_{g_j}$ in Fig.~\ref{Fig11} for different $E_{J_j}/E_{C_j}$ and $\gamma_j(\varphi_{ext})/E_{C_j}$. It shows that as long as we keep $E_{J_j}/E_{C_j}$ in charge regime, the increase of $\gamma_j(\varphi_{ext})$ will not destroy the anharmonicity of the subsystem, and level of anharmonicity still depends on ratio $E_{J_j} /E_{C_j}$.
Thus, in the following discussion, we assume both CPBs working in charge regime, and write the Hamiltonian in Eq.~(\ref{EqA8}) in the number basis with $\hat{n}_j=\sum_m m \ket{m_j}\bra{m_j}$, $\cos{(\hat{\varphi}_j)}=\frac{1}{2}(\sum_{m}|m_j\rangle\langle m_j+1|+\textrm{H.C})$ and $\sin{(\hat{\varphi}_j)}=-\frac{i}{2}(\sum_{m}|m_j\rangle \langle m_j+1|-\textrm{H.C})$,  where $|m_j\rangle$ is the $m$th exited state of the $j$th subsystem. After we perform the two-level approximation, the operator $\sin{(\hat{\varphi}_j)}$ and the nonlinear term $\gamma_j(\varphi_{ext})\sin{(\hat{\varphi}_j)}^2$ in the subsystem basis read
\begin{eqnarray}
\sin{(\hat{\varphi}_j)}=\frac{1}{2}\sigma^y_j,\quad \gamma_j(\varphi_{ext})\sin{(\hat{\varphi}_j)}^2=\frac{\gamma_j(\varphi_{ext})}{4}\mathbb{I},
\end{eqnarray}
where $\sigma_j^\alpha$ is the Pauli matrix, $\mathbb{I}_j$ is the identity operator. Thus, we can neglect the term proportional to $\sin{(\hat{\varphi}_j)}^2$, as it only provides a shift to the qubit frequency. Finally, we obtain the simplified Hamiltonian as follows
\begin{eqnarray}
\hat{\mathcal{H}}=\frac{\omega_1}{2}\sigma_1^z+\frac{\omega_2}{2}\sigma_2^z+\frac{\gamma_{12}(\varphi_{ext})}{4}\sigma_1^y\sigma_2^y,
\label{EqA14}
\end{eqnarray}
Next, we consider the external flux $\varphi_{ext}$ to be composed of a DC signal and a small AC signal as $\varphi_{ext}=\varphi_{ext}(t)=\varphi_{DC}+\varphi_{AC}(t)$, where $\varphi_{AC}(t)=A_1\cos{(\nu_1 t +\tilde{\varphi}_1)}+A_2\cos{(\nu_2 t +\tilde{\varphi}_2)}$ and $|A_{1}|, |A_{2}|\ll |\varphi_{DC}|$, with which we can approximate
\begin{eqnarray}
\frac{1}{E_{J_s}^{\textrm{eff}}}\approx\frac{1}{\bar{E}_{J_s}}\left[1+\frac{\sin{({\varphi}_{DC})}}{\cos{({\varphi}_{DC})}}{\varphi}_{AC}(t)\right].
\label{EqA15}
\end{eqnarray}
Here, $\bar{E}_{J_s}=2E_{J_s}\cos{{\varphi}_{DC}}$. By replacing Eq.~(\ref{EqA15}) in the Hamiltonian Eq. (\ref{EqA14}), we obtain
\begin{eqnarray}
\hat{\mathcal{H}}=\frac{\omega_1}{2}\sigma_1^z+\frac{\omega_2}{2}\sigma_2^z+\left[g_0+g_1\varphi_{AC}(t)\right]\sigma_1^y\sigma_2^y,
\label{EqA16}
\end{eqnarray}
where the coupling strength
\begin{align}
g_0=\frac{C_{c}^2E_{J_1}E_{J_2}}{4(C_{1}+C_c)(C_{2}+C_c)\bar{E}_{J_s}},\quad g_1=\frac{C_{c}^2E_{J_1}E_{J_2}}{4(C_{1}+C_c)(C_{2}+C_c)\bar{E}_{J_s}}\frac{\sin{({\varphi}_{DC})}}{\cos{({\varphi}_{DC})}}.
\label{EqA17}
\end{align}
To visualize the dynamics of our system, we go to the interaction picture characterized by the free Hamiltonian $\hat{\mathcal{H}}_0=\frac{\omega_1}{2}\sigma_1^z+\frac{\omega_2}{2}\sigma_2^z$ and perform the rotating wave approximation (RWA) obtaining
\begin{align}\nonumber
\hat{\mathcal{H}}_I \approx & -\frac{g_1}{2}\sigma_1^-\sigma_2^-\bigg(A_1e^{i\tilde{\varphi}_1}e^{i(\nu_1-\mu_{12})t}+A_2e^{i\tilde{\varphi}_2}e^{i(\nu_2-\mu_{12})t}\bigg)\\\nonumber
&+\frac{g_1}{2}\sigma_1^-\sigma_2^+\bigg(A_1e^{i\tilde{\varphi}_1}e^{i(\nu_1-\Delta_{12})t}+A_2e^{i\tilde{\varphi}_2}e^{i(\nu_2-\Delta_{12})t}\bigg)\\\nonumber
& + \frac{g_1}{2}\sigma_1^+\sigma_2^-\bigg(A_1e^{-i\tilde{\varphi}_1}e^{-i(\nu_1-\Delta_{12})t}+A_2e^{-i\tilde{\varphi}_2}e^{-i(\nu_2-\Delta_{12})t}\bigg)\\
&-\frac{g_1}{2}\sigma_1^+\sigma_2^+\bigg(A_1e^{-i\tilde{\varphi}_1}e^{-i(\nu_1-\mu_{12})t}+A_2e^{-i\tilde{\varphi}_2}e^{-i(\nu_2-\mu_{12})t}\bigg).
\label{EqA17}
\end{align}
where $\Delta_{12}=\omega_1-\omega_2$, $\mu_{12}=\omega_1+\omega_2$, and we neglected the fast oscillating terms proportional to exp(${\pm i(\Delta_{12}+\nu_{1(2)})t}$), exp(${\pm i(\mu_{12}+\nu_{1(2)})t}$), exp(${\pm i\Delta_{12}t}$), and exp$({\pm i\mu_{12}t}$), since we consider the qubits are far from resonance and the coupling strength $\{g_0, {g_1A_1}/{2}, {g_1A_2}/{2}\}\ll\{\Delta_{12}, \mu_{12}, \nu_1, \nu_2\}$. Next, we assume that $\nu_1=\Delta_{12}$ and $\nu_2=\mu_{12}$, with which we can perform the second RWA neglecting the fast oscillating terms proportional to exp(${\pm i(\Delta_{12}-\nu_{2})t}$) and exp(${\pm i(\mu_{12}-\nu_{1})t}$) in Eq.~(\ref{EqA17}) obtaining
\begin{align}
\hat{\mathcal{H}}_I&\approx\frac{g_1}{2}A_1\bigg(e^{i\tilde{\varphi}_1}\sigma_1^-\sigma_2^++e^{-i\tilde{\varphi}_1}\sigma_1^+\sigma_2^-\bigg)-\frac{g_1}{2}A_2\bigg(e^{i\tilde{\varphi}_2}\sigma_1^-\sigma_2^-+ e^{-i\tilde{\varphi}_2}\sigma_1^+\sigma_2^+\bigg).
\label{EqA19}
\end{align}
\begin{figure}[t!]
\centering
\includegraphics[width=0.6\linewidth]{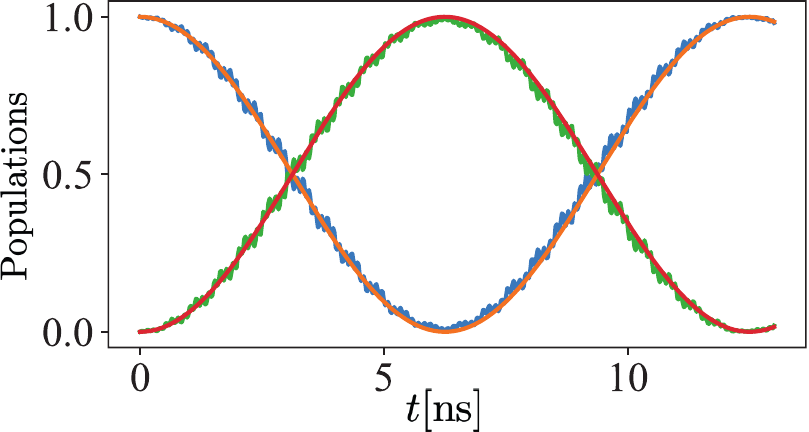}
\caption{Population inversions of the states $|01\rangle$ (red line) and $|10\rangle$ (orange line), numerically calculated from Eq.~(\ref{EqA19}), and the states $|01\rangle$ (green line) and $|10\rangle$ (blue line), numerically calculated from Eq.~(\ref{EqA16}). These calculations have been performed with parameters, $\omega_1/2\pi=9$~(GHz), $\omega_2/2\pi=1$~(GHz), $g_0/2\pi=0.2$~(GHz), $A_{1(2)}g_1/2\pi$=0.08~(GHz), $\nu_1/2\pi=8$~(GHz), $\nu_2/2\pi=10$~(GHz), and $\tilde{\varphi}_{1}=2\pi$, and $\tilde{\varphi}_{2}=\pi$.}
\label{Fig12}
\end{figure}
To prove the justification of the RWA we applied, we plot the population inversion between the states $\ket{01}$ and $\ket{10}$ in Fig.~\ref{Fig12}. It shows that the numerical results calculated from the Hamiltonian in Eq.~(\ref{EqA19}) (red line and orange line) coincide well with the results calculated Eq. (\ref{EqA16}) (green line and blue line), which proves the validity of the approximation we applied. Finally, by replacing $\sigma_j^+=(\sigma_j^x+i\sigma_j^y)/2$ and $\sigma_j^-=(\sigma_j^x-i\sigma_j^y)/2$ in Eq. (\ref{EqA19}), we obtain the interaction Hamiltonian in terms of Pauli matrices as follows
\begin{eqnarray}
\nonumber
\hat{\mathcal{H}}_I \approx & \frac{g_1}{4}\bigg((A_1\cos{\tilde{\varphi}_1}-A_2\cos{\tilde{\varphi}_2})\sigma_1^x\sigma_2^x-(A_1\sin{\tilde{\varphi}_1} + A_2\sin{\tilde{\varphi}_2})\sigma_1^x\sigma_2^y\\
& + (A_1\sin{\tilde{\varphi}_1}-A_2\sin{\tilde{\varphi}_2})\sigma_1^y\sigma_2^x+(A_1\cos{\tilde{\varphi}_1}+A_2\cos{\tilde{\varphi}_2})\sigma_1^y\sigma_2^y\bigg), 
\label{EqA20}
\end{eqnarray}
and the interactions we can engineer with different phase $\tilde{\varphi}_1$ and $\tilde{\varphi}_2$ are shown in the Tab.~\ref{Tab01} of the main text.

\section{Simplified three-qubit Hamiltonian}
\label{AppB}
In this section, we derive the simplified Hamiltonian of the three-qubit system, where the three charge qubits coupled with each other through the grounded SQUIDs as shown in Fig.~\ref{Fig04} of the main text. The Lagrangian of the circuit reads
\begin{align}\nonumber
\mathcal{L} &=  \sum_{\ell=1}^3\bigg[\frac{C_{g_\ell}}{2}(\Phi^{\prime}_\ell-V_{g_\ell})^2+\frac{C_{J_\ell}}{2}\Phi^{\prime 2}_\ell+E_{J_\ell}\cos{(\varphi_\ell)}\bigg]+\sum_{j=1}^2\bigg[\frac{C_{s}}{2}{\Phi^{\prime}_{s_j}}^2+E_{J_{sj}}^{\textrm{eff}}\cos{(\varphi_{s_j})}\bigg]\\
&+\frac{C_{c}}{2}(\Phi^{\prime}_1-\Phi^{\prime}_{s_1})^2 + \frac{C_{c}}{2}(\Phi^{\prime}_{s_1}-\Phi^{\prime}_{2})^2+\frac{C_{c}}{2}(\Phi^{\prime}_2-\Phi^{\prime}_{s_2})^2+\frac{C_{c}}{2}(\Phi^{\prime}_{s_2}-\Phi^{\prime}_{3})^2,
\label{EqB1}
\end{align}
where $E_{J_{sj}}^{\textrm{eff}}=2E_{J_s}\cos{\big(\varphi^{(j)}_{ext}\big)}$ is the effective Josephson energy of the $j$th SQUID and we assume that the third CPB to be the same as the first one \ie $C_{g_3}=C_{g_1}$, $C_{J_3}=C_{J_1}$, and $V_{g_3}=V_{g_1}$. Now we calculate the conjugate momenta (node charge) $Q_{j}=\partial L/\partial  \Phi^{\prime}_{j}$
\begin{align}\nonumber
&Q_{1(3)}=C_{g_1}(\Phi^{\prime}_{1(3)}-V_{g_1})+C_{J_1}\Phi^{\prime}_{1(3)}+C_{c}(\Phi^{\prime}_{1(3)}-\Phi^{\prime}_{s_{1(2)}}),\\\nonumber
&Q_2=C_{g_2}(\Phi^{\prime}_2-V_{g_2})+C_{J_2}\Phi^{\prime}_2-C_{c}(\Phi^{\prime}_{s_1}-\Phi^{\prime}_{2})+C_{c}(\Phi^{\prime}_{2}-\Phi^{\prime}_{s_2}),\\
&Q_{s_{1(2)}}=C_s\Phi^{\prime}_{s_{1(2)}}-C_{c}(\Phi^{\prime}_{1(2)}-\Phi^{\prime}_{s_{1(2)}})+C_{c}(\Phi^{\prime}_{s_{1(2)}}-\Phi^{\prime}_{2(3)}).
\label{EqB2}
\end{align}
By applying the Legendre transformation $\mathcal{H}(\Phi_k,Q_k)=\sum_k Q_k\Phi^{\prime}_k-\mathcal{L}$, we obtain the Hamiltonian
\begin{align}\nonumber
\mathcal{H} = & \sum_{\ell=1}^3\bigg[\frac{(Q_\ell-2e\tilde{n}_{g_\ell})^2}{2\tilde{C}_{J_\ell}}-E_{J_\ell}\cos{(\varphi_\ell)}\bigg]+\sum_{j=1}^2\bigg[\frac{(Q_{s_1}-2e\tilde{n}_{g_{s_1}})^2}{2\tilde{C}_{J_{s_1}}}-E_{J_{sj}}^{\textrm{eff}}\cos{(\varphi_{s_j})}\bigg]\\\nonumber
&+g_{12}Q_1Q_2+g_{13}Q_1Q_3 + g_{{1s}_1}Q_1Q_{s_1}+g_{1s_2}Q_1Q_{s_2}+g_{23}Q_{2}Q_{3}\\
&+g_{2s_1}Q_2Q_{s_1}+g_{2s_2}Q_{2}Q_{s_2}+g_{3s_1}Q_{3}Q_{s_1}+g_{3s_2}Q_{3}Q_{s_2}+g_{s_1s_2}Q_{s_1}Q_{s_2},
\label{EqB3}
\end{align}
where the effective Josephson capacitances 
\begin{align}\nonumber
&\tilde{C}_{J_{1(3)}}={C_\star^5}/\bigg(C_c^3(4C_1+2C_2+4C_s+C_c)+4C_2C_1C_c(C_s+C_c)\\\nonumber
&+C_sC_c(C_2+2C_1)(C_s+3C_c)+C_s^2(2C_c^2+C_1C_2)\bigg),\\\nonumber
&\tilde{C}_{J_2}=\frac{C_\star^5}{[C_c^2+C_1C_s+C_c(2C_1+C_s)]^2},\\
&\tilde{C}_{J_{{s_1}(s_2)}}=\frac{C_\star^5}{(C_c+C_1)[(C_1+C_c)C_2C_s+C_c^2(3C_1+C_2+2C_s+C_c)+2C_cC_1(C_2+C_s)]},
\label{EqB5}
\end{align}
the coupling strength
\begin{align}\nonumber
&g_{1s_1(3s_2)}=\frac{C_c[C_1C_2(2C_c+C_s)+C_c^2(3C_1+C_2+C_c+2C_s)+C_cC_s(2C_1+C_2)]}{C^5_\star},\\\nonumber
&g_{1s_2(3s_1)}=\frac{C^3_c(C_{1} + C_c )}{C^5_\star},g_{13}=\frac{C_c^4}{C^5_\star},\\\nonumber
&g_{2s_1(2s_2)}=\frac{C_c(C_c+C_1)[C^2_c+C_1C_s+C_c(2C_1+C_s)]}{C^5_\star},\\
&g_{12(23)}=\frac{C_c^2[C_c^2+C_1C_s+C_c(2C_1+C_s)]}{C^5_\star},
g_{s_1s_2}=\frac{C_c^2(C_c+C_1)^2}{C^5_\star},
\label{EqB6}
\end{align}
and the gate-charge numbers
\begin{subequations}
\begin{align}\nonumber
\tilde{n}_{g_{1(3)}}&=-\frac{\tilde{C}_{J_1}[C_c^2+C_1C_s+C_c(2C_1+C_s)][2C_c^2+C_2C_s+2C_c(C_2+C_s)]}{2eC_\star^5}{C_{g_1}V_{g_1}}\\
&-\frac{\tilde{C}_{J_1}C_c^2[C_c^2+C_1C_s+C_c(2C_1+C_s)]}{2eC_{\star}^5}{C_{g_2}V_{g_2}},
\end{align}
\begin{align}\nonumber
\tilde{n}_{g_{2}}&=-\frac{2\tilde{C}_{J_2}C_c^2[C_c^2+C_1C_s+C_c(2C_1+C_s) ]}{2eC_{\star}^5}{C_{g_1}V_{g_1}}\\
&-\frac{\tilde{C}_{J_2}[C_c^2+C_1C_s+C_c(2C_1+C_s)]^2}{2eC_{\star}^5}{C_{g_2}V_{g_2}},
\end{align}
\begin{align}\nonumber
\tilde{n}_{g_{s_1(s_2)}}&=-\frac{\tilde{C}_{J_s}C_c(2C_c+C_2)[C_1(C_s+2C_c)+C_c(C_s+C_c)]}{2eC_{\star}^5}{C_{g_1}V_{g_1}}\\
&-\frac{ \tilde{C}_{J_s}C_c(C_c+C_1)[C_cC_1+(C_1+C_c)(C_c+C_s)]}{2eC_{\star}^5}{C_{g_2}V_{g_2}},
\label{EqB7}
\end{align}
\end{subequations}
with $C_\star^5= [C_c^2+C_1C_s+C_c(2C_1+C_s)][C_1C_2C_s+C_c^2(2C_1+C_2+2C_s)+C_c(2C_2C_1+C_2C_s+2C_sC_1)]$, and $C_j=C_{J_j}+C_{g_j}$ ($j=\{1,2\}$). Next, we calculate the effective Hamiltonian of this three-qubit model, by applying the approximations $\Phi^{\prime}_{s_{1(2)}}\ll\Phi^{\prime}_{1(2,3)}$ ($\Phi^{\prime\prime}_{s_{1(2)}}\ll\Phi^{\prime\prime}_{1(2,3)}$), and $\Phi_{s_{1(2)}}\ll\Phi_{1(2,3)}$, where we assume the both SQUIDs in phase regime with high plasma frequency and low impedance. With the first approximation, we can neglect the terms proportional to $\Phi^{\prime}_{s_{1(2)}}$ in Eq.~{(\ref{EqB2})} obtaining the relation between nodes charge as follows
\begin{align}
Q_{s_{1(2)}}&=-C_{c}\left(\frac{Q_{1(3)}+C_{g_1}V_{g_1}}{C_{1}+C_c}+\frac{Q_2+C_{g_2}V{g_2}}{C_{2}+2C_c}\right).
\label{EqB7}
\end{align}
Moreover, to obtain the relation between nodes flux, we calculate the E-L equations, and by applying the two approximations, we can neglect the terms proportional to $\Phi^{\prime\prime}_{s_{1(2)}}$ and approximate $\sin{(\varphi_{s_{1(2)}})}=\varphi_{s_{1(2)}}$ obtaining
\begin{align}
&{\varphi_{s_{1(2)}}}=-\frac{C_cE_{J_{1}}\sin{(\varphi_{1(3)}})}{E_{J_{s1(s2)}}^{\textrm{eff}}(C_{1}+C_c)}-\frac{C_cE_{J_2}\sin{(\varphi_2})}{E_{J_{s1(s2)}}^{\textrm{eff}}(C_{2}+2C_c)},
\label{EqB8}
\end{align}
Meanwhile, with the condition $\Phi_{s_{1(2)}}\ll\Phi_{1(2,3)}$, we can approximate $\cos{\varphi_{s_{1(2)}}}\approx(1-{\varphi_{s_{1(2)}}^2/2})$, where we can keep the potential energy of the SQUID up to the second-order and by replacing Eq.~(\ref{EqB7}) and Eq.~(\ref{EqB8}) in the Hamiltonian in Eq. (\ref{EqB3}), we obtain
\begin{align}\nonumber
{\mathcal{H}} &=  \frac{1}{2\bar{C}_{J_1}}(Q_1-\bar{n}_{g_1})^2-E_{J_1}\cos{(\varphi_1)}+\gamma_1\left(\varphi^{(1)}_{ext}\right) \sin{(\varphi_1)}^2+\frac{1}{2\bar{C}_{J_2}}(Q_2-\bar{n}_{g_2})^2\\\nonumber
&-E_{J_2}\cos{(\varphi_2)}+\gamma_2\left(\varphi^{(1)}_{ext}, \varphi^{(2)}_{ext}\right) \sin{(\varphi_2)}^2 + \frac{1}{2\bar{C}_{J_3}}(Q_3-\bar{n}_{g_3})^2-E_{J_3}\cos{(\varphi_3)}\\
&+\gamma_3\left(\varphi^{(2)}_{ext}\right) \sin{(\varphi_3)}^2+\gamma_{12}\left (\varphi^{(1)}_{ext}\right)\sin{(\varphi_1)}\sin{(\varphi_2)}+\gamma_{23}\left(\varphi^{(2)}_{ext}\right) \sin{(\varphi_2)}\sin{(\varphi_3)},
\label{EqB9}
\end{align}
where 
\begin{align}\nonumber
&\bar{C}_{J_{1(3)}}=C_1+C_c, \quad \bar{C}_{J_2}=C_2+2C_c,\quad\bar{n}_{g_{1(3)}}=-\frac{C_{g_1}V_{g_1}}{2e}, \quad \bar{n}_{g_2}=-\frac{C_{g_2}V_{g_2}}{2e},\\\nonumber
&\gamma_1\left(\varphi^{(1)}_{ext}\right)=\frac{C_{c}^2E_{J_1}^2}{2(C_{1}+C_c)^2E_{J_{s1}}^{\textrm{eff}}},\quad\gamma_3\left(\varphi^{(2)}_{ext}\right)=\frac{C_{c}^2E_{J_1}^2}{2(C_{1}+C_c)^2E_{J_{s2}}^{\textrm{eff}}},\\\nonumber
&\gamma_2\left(\varphi^{(1)}_{ext}, \varphi^{(2)}_{ext}\right)=\frac{C_{c}^2E_{J_2}^2\Big(\cos{\left(\varphi^{(1)}_{ext}\right)}+\cos{\left(\varphi^{(2)}_{ext}\right)}\Big)}{4(C_{2}+2C_c)^2E_{J_s}\cos{\left(\varphi^{(1)}_{ext}\right)}\cos{\left(\varphi^{(2)}_{ext}\right)}},\\
&\gamma_{12}\left(\varphi^{(1)}_{ext}\right)=\frac{C_{c}^2E_{J_1}E_{J_2}}{(C_{1}+C_c)(C_{2}+2C_c)E_{J_{s1}}^{\textrm{eff}}},\quad\gamma_{23}\left(\varphi^{(2)}_{ext}\right)=\frac{C_{c}^2E_{J_1}E_{J_2}}{(C_{1}+C_c)(C_{2}+2C_c)E_{J_{s2}}^{\textrm{eff}}}.
\end{align}
By promoting the classical variables to quantum operators, \ie $Q_j\rightarrow \hat{Q}_j=2e\hat{n}_j$ and $\varphi_j\rightarrow \hat{\varphi}_j$ with the commutation relation $[e^{i\hat{\varphi}_{j}},\hat{n}_{j}]=e^{i\hat{\varphi}_{j}}$ in Eq.~(\ref{EqB9}), we obtain the quantum Hamiltonian 
\begin{align}
\hat{\mathcal{H}}&=\sum_{j=1}^3\hat{\mathcal{H}}_{sub}^j+\gamma_{12}\left(\varphi^{(1)}_{ext}\right)\sin{(\hat{\varphi}_1)}\sin{(\hat{\varphi}_2)}+\gamma_{23}\left(\varphi^{(2)}_{ext}\right)\sin{(\hat{\varphi}_2)}\sin{(\hat{\varphi}_3)},
\label{EqB11}
\end{align}
where the Hamiltonian of the subsystem reads
\begin{align}\nonumber
\hat{\mathcal{H}}_{sub}^{1(3)}&=4E_{C_1}(\hat{n}_{1(3)}-\bar{n}_{g_{1}})^2-E_{J_1}\cos{\left(\hat{\varphi}_{1\left(3\right)}\right)}+\gamma_{1(3)}\left(\varphi^{(1)((2))}_{ext}\right)\sin{\left(\hat{\varphi}_{1\left(3\right)}\right)}^2,\\
\hat{\mathcal{H}}_{sub}^{2}&=4E_{C_2}(\hat{n}_{2}-\bar{n}_{g_2})^2-E_{J_2}\cos{(\hat{\varphi}_2)}+\gamma_2\left(\varphi^{(1)}_{ext},\varphi^{(2)}_{ext}\right)\sin{(\hat{\varphi}_2)}^2 ,
\label{EqB12}
\end{align}
and the charge energy $E_{C_{j}}={e^2}/({2\bar{C}_{J_j}})$. In the following discussion, we consider $\bar{n}_{g_1}=\bar{n}_{g_2}=0.5$ and $\hbar=1$. As mentioned in appendix~\ref{AppA}, the term proportional to $(\sin{\hat{\varphi}_j})^2$ in the subsystem Hamiltonian in Eq.~(\ref{EqB12}) does not destroy the anharmonicity of the system. Thus, in charge regime, we can safely perform the two-level approximation and write the Hamiltonian in Eq.~{(\ref{EqB11})} in the subsystem basis, where the operator $\sin{(\hat{\varphi}_j)}=\sigma_j^y/2$, and the nonlinear term proportional to $\sin{(\hat{\varphi}_j)}^2$ can be regarded as a shift to the qubit frequency obtaining
\begin{align}
\hat{\mathcal{H}}=\frac{\omega_1}{2}\sigma_1^z+\frac{\omega_2}{2}\sigma_2^z+\frac{\omega_1}{2}\sigma_3^z+\frac{\gamma_{12}\left(\varphi^{(1)}_{ext}\right)}{4}\sigma_1^y\sigma_2^y+\frac{\gamma_{23}\left(\varphi^{(2)}_{ext}\right)}{4}\sigma_2^y\sigma_3^y,
\label{EqB13}
\end{align}
where $\omega_1=E_{J_1}$, $\omega_2=E_{J_2}$, and the coupling strength ${\gamma_{12}\big(\varphi^{(1)}_{ext}\big)}/{4}$, ${\gamma_{23}\big(\varphi^{(2)}_{ext}\big)}/{4}$ depend on the external flux through the first, and the second SQUID, respectively. Here we consider the external flux $\varphi^{(j)}_{ext}$ $(j=\{1,2\})$ to be composed of a DC signal and a small AC signal as $\varphi^{(j)}_{ext}=\varphi^{(j)}_{DC}+\varphi^{(j)}_{AC}(t)$, where $\varphi^{(j)}_{AC}(t)=A^{(j)}_1\cos{\left(\nu^{(j)}_1 t +\tilde{\varphi}^{(j)}_1\right)}+A^{(j)}_2\cos{\left(\nu^{(j)}_2 t +\tilde{\varphi}^{(j)}_2\right)}$, and $|A_{1}|, |A_{2}|\ll |\varphi^{(j)}_{DC}|$. Thus, we can approximate
\begin{align}
\frac{1}{E_{J_{sj}}^\textrm{eff}}\approx\frac{1}{\bar{E}^{(j)}_{J_s}}\left[1+\frac{\sin{\big({\varphi}^{(j)}_{DC}}\big)}{\cos{\big({\varphi}^{(j)}_{DC}}\big)}{\varphi}^{(j)}_{AC}(t)\right],
\label{EqB14}
\end{align}
where $\bar{E}^{(j)}_{J_s}=2E_{J_s}\cos{\left({\varphi}^{(j)}_{DC}\right)}$. By replacing Eq.~({\ref{EqB14}}) in the Hamiltonian in Eq. (\ref{EqB13}), we obtain
\begin{align}
\hat{\mathcal{H}}=\frac{\omega_1}{2}\sigma_1^z+\frac{\omega_2}{2}\sigma_2^z+\frac{\omega_1}{2}\sigma_3^z+\left[g^{(1)}_0+g^{(1)}_1\varphi^{(1)}_{AC}(t)\right]\sigma_1^y\sigma_2^y+\left[g^{(2)}_0+g^{(2)}_1\varphi^{(2)}_{AC}(t)\right]\sigma_2^y\sigma_3^y,
\label{EqB15}
\end{align}
where the coupling strength
\begin{align}
g^{(j)}_0=\frac{C_{c}^2E_{J_1}E_{J_2}}{4(C_{1}+C_c)(C_{2}+2C_c)\bar{E}^{(j)}_{J_s}},\quad
g^{(j)}_1=\frac{C_{c}^2E_{J_1}E_{J_2}}{4(C_{1}+C_c)(C_{2}+2C_c)\bar{E}^{(j)}_{J_s}}\frac{\sin{\left({\varphi}^{(j)}_{DC}\right)}}{\cos{\left({\varphi}^{(j)}_{DC}\right)}}.
\label{EqB16}
\end{align}
To visualize the dynamics of the system, we go to the interaction picture characterized by the free Hamiltonian $\hat{\mathcal{H}}_0={\omega_1}\sigma_1^z/{2}+{\omega_2}\sigma_2^z/{2}+{\omega_1}\sigma_3^z/{2}$. Moreover, we consider the resonant conditions $\nu^{(j)}_1=\Delta_{12}=\omega_1-\omega_2$ and $\nu^{(j)}_2=\mu_{12}=\omega_1+\omega_2$ and perform the RWA obtaining
\begin{align}
\nonumber
\hat{\mathcal{H}}_I \approx & \frac{g^{(1)}_1A^{(1)}_1}{2}\bigg(\sigma_1^-\sigma_2^+e^{i\tilde{\varphi}^{(1)}_1}+\sigma_1^+\sigma_2^-e^{-i\tilde{\varphi}^{(1)}_1}\bigg)-\frac{g^{(1)}_1A^{(1)}_2}{2}\bigg(\sigma_1^-\sigma_2^-e^{i\tilde{\varphi}^{(1)}_2} + \sigma_1^+\sigma_2^+A_2e^{-i\tilde{\varphi}^{(1)}_2}\bigg)\\
& + \frac{g^{(2)}_1A^{(2)}_1}{2}\bigg(\sigma_2^-\sigma_3^+e^{-i\tilde{\varphi}^{(2)}_1}+\sigma_2^+\sigma_3^-e^{i\tilde{\varphi}^{(2)}_1}\bigg)-\frac{g^{(2)}_1A^{(2)}_2}{2}\bigg(\sigma_2^-\sigma_3^-e^{i\tilde{\varphi}^{(2)}_2}+\sigma_2^+\sigma_3^+e^{-i\tilde{\varphi}^{(2)}_2}\bigg),
\label{EqB17}
\end{align}
where we neglected the fast oscillating terms proportional to exp$\big({\pm i(\Delta_{12}+\nu^{(j)}_{1(2)})t}\big)$, exp$\big({\pm i(\mu_{12}+\nu^{(j)}_{1(2)})t}\big)$, exp$\left({\pm i\Delta_{12}t}\right)$, exp$\left({\pm i\mu_{12}t}\right)$, exp$\big({\pm i(\Delta_{12}-\nu^{(j)}_{2})t}\big)$, and exp$\big({\pm i(\mu_{12}-\nu^{(j)}_{1})t}\big)$, as we consider the qubits are far from resonance, and the coupling strength $\{g^{(j)}_0, {g^{(j)}_1A^{(j)}_{1(2)}}/{2}\}\ll\{\Delta_{12}, \mu_{12}, \nu^{(j)}_{1(2)}\}$ ($j=\{1,2\}$).
\begin{figure}[t!]
\centering
\includegraphics[width=0.6\linewidth]{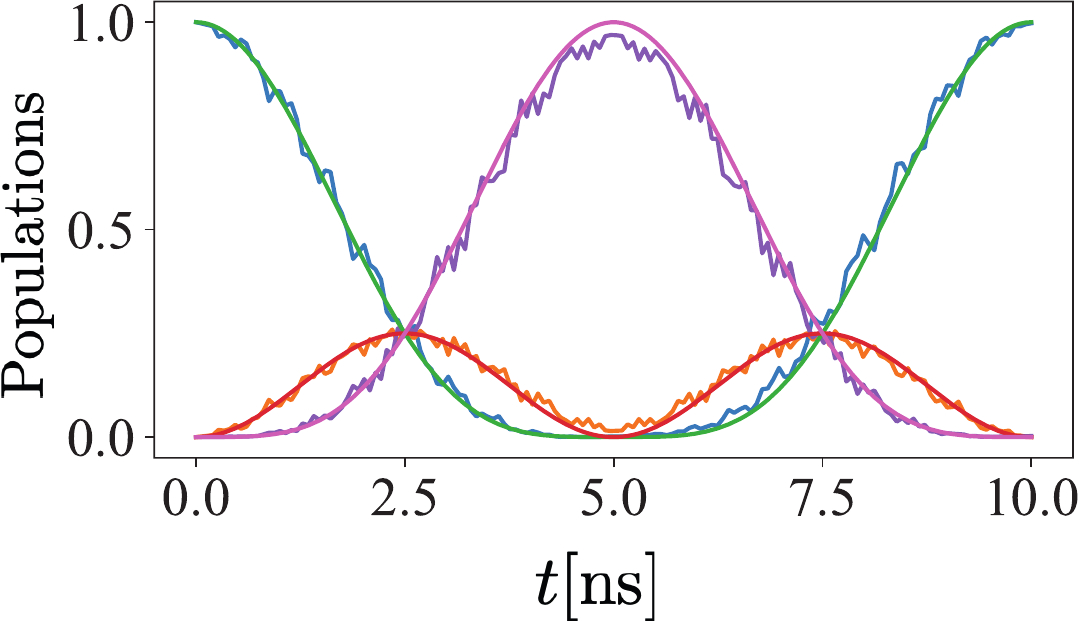}
\caption{Population of the states $|000\rangle$ (blue line), $|011\rangle$ (orange line), and $|101\rangle$ (purple line), numerically calculated from Eq.~(\ref{EqB13}), and the population of the states $|000\rangle$ (green line), $|011\rangle$ (red line), and $|101\rangle$ (pink line), numerically calculated from Eq.~(\ref{EqB17}), with physical parameters $\omega_1/2\pi=9$~(GHz), $\omega_2/2\pi=1$~(GHz), $g^{1(2)}_0/2\pi=0.2$~(GHz), $A^{(1)}_{1(2)}g^{(1)}_1/2\pi=A^{(2)}_{1(2)}g^{(2)}_1/2\pi=0.1$~(GHz), $\nu_1/2\pi=8$~(GHz), $\nu_2/2\pi=10$~(GHz), and the phase $\tilde{\varphi}^{(1)}_{1}=\tilde{\varphi}^{(2)}_{1}=2\pi$, $\tilde{\varphi}^{(1)}_{2}=\tilde{\varphi}^{(2)}_{2}=\pi$.}
\label{Fig13}
\end{figure}
To prove the justification of the RWA we applied, in Fig.~\ref{Fig13} we plot the population of the states $|000\rangle$ (blue line), $|011\rangle$ (orange line), and $|101\rangle$ (purple line), numerically calculated from Eq.~(\ref{EqB13}), and the population of the states $|000\rangle$ (green line), $|011\rangle$ (red line), and $|101\rangle$ (pink line), numerically calculated from Eq.~(\ref{EqB17}). Despite the slight fluctuations,  the results calculated from Eq.~(\ref{EqB13}) still coincide with the ones calculated from the Hamiltonian in Eq.~(\ref{EqB17}), which proves the validity of the RWA we applied. Finally, to visualize the types of interaction we can engineer, we write the Hamiltonian in Eq.~(\ref{EqB17}) in terms of Pauli matrices as follows
\begin{align}
\hat{\mathcal{H}}_I\approx\hat{\mathcal{H}}^{1,2}_I+\hat{\mathcal{H}}^{2,3}_I,
\label{EqB18}
\end{align}
where
\begin{align}\nonumber
\hat{\mathcal{H}}^{j,j+1}_I &=  \frac{g^{(j)}_1}{4}\bigg[\bigg(A^{(j)}_1\cos{\tilde{\varphi}^{(j)}_1}-A^{(j)}_2\cos{\tilde{\varphi}^{(j)}_2}\bigg)\sigma_j^x\sigma_{j+1}^x\\\nonumber
&+\bigg((-1)^jA^{(j)}_1\sin{\tilde{\varphi}^{(j)}_1}-A_2\sin{\tilde{\varphi}^{(j)}_2}\bigg)\sigma_j^x\sigma_{j+1}^y\\\nonumber
& + \bigg((-1)^{j+1}A^{(1)}_1\sin{\tilde{\varphi}^{(j)}_1}-A^{(j)}_2\sin{\tilde{\varphi}^{(j)}_2}\bigg)\sigma_j^y\sigma_{j+1}^x\\
&+\bigg(A^{(j)}_1\cos{\tilde{\varphi}^{(j)}_1}+A^{(j)}_2\cos{\tilde{\varphi}^{(j)}_2}\bigg)\sigma_j^y\sigma_{j+1}^y\bigg].
\label{EqB19}
\end{align}
Notice that the phase $\tilde{\varphi}^{(j)}_1$ required to achieve $\pm\sigma_j^x\sigma_{j+1}^y$ and $\pm\sigma_j^y\sigma_{j+1}^x$ is different for odd and even $j$, and the interactions we can engineer are shown in Tab.~{\ref{Tab02}}.

\section{Mapping Fermion Hubbard model to spin model}
\label{AppC}
For completeness, we derive the Hamiltonian of an $h\times \ell$ fermion-lattice (see Fig.~\ref{Fig06}~(a)) in terms of spin operators by applying the Wigner-Jordan transformation. The Hamiltonian can be expressed as
\begin{align}
\mathcal{H}_{\textrm{Hubb}}=\mathcal{A}\sum_{\alpha=\{\uparrow,\downarrow\}}\sum_{\langle j,k\rangle}\big(c^{\dagger}_{j,\alpha}c_{k,\alpha}+c^{\dagger}_{k,\alpha}c_{j,\alpha}\big)+\mathcal{B}\sum_{j}n_{j,\uparrow}n_{j,\downarrow},
\end{align}
where $\mathcal{A}$ is the kinetic energy, $\mathcal{B}$ is the on-site repulsion, $c_{j,\alpha}^{\dagger}$ ($c_{j,\alpha}$) are the creation (annihilation) operators that act over the $jth$ site, $n_{j,\uparrow(\downarrow)}=c^{\dagger}_{j,\uparrow(\downarrow)}c_{j,\uparrow(\downarrow)}$ is the number operator, and $\alpha=\uparrow,\downarrow$ is the spin component. To suppress the index $\alpha$, we map this lattice to an equivalent $2\ell\times h$ lattice as shown in Fig.~\ref{Fig06}~(b), with $c_{j,\uparrow}^{\dagger}=b_{2j-1}^{\dagger}$, $c_{j,\downarrow}^{\dagger}=b_{2j}^{\dagger}$, where $b_{k}^{\dagger}(b_{k})$ are the creation (annihilation) operation over the site $k$ for the lattice. Now the Hubbard Hamiltonian can be written in terms of
\begin{align}\nonumber
\mathcal{H}_{\textrm{Hubb}}&=\mathcal{A}\sum_{k=0}^{h-1}\sum_{j=1}^{\ell-1}\bigg[\left(b_{2k\ell+2j-1}^{\dagger}b_{2k\ell+2j+1}+b_{2k\ell+2j+1}^{\dagger}b_{2k\ell+2j-1}\right)\\\nonumber
&+\left(b_{2k\ell+2j}^{\dagger}b_{2k\ell+2(j+1)}+b_{2k\ell+2(j+1)}^{\dagger}b_{2k\ell+2j}\right)\bigg]\\\nonumber
&+\mathcal{A}\sum_{k=0}^{h-2}\sum_{j=1}^{2\ell}\bigg[b_{2k\ell+j}^{\dagger}b_{2(k+1)\ell+j}+b_{2(k+1)\ell+j}^{\dagger}b_{2k\ell+j}\bigg]\\
&+\mathcal{B}\sum_{j=1}^{k\ell}\bigg(b^{\dagger}_{2j-1}b_{2j-1}b^{\dagger}_{2j}b_{2j}\bigg),
\label{EqC3}
\end{align}
where the three terms correspond to the horizontal hopping Hamiltonian, the vertical hopping Hamiltonian, and the Coulomb interaction, respectively. By applying the Wigner-Jordan transformation, we map the operator $b_j^{\dagger}(b_j)$ to the combination of Pauli matrices as follows
\begin{align}\nonumber
b_j^{\dagger}&=\bigg[\prod_{l=1}^{j-1}(-\sigma_{l}^z)\bigg]\sigma_j^{\dagger}=\frac{(-1)^{j-1}}{2}\bigg[\prod_{l=1}^{j-1}\sigma_{\ell}^z\bigg](\sigma_j^x+i\sigma_j^y),\\
b_j&=\bigg[\prod_{l=1}^{j-1}(-\sigma_{l}^z)\bigg]\sigma_j=\frac{(-1)^{j-1}}{2}\bigg[\prod_{l=1}^{j-1}\sigma_{l}^z\bigg](\sigma_j^x-i\sigma_j^y),
\label{EqC4}
\end{align}
where $\sigma_j^k$ is the $k$-Pauli-matrix associated with the spin$-1/2$ of the $j${th} position of the chain as shown in Fig.~\ref{Fig06}~(c). Using the Eq. (\ref{EqC4}) and with $k>j$, we obtain
\begin{align}
b_j^{\dagger}b_k+b_k^{\dagger}b_j=\frac{(-1)^{k-j+1}}{2}\left(\sigma_j^x\mathcal{Z}_{j+1}^{k-1}\sigma_k^x+\sigma_j^y\mathcal{Z}_{j+1}^{k-1}\sigma_k^y\right),
\label{EqC5}
\end{align}
where $\mathcal{Z}_{j}^k=\otimes_{\ell=j}^{k}\sigma_{\ell}^z$.
\begin{figure}[t]
	\centering
	\includegraphics[width=0.6\linewidth]{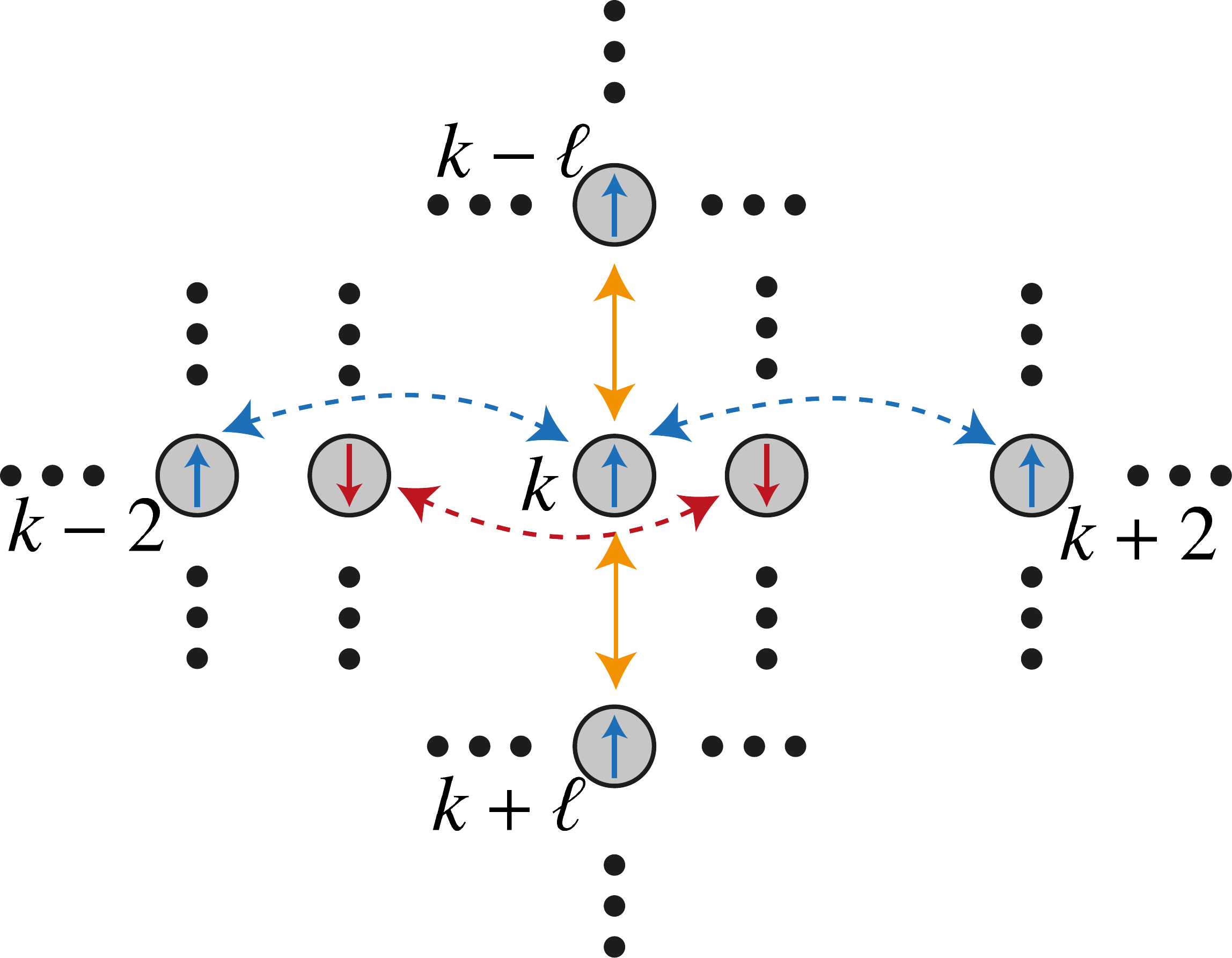}
	\caption{Interactions of the site $k$. Blue and red dashed arrows refer to the interactions in the rows for up and down spin respectively. Yellow solid arrows refer to the interactions in the columns.}
	\label{Fig14}
\end{figure}
And we can see from Fig.~\ref{Fig14} that $k-j$ is an even number for all $j$ and $k$ involved in the interaction, therefore
\begin{equation}
b_j^{\dagger}b_k+b_k^{\dagger}b_j=-\frac{1}{2}\bigg(\sigma_j^x\mathcal{Z}_{j+1}^{k-1}\sigma_k^x+\sigma_j^y\mathcal{Z}_{j+1}^{k-1}\sigma_k^y\bigg).
\label{EqC6}
\end{equation}
Now, we simulate this multi-body interaction using only two-body gates. Before everything, we note that
\begin{subequations}
\begin{align}
&{U}_{j}^{y}\sigma_{j}^x{U}_{j}^{y^{\dagger}}=-\sigma_{j}^z\sigma_{j+1}^y,\quad
{U}_{j}^{x}\sigma_{j}^y{U}_{j}^{x^{\dagger}}=\sigma_{j}^z\sigma_{j+1}^x,\label{EqC7a}\\
&U_{(j,k)}^{y,y}\sigma_j^x\sigma_{k}^xU_{(j,k)}^{y,y\dagger}=\sigma_{j-1}^y\sigma_{j}^z\sigma_{k}^z\sigma_{k+1}^y,\quad
U_{(j,k)}^{x,x}\sigma_j^y\sigma_{k}^yU_{(j,k)}^{x,x\dagger}=\sigma_{j-1}^x\sigma_{j}^z\sigma_{k}^z\sigma_{k+1}^x, 
\label{EqC7b}
\end{align}
\end{subequations}
with ${U}_{j}^{\alpha}=\textrm{exp}({-i\frac{\pi}{4}\sigma_{j}^{\alpha}\sigma_{j+1}^{\alpha}})$ and $U_{(j,k)}^{\alpha,\beta}={U}_{j-1}^{\alpha}{U}_{k}^{\beta}$ $(j\ne k)$, which allow us construct all multi-body operators of the form given by Eq. (\ref{EqC6}). In the following discussion, we derive the Hamiltonian corresponding to the horizontal hopping $\mathcal{H}_{\textrm{hori}}$, vertical hopping $\mathcal{H}_{\textrm{verti}}$, and Coulomb interaction $\mathcal{H}_{\textrm{coul}}$ in terms of Pauli matrices, respectively.
\subsection{Horizontal hopping}
We first calculate the horizontal hopping Hamiltonian, which involves eight types of interactions. And all these interactions  can be divided into two directions (see Fig.~\ref{Fig07}), \ie the forward hopping (solid arrows) $b^\dagger_jb_{j+2}$ and the backward hopping (dashed arrows) $b^\dagger_{j+2}b_{j}$, where we can write both of them in terms of Pauli matrices with Eq.~({\ref{EqC6}}) and Eq.~({\ref{EqC7a}}),
\begin{align}\nonumber
b^\dagger_jb_{j+2}&=-\frac{1}{2}\sigma_j^x\sigma_{j+1}^z\sigma_{j+2}^x=\frac{1}{2}U_{j+1}^{x^\dagger}\sigma_{j}^x\sigma_{j+1}^yU_{j+1}^{x},\\
b^\dagger_{j+2}b_{j}&=-\frac{1}{2}\sigma_j^y\sigma_{j+1}^z\sigma_{j+2}^y=\frac{1}{2}U_{j+1}^y\sigma_{j}^y\sigma_{j+1}^xU_{j+1}^{y^\dagger}.
\label{EqC7}
\end{align}
Now we calculate the Hamiltonian corresponding to the blue solid arrows (see Fig.~\ref{Fig07}), which contains the hopping terms
\begin{align}
\mathcal{H}_{\textrm{Blue}}^s=b_{1}^\dagger b_{3}+b_{5}^\dagger b_{7}+b_{9}^\dagger b_{11}+b_{13}^\dagger b_{15}.....+b_{p}^\dagger b_{p+2},
\label{EqC8}
\end{align}
where $p=2\ell-4-(-1)^{\ell+1}$. By replacing Eq.~(\ref{EqC7}) in Eq.~(\ref{EqC8}), we obtain
\begin{align}
\mathcal{H}_{\textrm{Blue}}^s&=b_{1}^\dagger b_{3}+b_{5}^\dagger b_{7}+b_{9}^\dagger b_{11}+b_{13}^\dagger b_{15}.....+b_{p}^\dagger b_{p+2}=\frac{1}{2}\bigg[U_{(1,2)}^{x^\dagger} H_{1,2}^{(x,y)}U_{(1,2)}^{x} \bigg],
\end{align}
where 
\begin{align}
U_{(n,i)}^{a}=\prod_{k=0}^{h-1} \prod_{j=1}^{m_n} U^a_{2k\ell+4(j-1)+i},\quad H_{n,i}^{(a,b)}=\sum_{k=0}^{h-1}\sum_{j=1}^{m_n}\sigma_{2k\ell+4j-5+i}^a \sigma_{2k\ell+4(j-1)+i}^b,
\end{align}
with $m_1=\big[2\ell-1-(-1)^{\ell+1}\big]/{4}$, and $m_2=\big[2\ell-3+(-1)^{\ell+1}\big]/{4}$, which corresponds to the number of the hopping terms of the blue(red) solid/dashed arrows, and green(brown) solid/dashed arrows respectively (see Fig.~\ref{Fig07}), where $m_1 + m_2 = \ell-1$. As for the blue dashed arrows, we have
\begin{align}
\mathcal{H}_{\textrm{Blue}}^d=b_{3}^\dagger b_{1}+b_{7}^\dagger b_{5}+b_{11}^\dagger b_{9}+b_{15}^\dagger b_{13}.....+b_{p+2}^\dagger b_{p}=\frac{1}{2}\bigg[U_{(1,2)}^{y} H_{1,2}^{(y,x)}U_{(1,2)}^{y^\dagger} \bigg].
\label{EqC11}
\end{align}
Thus for the horizontal hopping corresponding to the blue arrows, we have
\begin{align}
\mathcal{H}_{\textrm{Blue}}&=\mathcal{H}_{\textrm{Blue}}^s+\mathcal{H}_{\textrm{Blue}}^d=\frac{1}{2}\bigg[U_{(1,2)}^{x^\dagger} H_{1,2}^{(x,y)}U_{(1,2)}^{x} +U_{(1,2)}^{y} H_{1,2}^{(y,x)}U_{(1,2)}^{y^\dagger} \bigg].
\end{align}
Following the previous procedure, we obtain the Hamiltonian corresponding to the red arrows, green arrows, and brown arrows (see Fig.~{\ref{Fig07}}) as follows 
\begin{align}\nonumber
\mathcal{H}_{\textrm{Red}}&=\frac{1}{2}\bigg[U_{(1,3)}^{x^\dagger} H_{1,3}^{(x,y)}U_{(1,3)}^{x} +U_{(1,3)}^{y} H_{1,3}^{(y,x)}U_{(1,3)}^{y^\dagger} \bigg],\\\nonumber
\mathcal{H}_{\textrm{Green}}&=\frac{1}{2}\bigg[U_{(2,4)}^{x^\dagger} H_{2,4}^{(x,y)}U_{(2,4)}^{x}+U_{(2,4)}^{y} H_{2,4}^{(y,x)}U_{(2,4)}^{y^\dagger} \bigg],\\
\mathcal{H}_{\textrm{Brown}}&=\frac{1}{2}\bigg[U_{(2,5)}^{x^\dagger} H_{2,5}^{(x,y)}U_{(2,5)}^{x} +U_{(2,5)}^{y} H_{2,5}^{(y,x)}U_{(2,5)}^{y^\dagger} \bigg],
\end{align}
where the first term and the second term correspond to the forward hopping and the backward hopping, respectively. Finally, the Hamiltonian of the horizontal hopping in terms of Pauli matrices read
\begin{align}
\nonumber
&\mathcal{H}_{\textrm{hori}} = \mathcal{H}_{\textrm{Blue}}+\mathcal{H}_{\textrm{Red}}+\mathcal{H}_{\textrm{Green}}+\mathcal{H}_{\textrm{Brown}} \\ \nonumber
&=  \frac{\mathcal{A}}{2}\bigg[U_{(1,2)}^{x^\dagger} H_{1,2}^{(x,y)}U_{(1,2)}^{x}+U_{(1,2)}^{y} H_{1,2}^{(y,x)}U_{(1,2)}^{y^\dagger}+U_{(1,3)}^{x^\dagger} H_{1,3}^{(x,y)}U_{(1,3)}^{x} +U_{(1,3)}^{y} H_{1,3}^{(y,x)}U_{(1,3)}^{y^\dagger}\\
& + U_{(2,4)}^{x^\dagger} H_{2,4}^{(x,y)}U_{(2,4)}^{x}+U_{(2,4)}^{y} H_{2,4}^{(y,x)}U_{(2,4)}^{y^\dagger} +U_{(2,5)}^{x^\dagger} H_{2,5}^{(x,y)}U_{(2,5)}^{x}+U_{(2,5)}^{y} H_{2,5}^{(y,x)}U_{(2,5)}^{y^\dagger} \bigg].
\label{EqC16}
\end{align}

\subsection{Vertical hopping}
In this subsection, we calculate the vertical Hamiltonian in terms of Pauli matrices. For the vertical hopping, each term in the Hamiltonian in Eq.~(\ref{EqC3}) reads
\begin{align}
b^{\dagger}_jb_{2\ell+j}+b^{\dagger}_{2\ell+j}b_{j}&=-\frac{1}{2}\left(\sigma_j^x\mathcal{Z}_{j+1}^{2\ell+j-1}\sigma_{2\ell+j}^x+\sigma_j^y\mathcal{Z}_{j+1}^{2\ell+j-1}\sigma_{2\ell+j}^y\right),
\label{EqC17}
\end{align}
where the first term corresponding to downward hopping (dashed arrows), and the second term corresponding to upward hopping (solid arrows) as shown in Fig.~\ref{Fig08}. To construct the term $\sigma_j^xZ_{j+1}^{2\ell+j-1}\sigma_{2\ell+j}^x$ in Eq.~(\ref{EqC17}), we insert Eq.~(\ref{EqC7b}) on both sides of $\sigma_{j+\ell-1}^x\sigma_{j+\ell}^y$ as follows
\begin{align}\nonumber
\sigma_j^xZ_{j+1}^{2\ell+j-1}\sigma_{2\ell+j}^x&=\left(U_{j+1,j+2\ell-1}^{(x,x)}U_{j,j+2\ell-2}^{(y,y)}...U_{j+\ell-1,j+\ell+1}^{(y,y)}U_{j+\ell}^{x}\right)\cdot\sigma_{j+\ell-1}^x\sigma_{j+\ell}^y\\
&\cdot\left(U_{j+\ell}^{x^\dagger} U_{j+\ell-1,j+\ell+1}^{(y,y)^\dagger} ...U_{j,j+2\ell-2}^{(y,y)^\dagger}U_{j+1,j+2\ell-1}^{(x,x)^\dagger}\right).
\label{EqC18}
\end{align}
And for $\sigma_j^yZ_{j+1}^{2\ell+j-1}\sigma_{2\ell+j}^y$, it can be written in terms of
\begin{align}\nonumber
\sigma_j^yZ_{j+1}^{2\ell+j-1}\sigma_{2\ell+j}^y&=-\left(U_{j+1,j+2\ell-1}^{(y,y)}U_{j,j+2\ell-2}^{(x,x)}...U_{j+\ell-1,j+\ell+1}^{(x,x)}U_{j+\ell}^{y}\right)\\
&\cdot\sigma_{j+\ell-1}^y\sigma_{j+\ell}^x\cdot\left(U_{j+\ell}^{y^\dagger} U_{j+\ell-1,j+\ell+1}^{(x,x)^\dagger}...U_{j,j+2\ell-2}^{(x,x)^\dagger}U_{j+1,j+2\ell-1}^{(y,y)^\dagger}\right).
\label{EqC19}
\end{align}
By replacing Eq.~(\ref{EqC18}) and Eq.~(\ref{EqC19}) in Eq.~(\ref{EqC17}), we obtain the vertical hopping Hamiltonian between $j$th and $(j+2\ell)$th fermion as follows
\begin{align}\nonumber
b_j^\dagger b_{2\ell+j}+b_{2\ell+j}^\dagger b_j &= \frac{1}{2}\bigg(U_{j+1,j+2\ell-1}^{(x,x)}U_{j,j+2\ell-2}^{(y,y)}...U_{j+\ell-1,j+\ell+1}^{(y,y)}U_{j+\ell}^{x^\dagger}\bigg)\cdot\sigma_{j+\ell-1}^x\sigma_{j+\ell}^y\\\nonumber
&\cdot\bigg(U_{j+\ell}^{x} U_{j+\ell-1,j+\ell+1}^{(y,y)^\dagger} ...U_{j,j+2\ell-2}^{(y,y)^\dagger}U_{j+1,j+2\ell-1}^{(x,x)^\dagger}\bigg)\\\nonumber
& + \frac{1}{2}\bigg(U_{j+1,j+2\ell-1}^{(y,y)}U_{j,j+2\ell-2}^{(x,x)}...U_{j+\ell-1,j+\ell+1}^{(x,x)}U_{j+\ell}^{y}\bigg)\cdot\sigma_{j+\ell-1}^y\sigma_{j+\ell}^x\\
&\cdot\bigg(U_{j+\ell}^{y^\dagger} U_{j+\ell-1,j+\ell+1}^{(x,x)^\dagger} ...U_{j,j+2\ell-2}^{(x,x)^\dagger}U_{j+1,j+2\ell-1}^{(y,y)^\dagger}\bigg),
\end{align}
with which we can write the hopping terms in each column together, obtaining the vertical hopping hamiltonian as follows
\begin{align}\nonumber
\mathcal{H}_{\textrm{verti}} &= \mathcal{A}\sum_{j=1}^{2\ell}\sum_{k=0}^{h-2}b_{2k\ell+j}^\dagger b_{2\ell+2k\ell+j}+b_{2\ell+2k\ell+j}^\dagger b_{2k\ell+j}\\ \nonumber
&=  \frac{\mathcal{A}}{2}\sum_{j=1}^{2\ell}\Bigg[\bigg(\tilde{U}_{j,1}^{(x,x)}\tilde{U}_{j,2}^{(y,y)}...\tilde{U}_{j,\ell-1}^{(y,y)}\tilde{U}_{j}^{x^{\dagger}}\bigg)\cdot\Theta_j^{x,y}\cdot \bigg(\tilde{U}_{j}^{x}U_{j,\ell-1}^{(y,y)^\dagger}...U_{j,2}\nonumber^{(y,y)^\dagger}U_{j,1}^{(x,x)^\dagger}\bigg)\\
& + \bigg(U_{j,1}^{(y,y)}U_{j,2}^{(x,x)}...U_{j,\ell-1}^{(x,x)}U_{j}^{y}\bigg)\cdot\Theta_j^{y,x} \cdot\bigg(\tilde{U}_{j}^{y^\dagger} \tilde{U}_{j,\ell-1}^{(x,x)^\dagger}...U_{j,2}^{(x,x)^\dagger}U_{j,1}^{(y,y)^\dagger}\bigg)\Bigg],
\end{align}
where we define
\begin{align}
\Theta_j^{a,b}=\sum_{k=0}^{h-2}\sigma_{2k\ell+j+\ell-1}^a\sigma_{2k\ell+j+\ell}^b,~
\tilde{U}_{j,i}^{(a,a)}=\prod_{k=0}^{h-2}U_{2k\ell+j+i,2k\ell+j+2\ell-i}^{(a,a)},~
\tilde{U}_{j}^{a}=\prod_{k=0}^{h-2}U_{2k\ell+j+\ell}^{a}.
\end{align}
\subsection{Coulomb interaction}
For the Coulomb interaction, each term in the last sum in the Hamiltonian in Eq.~(\ref{EqC3}) reads
\begin{align}
b_j^\dagger b_j=\sigma_j^\dagger\sigma_j=\frac{1}{2}(\sigma_j^z+\mathbb{I}),
\end{align}
where $\mathbb{I}_j$ is the identity operator and the Hamiltonian of the coulomb interaction reads
\begin{align}
\mathcal{H}_{\textrm{coul}}=\mathcal{B}\sum_{j=1}^{k\ell}b_{2j-1}^\dagger b_{2j-1}b_{2j}^\dagger b_{2j}=\frac{\mathcal{B}}{4}\sum_{j=1}^{k\ell}(\sigma_{2j-1}^z+\mathbb{I})(\sigma_{2j}^z+\mathbb{I}).
\end{align}
Finally, we write the Hamiltonian of the Hubbard model in terms of spin-1/2 operators 
\begin{align}\nonumber
\mathcal{H}_{\textrm{Hubb}} = & \mathcal{H}_{\textrm{hori}}+\mathcal{H}_{\textrm{verti}}+\mathcal{H}_{\textrm{coul}}\\\nonumber
= & \frac{\mathcal{A}}{2}\bigg[U_{(1,2)}^{x^\dagger} H_{1,2}^{(x,y)}U_{(1,2)}^{x} +U_{(1,2)}^{y} H_{1,2}^{(y,x)}U_{(1,2)}^{y^\dagger} +U_{(1,3)}^{x^\dagger} H_{1,3}^{(x,y)}U_{(1,3)}^{x} +U_{(1,3)}^{y} H_{1,3}^{(y,x)}U_{(1,3)}^{y^\dagger} \\\nonumber
&+U_{(2,4)}^{x^\dagger} H_{2,4}^{(x,y)}U_{(2,4)}^{x}+U_{(2,4)}^{y} H_{2,4}^{(y,x)}U_{(2,4)}^{y^\dagger}  + U_{(2,5)}^{x^\dagger} H_{2,5}^{(x,y)}U_{(2,5)}^{x} +U_{(2,5)}^{y} H_{2,5}^{(y,x)}U_{(2,5)}^{y^\dagger} \bigg]\\\nonumber
&+\frac{\mathcal{A}}{2}\sum_{j=1}^{2\ell}\bigg[\left(\tilde{U}_{j,1}^{(x,x)}\tilde{U}_{j,2}^{(y,y)}...\tilde{U}_{j,\ell-2}^{(x,x)}\tilde{U}_{j,\ell-1}^{(y,y)}\tilde{U}_{j}^{x^\dagger}\right)\cdot\Theta_j^{x,y}\cdot \left(\tilde{U}_{j}^{x}U_{j,\ell-1}^{(y,y)^\dagger}  U_{j,\ell-2}^{(x,x)^\dagger}...U_{j,2}\nonumber^{(y,y)^\dagger}U_{j,1}^{(x,x)^\dagger}\right)\\\nonumber
&+\left(U_{j,1}^{(y,y)}U_{j,2}^{(x,x)}...U_{j,\ell-2}^{(y,y)}U_{j,\ell-1}^{(x,x)}U_{j}^{y}\right)\cdot\Theta_j^{y,x}\cdot \left(\tilde{U}_{j}^{y^\dagger} \tilde{U}_{j,\ell-1}^{(x,x)^\dagger}\tilde{U}_{j,\ell-2}^{(y,y)^\dagger}...U_{j,2}^{(x,x)^\dagger}U_{j,1}^{(y,y)^\dagger}\right)\bigg] \\
&+\frac{\mathcal{B}}{4}\sum_{j=1}^{h\ell}\left(\sigma_{2j-1}^z+\mathbb{I}\right)\left(\sigma_{2j}^z+\mathbb{I}\right).
\end{align}

\section{Digital decomposition of the hopping Hamiltonian for a $2\times3$ Fermion Hubbard model}
\label{AppD}
In this section, we decompose the exact evolution of the horizontal hopping and vertical hopping of a $2\times3$ Fermion Hubbard model respectively into a sequence of discrete gates by applying Trotter expansion. 
\subsection{Horizontal hopping}
For a $2\times3$ fermion lattice, we first consider the horizontal hopping Hamiltonian $\mathcal{H}^*_{\textrm{hori}}$ as shown in Fig.~\ref{Fig09}(a)
\begin{align}
\mathcal{H}^*_{\textrm{hori}}=\mathcal{H}^*_{\textrm{up}}+\mathcal{H}^*_{\textrm{down}}
\label{EqD1}
\end{align}
where $\mathcal{H}^*_{\textrm{up}}=\mathcal{A}\big(b_1^\dagger b_3+b_3^\dagger b_1+b_5^\dagger b_7+b_7^\dagger b_5+b_9^\dagger b_{11}+b_{11}^\dagger b_9\big)$ and $\mathcal{H}^*_{\textrm{down}}=\mathcal{A}\big(b_2^\dagger b_4+b_4^\dagger b_2+b_6^\dagger b_8+b_8^\dagger b_6+b_{10}^\dagger b_{12}+b_{12}^\dagger b_{10}\big)$,
corresponding to the horizontal hopping for the spin-up fermion (blue solid/dashed arrows) and spin-down fermion (red solid/dashed arrows), respectively (see Fig.~{\ref{Fig09}}(a)).
By applying the JW transformation, we map the fermonic creation and annihilation operators onto spin operators, and finally obtain the horizontal Hamiltonian as follows
\begin{align}\nonumber
\mathcal{H}^*_{\textrm{hori}} &= \frac{\mathcal{A}}{2}U_2^yU_6^yU_{10}^y\left(\sigma_1^y\sigma_2^x+\sigma_5^y\sigma_6^x+\sigma_{9}^y\sigma_{10}^x\right)U_2^{y^\dagger}U_6^{y^\dagger}U_{10}^{y^\dagger}\\\nonumber
&+\frac{\mathcal{A}}{2}U_2^{x^\dagger}U_6^{x^\dagger}U_{10}^{x^\dagger}\left(\sigma_1^x\sigma_2^y+\sigma_5^x\sigma_6^y+\sigma_{9}^x\sigma_{10}^y\right)U_2^xU_6^xU_{10}^x\\\nonumber
& + \frac{\mathcal{A}}{2}U_3^yU_7^yU_{11}^y\left(\sigma_2^y\sigma_3^x+\sigma_6^y\sigma_7^x+\sigma_{10}^y\sigma_{11}^x\right)U_3^{y^\dagger}U_7^{y^\dagger}U_{11}^{y^\dagger}\\
&+\frac{\mathcal{A}}{2}U_3^{x^\dagger}U_7^{x^\dagger}U_{11}^{x^\dagger}\left(\sigma_2^x\sigma_3^y+\sigma_6^x\sigma_7^y+\sigma_{10}^x\sigma_{11}^y\right)U_3^xU_7^xU_{11}^x,
\label{EqD2}
\end{align}
where the four terms correspond to the blue solid arrows, blue dashed arrows, red solid arrows, and red dashed arrows, respectively in Fig.~{\ref{Fig09}}(a).
Now we approximate the time evolution of horizontal hopping Hamiltonian by applying the first-order Trotter expansion $e^{-iHt}\simeq\left(\prod_{\alpha=1}^Ne^{-iH_{\alpha}t/n}\right)^n$ obtaining
\begin{align}\nonumber
& U^*_{\textrm{hori}}(t/n) \\\nonumber
&\approx  \bigg[U_2^yU_6^yU_{10}^y\textrm{exp}\bigg({\frac{-i\mathcal{A}t}{2n}(\sigma_1^y\sigma_2^x+\sigma_5^y\sigma_6^x+\sigma_{9}^y\sigma_{10}^x)}\bigg)U_2^{y^\dagger}U_6^{y^\dagger}U_{10}^{y^\dagger}\\\nonumber
&\cdot U_2^{x^\dagger}U_6^{x^\dagger}U_{10}^{x^\dagger}\textrm{exp}\bigg({\frac{-i\mathcal{A}t}{2n}(\sigma_1^x\sigma_2^y+\sigma_5^x\sigma_6^y+\sigma_{9}^x\sigma_{10}^y)}\bigg)U_2^xU_6^xU_{10}^x  \\\nonumber
&\cdot U_3^yU_7^yU_{11}^y\textrm{exp}\bigg({\frac{-i\mathcal{A}t}{2n}(\sigma_2^y\sigma_3^x+\sigma_6^y\sigma_7^x+\sigma_{10}^y\sigma_{11}^x)}\bigg)U_3^{y^\dagger}U_7^{y^\dagger}U_{11}^{y^\dagger}\\
&\cdot U_3^{x^\dagger}U_7^{x^\dagger}U_{11}^{x^\dagger}\textrm{exp}\bigg({\frac{-i\mathcal{A}t}{2n}(\sigma_2^x\sigma_3^y+\sigma_6^x\sigma_7^y+\sigma_{10}^x\sigma_{11}^y)}\bigg)U_3^xU_7^xU_{11}^x\bigg]^n,
\label{EqD3}
\end{align}
where $U^*_{\textrm{word}}(t)=e^{-i\mathcal{H}^*_{\textrm{word}}t}$.

\subsection{Vertical hopping}
The Hamiloinian of the vertical hopping can be written in terms of 
\begin{align}
\mathcal{H}^*_{\textrm{verti}}=\sum_{\ell=1}^8 h_\ell,
\label{EqD4}
\end{align}
where 
\begin{align}\nonumber
&h_\ell=b_\ell^\dagger b_{\ell+4}+b_{\ell+4}^\dagger b_\ell\\
&=\frac{\mathcal{A}}{2}\left(U_{(\ell+1,\ell+3)}^{x,x}U_{\ell+2}^y\sigma_{\ell+1}^y\sigma_{\ell+2}^xU_{\ell+2}^{y^\dagger}U_{(\ell+1,\ell+3)}^{{x,x}^\dagger}+U_{(\ell+1,\ell+3)}^{y,y}U_{\ell+2}^{x^\dagger}\sigma_{\ell+1}^x\sigma_{\ell+2}^yU_{\ell+2}^{x}U_{(\ell+1,\ell+3)}^{{y,y}^\dagger}\right),
\label{EqD5}
\end{align}
represents the hopping between the $\ell$th qubit and $(\ell+4)$th qubit.

 \begin{figure}[t!]
\centering
\includegraphics[width=0.4\linewidth]{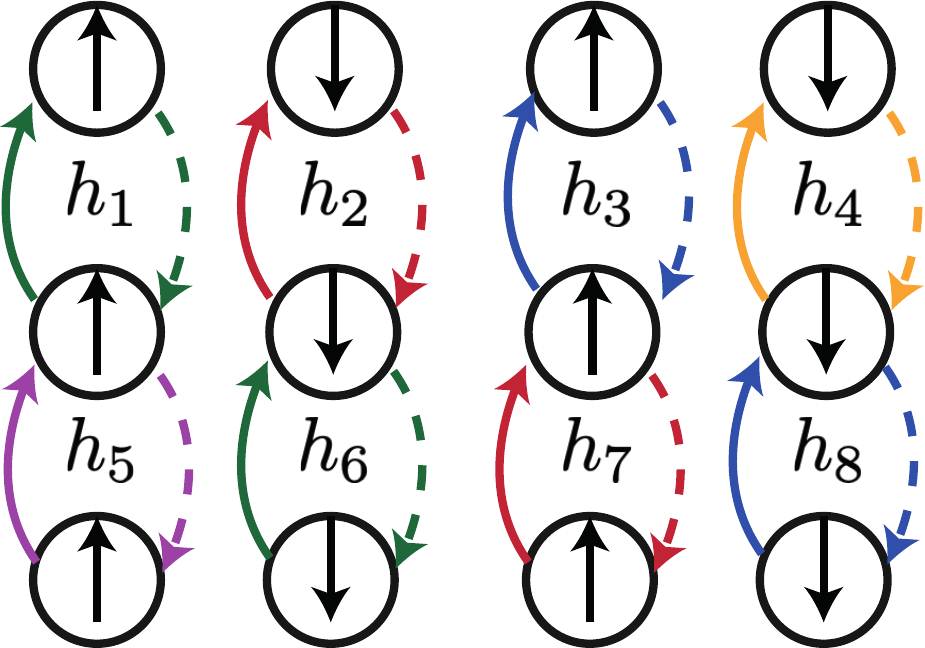}
\caption{Diagram for the different vertical hopping interactions in a $2\times3$ fermion lattice. Solid/dashed arrows with the same color correspond to the  interactions can be implemented at the same time in an analog way.}
\label{Fig15}
\end{figure}

To avoid the sub-gates required in the same interaction sharing the qubits, we define the eight terms in Eq.~(\ref{EqD4}) into five groups $\{h_1,h_6\}$, $\{h_2,h_7\}$, $\{h_3,h_8\}$, $\{h_4\}$, and $\{h_5\}$, where each group includes both upward (solid arrows) and downward (dashed arrows) hopping as shown in Fig.~\ref{Fig15}. Furthermore, all the interactions with the same color and same texture (solid/dashed) can be simulated at the same time \ie to simulate the vertical hopping of a $2\times3$ fermion lattice, it requires ten types of interactions, and each interaction needs ten gates (see Eq.~{(\ref{EqD5})}). Finally, the corresponding Trotter expansion for each group are shown as follows
\begin{subequations}
\begin{align}\nonumber
&\textrm{exp}(-i(h_1+h_6)t) \approx  \Bigg[U_{(2,4)}^{x,x}U_{(7,9)}^{x,x}U_3^yU_8^y \textrm{exp}\bigg(-\frac{i\mathcal{A}t}{2n}(\sigma_2^y\sigma_3^x+\sigma_7^y\sigma_8^x)\bigg)U_8^{y^\dagger}U_3^{y^\dagger}U_{(7,9)}^{{x,x}^\dagger}U_{(2,4)}^{{x,x}^\dagger}\label{D6a}
\\
& \cdot U_{(2,4)}^{y,y}U_{(7,9)}^{y,y}U_3^{x^\dagger}U_8^{x^\dagger} \textrm{exp}\bigg(-\frac{i\mathcal{A}t}{2n}(\sigma_2^x\sigma_3^y+\sigma_7^x\sigma_8^y)\bigg)U_8^{x}U_3^{x}U_{(7,9)}^{{y,y}^\dagger}U_{(2,4)}^{{y,y}^\dagger}\Bigg]^n,
\\\nonumber
&\textrm{exp}(-i(h_2+h_7)t) \approx  \Bigg[U_{(3,5)}^{x,x}U_{(8,10)}^{x,x}U_4^yU_9^y \textrm{exp}\bigg(-\frac{i\mathcal{A}t}{2n}(\sigma_3^y\sigma_4^x+\sigma_8^y\sigma_9^x)\bigg)U_9^{y^\dagger}U_4^{y^\dagger}U_{(8,10)}^{{x,x}^\dagger}U_{(3,5)}^{{x,x}^\dagger}\\\label{D6b}
& \cdot U_{(3,5)}^{y,y}U_{(8,10)}^{y,y}U_4^{x^\dagger}U_9^{x^\dagger} \textrm{exp}\bigg(-\frac{i\mathcal{A}t}{2n}(\sigma_3^x\sigma_4^y+\sigma_8^x\sigma_9^y)\bigg)U_9^{x}U_4^{x}U_{(8,10)}^{{y,y}^\dagger}U_{(3,5)}^{{y,y}^\dagger}\Bigg]^n,
\\\nonumber
&\textrm{exp}(-i(h_3+h_8)t) \approx  \Bigg[U_{(4,6)}^{x,x}U_{(9,11)}^{x,x}U_5^yU_{10}^y \textrm{exp}\bigg(-\frac{i\mathcal{A}t}{2n}(\sigma_4^y\sigma_5^x+\sigma_9^y\sigma_{10}^x)\bigg)U_{10}^{y^\dagger}U_5^{y^\dagger}U_{(9,11)}^{{x,x}^\dagger}\\\label{D6c}&U_{(4,6)}^{{x,x}^\dagger}\cdot U_{(4,6)}^{y,y}U_{(9,11)}^{y,y}U_5^{x^\dagger}U_{10}^{x^\dagger} \textrm{exp}\bigg(-\frac{i\mathcal{A}t}{2n}(\sigma_4^x\sigma_5^y+\sigma_9^x\sigma_{10}^y)\bigg)U_{10}^{x}U_5^{x}U_{(9,11)}^{{y,y}^\dagger}U_{(4,6)}^{{y,y}^\dagger}\Bigg]^n,
\\\nonumber
&\textrm{exp}(-ih_4t) \approx  \Bigg[U_{(5,7)}^{x,x}U_{6}^y \textrm{exp}\bigg(-\frac{i\mathcal{A}\sigma_5^y\sigma_6^xt}{2n}\bigg)U_{6}^{y^\dagger}U_{(5,7)}^{{x,x}^\dagger} \\
&\cdot U_{(5,7)}^{y,y}U_{6}^{x^\dagger}\textrm{exp}\bigg(-\frac{i\mathcal{A}\sigma_5^x\sigma_6^yt}{2n}\bigg)U_{6}^{x}U_{(5,7)}^{{y,y}^\dagger}\Bigg]^n,\label{D6d}
\\\nonumber
&\textrm{exp}(-ih_5t) \approx  \Bigg[U_{(6,8)}^{x,x}U_{7}^y \textrm{exp}\bigg(-\frac{i\mathcal{A}\sigma_6^y\sigma_7^xt}{2n}\bigg)U_{7}^{y^\dagger}U_{(6,8)}^{{x,x}^\dagger}\\
&\cdot U_{(6,8)}^{y,y}U_{7}^{x^\dagger} \textrm{exp}\bigg(-\frac{i\mathcal{A}\sigma_6^x\sigma_7^yt}{2n}\bigg)U_{7}^{x}U_{(6,8)}^{{y,y}^\dagger}\Bigg]^n,\label{D6e}
\end{align}
\end{subequations}
with which we obtain the Trotter expansion for the whole vertical hopping 
\begin{align}\nonumber
U^*_{\textrm{verti}}(t/n)&\approx \bigg[\textrm{exp}\left(-i(h_1+h_6)\frac{t}{n}\right)\cdot\textrm{exp}\left(-i(h_2+h_7)\frac{t}{n}\right)\cdot\textrm{exp}\left(-i(h_3+h_8)\frac{t}{n}\right)\\
&\cdot\textrm{exp}\left(-ih_4\frac{t}{n}\right)\cdot\textrm{exp}\left(-ih_5\frac{t}{n}\right)\bigg]^n.
\label{EqD7}
\end{align}
Now, we approximate the time evolution of the hopping Hamiltonian $\mathcal{H}^*_{\textrm{hop}}=\mathcal{H}^*_{\textrm{hori}}+\mathcal{H}^*_{\textrm{verti}}$ for a $2\times3$ Fermion lattice as follows
\begin{align}
U^*_{\textrm{hop}}(t/n)&\approx \left[e^{-i\mathcal{H}^*_{\textrm{hori}}t/n}e^{-i\mathcal{H}^*_{\textrm{verti}}t/n}\right]^n=\left[U^*_{\textrm{hori}}(t/n)U^*_{\textrm{verti}}(t/n)\right]^n,
\label{EqD8}
\end{align}
where $U^*_{\textrm{hori}}(t/n)$, and $U^*_{\textrm{verti}}(t/n)$ is defined in Eq.~(\ref{EqD3}), and Eq.~(\ref{EqD7}) respectively.

\section{Circuit QED implementation}
\label{AppE}
In this section, we present a cQED encoding of a $2\times3$ Fermi-Hubbard model with a 12-qubit system
\begin{align}
\hat{\mathcal{H}}=\sum_{j=1}^{12}\frac{\omega_j}{2}\sigma_j^z+\left[g^{(j)}_0+g^{(j)}_1\varphi^{(j)}_{AC}\right]\sigma_j^y\sigma_{j+1}^y,
\label{EqE1}
\end{align}
where $\omega_j=\omega_1$ for odd $j$, $\omega_j=\omega_2$ for even $j$ and $\varphi^{(j)}_{AC}=A^{(j)}_1\cos{(\nu^{(j)}_1 t +\tilde{\varphi}^{(j)}_1)}+A^{(j)}_2\cos{(\nu^{(j)}_2 t +\tilde{\varphi}^{(j)}_2)}$ is the time-dependent AC signal through the $j${th} SQUID. Moreover, the effectivce coupling strength $g^{(j)}_0$, and $g^{(j)}_1$ are defined in Eq.~(\ref{EqB16}). Now we write the Hamiltonian in Eq.~(\ref{EqE1}) in interaction picture concerning $\mathcal{\hat{H}}_0=\sum_{j=1}^{12}\omega_j\sigma_j^z/2$ and perform the RWA obtaining
\begin{align}
\hat{\mathcal{H}}_I=\sum_{j=1}^{11}\hat{\mathcal{H}}^{j,j+1}_{I},
\label{EqE2}
\end{align}
with the interaction Hamiltonian between the $j$th and $(j+1)$th qubit
\begin{align}\nonumber
\hat{\mathcal{H}}^{j,j+1}_{I} = & \frac{A^{(j)}_1g^{(j)}_1}{2} \left(e^{i(-1)^j(\Delta_{12}-\nu^{(j)}_1) t}e^{i(-1)^{j+1}\tilde{\varphi}^{(j)}_1}\sigma_j^-\sigma_{j+1}^++e^{i(-1)^{j+1}(\Delta_{12}-\nu^{(j)}_1) t}e^{i(-1)^j\tilde{\varphi}^{(j)}_1}\sigma_j^+\sigma_{j+1}^-\right) \\
& - \frac{A^{(j)}_2g^{(j)}_1}{2}\left(e^{-i(\mu_{12}-\nu^{(j)}_2) t}e^{i\tilde{\varphi}^{(j)}_2}\sigma_j^-\sigma_{j+1}^-+e^{i(\mu_{12}-\nu^{(j)}_2) t} e^{-i\tilde{\varphi}^{(j)}_2}\sigma_j^+\sigma_{j+1}^+\right).
\label{EqE3}
\end{align}

\begin{figure}[t!]
\centering
\includegraphics[width=0.9\linewidth]{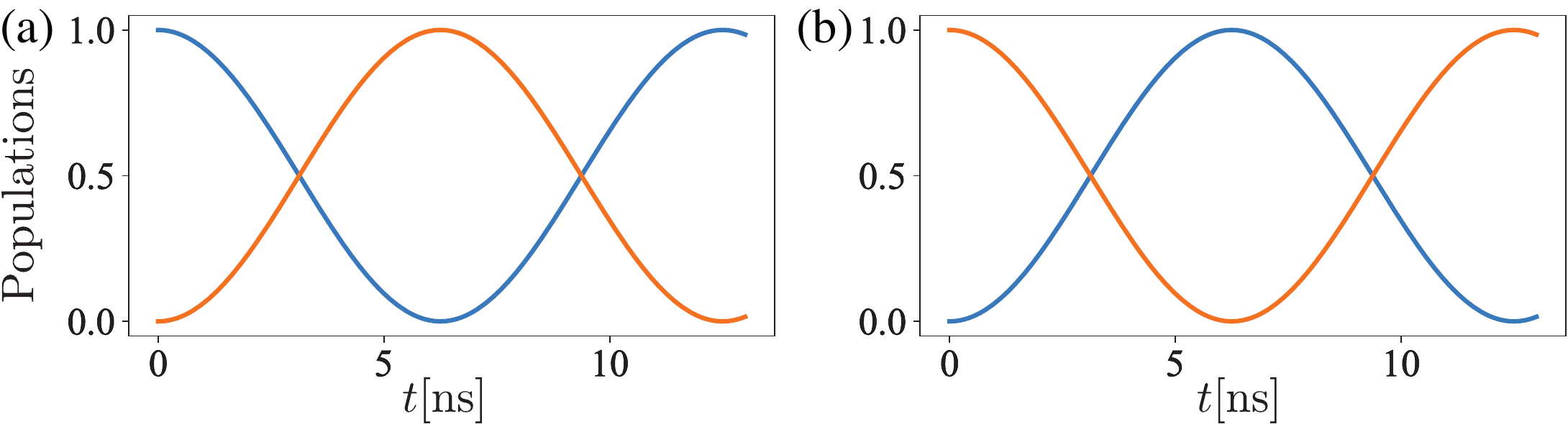}
\caption{Population evolution of the $j$th and ${(j+1)}$th qubit calculated from Eq.~(\ref{EqE3}). The physical parameters of the circuit we consider are shown in Tab. \ref{Tab03} of the main text. (a)  Population inversion of the states $|0\rangle_j|1\rangle_{j+1}$ (blue line) and $|1\rangle_j|0\rangle_{j+1}$ (orange line), where $\nu_1^{(j)}=\nu_{12}$ and $\nu_2^{(j)}=0$. (b) Population inversion of the states $|1\rangle_j|1\rangle_{j+1}$ (orange line) and $|0\rangle_j|0\rangle_{j+1}$ (blue line), where $\nu_2^{(j)}=\mu_{12}$ and $\nu_1^{(j)}=0$.}
\label{Fig16}
\end{figure}

Here, we neglected the fast oscillating terms proportional to exp(${\pm i(\Delta_{12}+\nu^{(j)}_{1(2)})t}$), exp(${\pm i(\mu_{12}+\nu^{(j)}_{1(2)})t}$), exp(${\pm i\Delta_{12}t}$), exp$({\pm i\mu_{12}t}$), as we assume that the adjacent qubits are far from resonance and the coupling strength $\{g^{(j)}_0, {g^{(j)}_1A^{(j)}_{1(2)}}/{2}\}\ll\{\Delta_{12}, \mu_{12}, \nu^{(j)}_{1(2)}\}$. In Fig.~\ref{Fig16}, we show that we can activate coupling terms $\{\sigma_j^+\sigma_{j+1}^-,\sigma_j^-\sigma_{j+1}^+\}$ and $\{\sigma_j^+\sigma_{j+1}^+,\sigma_i^-\sigma_{j+1}^-\}$ in Eq.~(\ref{EqE3}), respectively, and the physical parameters we consider are shown in Tab.~\ref{Tab03} of the main text.

Moreover, by considering the resonant conditions $\nu_1^{(j)}=\Delta_{12}$, and $\nu_2^{(j)}=\mu_{12}$, and we can neglect the fast oscillating terms proportional to exp$\big({\pm i(\Delta_{12}-\nu^{(j)}_{2})t}\big)$ and exp$\big({\pm i(\mu_{12}-\nu^{(j)}_{1})t}\big)$ in Eq.~(\ref{EqE3}) obtaining the interaction Hamiltonian given by Eq.~(\ref{EqB17}). And for a proper choice of the phase $\varphi_{1(2)}^{(j)}$, the operators we can engineer are summarized in Tab. \ref{Tab02}, with which we can simulate the time evolution of the hopping Hamiltonian (see Eq.~(\ref{EqD8})) in an analog way. The sequence of the gates and the corresponding signal parameters as shown as follows.
\begin{table}[H]
\centering
\begin{tabular*} {9.3cm}{llllllllllll}
\hline  
\hline  

Horizontal Hopping \\
\hline  
Operator &$\tilde{\varphi}^{(j)}_1$&$\tilde{\varphi}^{(j)}_2$\\ 
\hline \\
$U_3^xU_7^xU_{11}^x$      
&$\tilde{\varphi}^{(3),(7),(11)}_1= 2\pi$                &$\tilde{\varphi}^{(3),(7),(11)}_2=\pi$             \\    
\hline  \\
exp$\big(-{\frac{i\mathcal{A}}{2}(\sigma_2^x\sigma_3^y+\sigma_6^x\sigma_7^y+\sigma_{10}^x\sigma_{11}^y)\frac{t}{n}}\big)$      
&$\tilde{\varphi}^{(2),(6),(10)}_1= {1}/{2}\pi$              &$\tilde{\varphi}^{(2),(6),(10)}_2={3}/{2}\pi$           \\   
\hline  \\
$U_3^{x^\dagger}U_7^{x^\dagger}U_{11}^{x^\dagger}$         
&$\tilde{\varphi}^{(3),(7),(11)}_1= \pi$                &$\tilde{\varphi}^{(3),(7),(11)}_2=2\pi$            \\ 

\hline  \\
$U_3^{y^\dagger}U_7^{y^\dagger}U_{11}^{y^\dagger}$       
&$\tilde{\varphi}^{(3),(7),(11)}_1= \pi$                &$\tilde{\varphi}^{(3),(7),(11)}_2=\pi$         \\    
\hline  \\
{exp}$\big({\frac{-i\mathcal{A}}{2}(\sigma_2^y\sigma_3^x+\sigma_6^y\sigma_7^x+\sigma_{10}^y\sigma_{11}^x)\frac{t}{n}}\big)$
&$\tilde{\varphi}^{(2),(6),(10)}_1={3}/{2} \pi$                &$\tilde{\varphi}^{(2),(6),(10)}_2={3}/{2}\pi$        \\    
\hline  \\
$U_3^yU_7^yU_{11}^y$
&$\tilde{\varphi}^{(3),(7),(11)}_1=2\pi$                &$\tilde{\varphi}^{(3),(7),(11)}_2=2\pi$          \\    
\hline  \\
$U_2^xU_6^xU_{10}^x$
&$\tilde{\varphi}^{(2),(6),(10)}_1=2\pi$                &$\tilde{\varphi}^{(2),(6),(10)}_2=\pi$          \\    
\hline  \\
exp$\big(-{\frac{i\mathcal{A}}{2}(\sigma_1^x\sigma_2^y+\sigma_5^x\sigma_6^y+\sigma_{9}^x\sigma_{10}^y)\frac{t}{n}}\big)$
&$\tilde{\varphi}^{(1),(5),(9)}_1=3/2\pi$                &$\tilde{\varphi}^{(1),(5),(9)}_2=3/2\pi$         \\    
\hline  \\
$U_2^{x^\dagger}U_6^{x^\dagger}U_{10}^{x^\dagger}$
&$\tilde{\varphi}^{(2),(6),(10)}_1=\pi$                &$\tilde{\varphi}^{(2),(6),(10)}_2=2\pi$         \\   

\hline  \\
$U_2^{y^\dagger}U_6^{y^\dagger}U_{10}^{y^\dagger}$
&$\tilde{\varphi}^{(2),(6),(10)}_1=\pi$                &$\tilde{\varphi}^{(2),(6),(10)}_2=\pi$        \\  
\hline  \\
{exp}$\big({-\frac{i\mathcal{A}}{2}(\sigma_1^y\sigma_2^x+\sigma_5^y\sigma_6^x+\sigma_{9}^y\sigma_{10}^x)\frac{t}{n}}\big)$
&$\tilde{\varphi}^{(1),(5),(9)}_1=1/2\pi$                &$\tilde{\varphi}^{(1),(5),(9)}_2=3/2\pi$         \\    
\hline  \\
$U_2^yU_6^yU_{10}^y$
&$\tilde{\varphi}^{(2),(6),(10)}_1=2\pi$                &$\tilde{\varphi}^{(2),(6),(10)}_2=2\pi$            \\     
\hline 
\hline 
\end{tabular*}  
\caption{Phase parameters required to simulate the time evolution of horizontal hopping Hamiltonian, see Eq. (\ref{EqD3}).}
\label{TabE1}
\end{table}

\begin{table}[H]
\centering
\begin{tabular*} {5.3cm}{llllllllllll}
\hline  
\hline  
Vertical Hopping \\
\hline  
Operator &$\tilde{\varphi}_1$&$\tilde{\varphi}_2$\\ 
\hline \\
$U_{(6,8)}^{{y,y}^\dagger}$
&$\tilde{\varphi}^{5,8}_1= \pi$                &$\tilde{\varphi}^{5,8}_2=\pi$          \\ 
\hline  \\
$U_{7}^x$
&$\tilde{\varphi}^{7}_1= 2\pi$                &$\tilde{\varphi}^{7}_2=\pi$         \\    
\hline  \\
 exp$\bigg(-\frac{i\mathcal{A}}{2}\sigma_6^x\sigma_7^y\frac{t}{n}\bigg)$
&$\tilde{\varphi}^{6}_1= 1/2\pi$                &$\tilde{\varphi}^{6}_2=3/2\pi$          \\    
\hline  \\
$U_{7}^{x^\dagger}$
&$\tilde{\varphi}^{7}_1= \pi$              &$\tilde{\varphi}^{7}_2={2}\pi$        \\   

\hline  \\
$U_{(6,8)}^{y,y}$
&$\tilde{\varphi}^{5,8}_1={2} \pi$                &$\tilde{\varphi}^{5,8}_2={2}\pi$       \\    
\hline  \\
$U_{(6,8)}^{{x,x}^\dagger}$
&$\tilde{\varphi}^{5,8}_1=\pi$                &$\tilde{\varphi}^{5,8}_2=2\pi$         \\    
\hline  \\
$U_{7}^{y^\dagger}$
&$\tilde{\varphi}^{7}_1=\pi$                &$\tilde{\varphi}^{7}_2=\pi$       \\   
\hline  \\
exp$\bigg(-\frac{i\mathcal{A}}{2}\sigma_6^y\sigma_7^x\frac{t}{n}\bigg)$
&$\tilde{\varphi}^{6}_1=3/2\pi$                &$\tilde{\varphi}^{6}_2=3/2\pi$      \\    
\hline  \\
$U_{7}^y$
&$\tilde{\varphi}^{7}_1=2\pi$                &$\tilde{\varphi}^{7}_2=2\pi$          \\    
\hline  \\
$U_{(6,8)}^{x,x}$
&$\tilde{\varphi}^{5,8}_1=2\pi$                &$\tilde{\varphi}^{5,8}_2=\pi$         \\  
\hline  
\hline 
\end{tabular*}  
\caption{Phase parameters required to simulate the time evolution of the vertical hopping $h_5$, see Eq.~(\ref{D6e}).}
\label{TabE2}
\end{table}

\begin{table}[H]
\centering
\begin{tabular*} {5.7cm}{llllllllllll}
\hline  
\hline  
Vertical Hopping \\
\hline  
Operator &$\tilde{\varphi}^{(j)}_1$&$\tilde{\varphi}^{(j)}_2$\\ 
\hline \\
$U_{(5,7)}^{{y,y}^\dagger}$
&$\tilde{\varphi}^{(4),(7)}_1= \pi$                &$\tilde{\varphi}^{(4),(7)}_2=\pi$    \\ 
\hline  \\
$U_{6}^x$
&$\tilde{\varphi}^{(6)}_1= 2\pi$                &$\tilde{\varphi}^{(6)}_2=\pi$         \\    
\hline  \\
 exp$\big(-\frac{i\mathcal{A}}{2}\sigma_5^x\sigma_6^y\frac{t}{n}\big)$
&$\tilde{\varphi}^{(5)}_1= 3/2\pi$                &$\tilde{\varphi}^{(5)}_2=3/2\pi$       \\    
\hline  \\
$U_{6}^{x^\dagger}$
&$\tilde{\varphi}^{(6)}_1= \pi$              &$\tilde{\varphi}^{(6)}_2={2}\pi$       \\   
\hline  \\
$U_{(5,7)}^{y,y}$
&$\tilde{\varphi}^{(4),(7)}_1={2} \pi$                &$\tilde{\varphi}^{(4),(7)}_2={2}\pi$    \\    
\hline  \\
$U_{(5,7)}^{{x,x}^\dagger}$
&$\tilde{\varphi}^{(4),(7)}_1=\pi$                &$\tilde{\varphi}^{(4),(7)}_2=2\pi$          \\    
\hline  \\
$U_{6}^{y^\dagger}$
&$\tilde{\varphi}^{(6)}_1=\pi$                &$\tilde{\varphi}^{(6)}_2=\pi$         \\   
\hline  \\
exp$\big(-\frac{i\mathcal{A}}{2}\sigma_5^y\sigma_6^x\frac{t}{n}\big)$
&$\tilde{\varphi}^{(5)}_1=1/2\pi$                &$\tilde{\varphi}^{(5)}_2=3/2\pi$            \\    
\hline  \\
$U_{6}^y$
&$\tilde{\varphi}^{(6)}_1=2\pi$                &$\tilde{\varphi}^{(6)}_2=2\pi$        \\    
\hline  \\
$U_{(5,7)}^{x,x}$
&$\tilde{\varphi}^{(4),(7)}_1=2\pi$                &$\tilde{\varphi}^{(4),(7)}_2=\pi$        \\  
\hline  
\hline 
\end{tabular*}  
\caption{Phase parameters to simulate the time evolution of the vertical hopping $h_4$, see Eq.~(\ref{D6d}).}
\label{TabE3}
\end{table}

\begin{table}[H]
\centering
\begin{tabular*} {8.3cm}{llllllllllll}
\hline  
\hline  
Vertical Hopping \\
\hline  
Operator &$\tilde{\varphi}^{(j)}_1$&$\tilde{\varphi}^{(j)}_2$\\ 
\hline \\
$U_{(9,11)}^{{y,y}^\dagger}U_{(4,6)}^{{y,y}^\dagger}$ 
&$\tilde{\varphi}^{(3),(6),(8),(11)}_1= \pi$                &$\tilde{\varphi}^{(3),(6),(8),(11)}_2=\pi$             \\ 
\hline  \\
$U_{10}^{x}U_5^{x}$ 
&$\tilde{\varphi}^{(5),(10)}_1= 2\pi$       &$\tilde{\varphi}^{(5),(10)}_2=\pi$      \\   
\hline  \\
exp$\big(-\frac{i\mathcal{A}}{2}(\sigma_4^x\sigma_5^y+\sigma_9^x\sigma_{10}^y)\frac{t}{n}\big)$ 
&\tabincell{c}{$\tilde{\varphi}^{(4)}_1= 1/2\pi$\\$\tilde{\varphi}^{(9)}_1= 3/2\pi$}        &\tabincell{c}{$\tilde{\varphi}^{(4)}_2=3/2\pi$\\$\tilde{\varphi}^{(9)}_2=3/2\pi$}     \\    
\hline  \\
$U_5^{x^\dagger}U_{10}^{x^\dagger} $ 
&$\tilde{\varphi}^{(5),(10)}_1= \pi$             &$\tilde{\varphi}^{(5),(10)}_2=2\pi$      \\    
\hline  \\
$U_{(4,6)}^{y,y}U_{(9,11)}^{y,y}$ 
&$\tilde{\varphi}^{(3),(6),(8),(11)}_1={2} \pi$                &$\tilde{\varphi}^{(3),(6),(8),(11)}_2={2}\pi$          \\    
\hline  \\
$U_{(9,11)}^{{x,x}^\dagger}U_{(4,6)}^{{x,x}^\dagger}$ 
&$\tilde{\varphi}^{(3),(6),(8),(11)}_1=\pi$                &$\tilde{\varphi}^{(3),(6),(8),(11)}_2=2\pi$       \\    
\hline  \\
$U_{10}^{y^\dagger}U_5^{y^\dagger}$ 
&$\tilde{\varphi}^{(5),(10)}_1=\pi$                &$\tilde{\varphi}^{(5),(10)}_2=\pi$            \\   
\hline  \\
exp$\big(-\frac{i\mathcal{A}}{2}(\sigma_4^y\sigma_5^x+\sigma_9^y\sigma_{10}^x)\frac{t}{n}\big)$ 
&\tabincell{c}{$\tilde{\varphi}^{(4)}_1=3/2\pi$ \\$\tilde{\varphi}^{(9)}_1=1/2\pi$}               &\tabincell{c}{$\tilde{\varphi}^{(4)}_2=3/2\pi$\\$\tilde{\varphi}^{(9)}_2=3/2\pi$}            \\    
\hline  \\
$U_5^yU_{10}^y $ 
&$\tilde{\varphi}^{(5),(10)}_1=2\pi$                &$\tilde{\varphi}^{(5),(10)}_2=2\pi$            \\    
\hline  \\
$U_{(4,6)}^{x,x}U_{(9,11)}^{x,x}$ 
&$\tilde{\varphi}^{(3),(6),(8),(11)}_1=2\pi$                &$\tilde{\varphi}^{(3),(6),(8),(11)}_2=\pi$           \\  
\hline  
\hline 
\end{tabular*}  
\caption{Phase parameters required to simulate the time evolution of the vertical hopping $h_3$ and $h_8$, see Eq.~(\ref{D6c}).}
\label{TabE4}
\end{table}

\begin{table}[H]
\centering
\begin{tabular*} {8.3cm}{llllllllllll}
\hline  
\hline  
Vertical Hopping \\
\hline  
Operator &$\tilde{\varphi}^{(j)}_1$&$\tilde{\varphi}^{(j)}_2$\\ 
\hline \\
$U_{(8,10)}^{{y,y}^\dagger}U_{(3,5)}^{{y,y}^\dagger}$
&$\tilde{\varphi}^{(2),(5),(7),(10)}_1= \pi$                &$\tilde{\varphi}^{(2),(5),(7),(10)}_2=\pi$            \\ 
\hline  \\
$U_9^{x}U_4^{x}$ 
&$\tilde{\varphi}^{(4),(9)}_1=2 \pi$       &$\tilde{\varphi}^{(4),(9)}_2=\pi$        \\   
\hline  \\
 exp$\big(-\frac{i\mathcal{A}}{2}(\sigma_3^x\sigma_4^y+\sigma_8^x\sigma_9^y)\frac{t}{n}\big)$ 
&\tabincell{c}{$\tilde{\varphi}^{(3)}_1= 3/2\pi$\\$\tilde{\varphi}^{(8)}_1= 1/2\pi$}          &\tabincell{c}{$\tilde{\varphi}^{(3)}_2=3/2\pi$\\$\tilde{\varphi}^{(8)}_2=3/2\pi$}          \\    
\hline  \\
$U_4^{x ^\dagger}U_9^{x^\dagger} $  
&$\tilde{\varphi}^{(4),(9)}_1= \pi$           &$\tilde{\varphi}^{(4),(9)}_2=2\pi$   \\    
\hline  \\
$U_{(3,5)}^{y,y}U_{(8,10)}^{y,y}$  
&$\tilde{\varphi}^{(2),(5),(7),(10)}_1={2} \pi$                &$\tilde{\varphi}^{(2),(5),(7),(10)}_2={2}\pi$       \\    
\hline  \\
$U_{(8,10)}^{{x,x}^\dagger}U_{(3,5)}^{{x,x}^\dagger}$ 
&$\tilde{\varphi}^{(2),(5),(7),(10)}_1=\pi$                &$\tilde{\varphi}^{(2),(5),(7),(10)}_2=2\pi$           \\    
\hline  \\
$U_9^{y^\dagger}U_4^{y^\dagger}$ 
&$\tilde{\varphi}^{(4),(9)}_1=\pi$                &$\tilde{\varphi}^{(4),(9)}_2=\pi$            \\   
\hline  \\
 exp$\big(-\frac{i\mathcal{A}}{2}(\sigma_3^y\sigma_4^x+\sigma_8^y\sigma_9^x)\frac{t}{n}\big)$ 
&\tabincell{c}{$\tilde{\varphi}^{(3)}_1=1/2\pi$\\$\tilde{\varphi}^{(8)}_1=3/2\pi$}                &\tabincell{c}{$\tilde{\varphi}^{(3)}_2=3/2\pi$ \\$\tilde{\varphi}^{(8)}_2=3/2\pi$}     \\    
\hline  \\
$U_4^yU_9^y $  
&$\tilde{\varphi}^{(4),(9)}_1=2\pi$                &$\tilde{\varphi}^{(4),(9)}_2=2\pi$      \\    
\hline  \\
$U_{(3,5)}^{x,x}U_{(8,10)}^{x,x}$  
&$\tilde{\varphi}^{(2),(5),(7),(10)}_1=2\pi$                &$\tilde{\varphi}^{(2),(5),(7),(10)}_2=\pi$            \\  
\hline  
\hline 
\end{tabular*}  
\caption{Phase parameters required to simulate the time evolution of the vertical hopping $h_2$ and $h_7$, see Eq.~(\ref{D6b}).}
\label{TabE5}
\end{table}

\begin{table}[H]
\centering
\begin{tabular*} {8.3cm}{llllllllllll}
\hline  
\hline  
Vertical Hopping \\
\hline  
Operator &$\tilde{\varphi}^{(j)}_1$&$\tilde{\varphi}^{(j)}_2$\\ 
\hline \\
$U_{(7,9)}^{{y,y}^\dagger}U_{(2,4)}^{{y,y}^\dagger}$  
&$\tilde{\varphi}^{(1),(4),(6),(9)}_1= \pi$                &$\tilde{\varphi}^{(1),(4),(6),(9)}_2=\pi$          \\ 
\hline  \\
$U_8^{x}U_3^{x}$ 
&$\tilde{\varphi}^{(3),(8)}_1= 2\pi$        &$\tilde{\varphi}^{(3),(8)}_2=\pi$    \\   
\hline  \\
exp$\big(-\frac{i\mathcal{A}}{2}(\sigma_2^x\sigma_3^y+\sigma_7^x\sigma_8^y)\frac{t}{n}\big)$
&\tabincell{c}{$\tilde{\varphi}^{(2)}_1= 1/2\pi$ \\$\tilde{\varphi}^{(7)}_1= 3/2\pi$ }              &\tabincell{c}{$\tilde{\varphi}^{(2)}_2=3/2\pi$ \\ $\tilde{\varphi}^{(7)}_2=3/2\pi$}         \\    
\hline  \\
$U_3^{x^\dagger}U_8^{x^\dagger} $ 
&{$\tilde{\varphi}^{(3),(8)}_1= \pi$  }            &{$\tilde{\varphi}^{(3)}_2=2\pi$   }      \\    
\hline  \\
$U_{(2,4)}^{y,y}U_{(7,9)}^{y,y}$  
&$\tilde{\varphi}^{(1),(4),(6),(9)}_1={2} \pi$                &$\tilde{\varphi}^{(1),(4),(6),(9)}_2={2}\pi$      \\    
\hline  \\
$U_{(7,9)}^{{x,x}^\dagger}U_{(2,4)}^{{x,x}^\dagger}$ 
&$\tilde{\varphi}^{(1),(4),(6),(9)}_1=\pi$                &$\tilde{\varphi}^{(1),(4),(6),(9)}_2=2\pi$           \\    
\hline  \\
$U_8^{y^\dagger}U_3^{y^\dagger}$ 
&$\tilde{\varphi}^{(3),(8)}_1=\pi$                &$\tilde{\varphi}^{(3),(8)}_2=\pi$            \\   
\hline  \\
exp$\big(-\frac{i\mathcal{A}}{2}(\sigma_2^y\sigma_3^x+\sigma_7^y\sigma_8^x)\frac{t}{n}\big)$ 
&\tabincell{c}{$\tilde{\varphi}^{(2)}_1=3/2\pi$    \\ $\tilde{\varphi}^{(7)}_1=1/2\pi$ }            &\tabincell{c}{$\tilde{\varphi}^{(2)}_2=3/2\pi$    \\ $\tilde{\varphi}^{(7)}_2=3/2\pi$}       \\    
\hline  \\
$U_3^yU_8^y$   
&$\tilde{\varphi}^{(3),(8)}_1=2\pi$                &$\tilde{\varphi}^{(3),(8)}_2=2\pi$          \\    
\hline  \\
$U_{(2,4)}^{x,x}U_{(7,9)}^{x,x} $  
&$\tilde{\varphi}^{(1),(4),(6),(9)}_1=2\pi$                &$\tilde{\varphi}^{(1),(4),(6),(9)}_2=\pi$         \\  
\hline  
\hline 
\end{tabular*}  
\caption{Phase parameters required to simulate the time evolution of the vertical hopping $h_1$ and $h_6$, see Eq.~(\ref{D6a}).}
\label{TabE6}
\end{table}


\begin{backmatter}

\section*{Acknowledgements}
Not applicable

\section*{Funding}
The authors acknowledge support from Spanish MCIU/AEI/FEDER (PGC2018-095113-B-I00), Basque Government IT986- 16, projects QMiCS (820505) and OpenSuperQ (820363) of EU Flagship on Quantum Technologies, EU FET Open Grants Quromorphic and EPIQUS, Shanghai STCSM (Grant No. 2019SHZDZX01-ZX04), Chilean Government Financiamiento Basal para Centros Cient\'ificos y Tecnol\'ogicos de Excelencia (Grant No. FB0807) and Proyecto AP\_539SF, DICYT (USA-2055 Dicyt), Universidad de Santiago de Chile.

\section*{Abbreviations}
QS: Quantum Simulation.\\
AQS: Analog Quantum Simulation.\\
DQS: Digital Quantum Simulation.\\
DAQS: Digital-Analog Quantum Simulation.\\
DQC: Digital-Quantum Computing.\\
DAQC: Digital-Analog Quantum Computing.\\
SQUIDs: Superconducting Quantum Interference Devices.\\
RWA: Rotating Wave Approximation.\\
2D: Two-Dimensional.\\

\section*{Availability of data and materials}
The datasets generated during and/or analysed during the current study are available from the corresponding author on
reasonable request.

\section*{Competing interests}
The authors declare that they have no competing interests.

\section*{Authors' contributions}
J. Y and J. C. R. are the responsible of the analytical calculations of the circuit quantization and numerical calculations of the quantum simulation, F. A.-A. is the responsible of the circuit design and quantum simulation algorithm, M. S. and E. S. supervised the work. All the authors help in the review and writing of the article.


\bibliographystyle{bmc-mathphys} 
\bibliography{bmc_article}      

\begin{thebibliography}{999}
\bibitem{Georgescu2014RMP}Georgescu IM, Ashhab S, Nori F. Quantum simulation. \href{https://journals.aps.org/rmp/abstract/10.1103/RevModPhys.86.153}{Rev Mod Phys. 2014;86:153}.

\bibitem{Cirac2012NatPhys}Cirac JI, Zoller P. Goals and opportunities in quantum simulation. \href{https://www.nature.com/articles/nphys2275?page=15}{Nat Phys. 2012;8:264}.

\bibitem{Aedo2018PRA} Aedo I, Lamata L. Analog quantum simulation of generalized Dicke models in trapped ions. \href{https://journals.aps.org/pra/abstract/10.1103/PhysRevA.97.042317}{Phys Rev A. 2018;97:042317}.

\bibitem{Braumuller2017NatCommun}Braum\"uller J, Marthaler M, Schneider A, Stehli A, Rotzinger H, Weides M, Ustinov AV. Analog quantum simulation of the Rabi model in the ultra-strong coupling regime. \href{https://www.nature.com/articles/s41467-017-00894-w}{Nat Commun. 2017;8:779}.

\bibitem{Lloyd1996Science}Lloyd S. Universal quantum simulators. \href{https://www.jstor.org/stable/2899535}{Science. 1996;273:1073}.

\bibitem{Preskill2017Quantum}Preskill J. Quantum Computing in the NISQ era and beyond. \href{https://quantum-journal.org/papers/q-2018-08-06-79/}{Quantum. 2017,2:79}.

\bibitem{ParraRodriguez2020PRA}Parra-Rodriguez A, Lougovski P, Lamata L, Solano E, Sanz M. Digital-analog quantum computation. \href{https://journals.aps.org/pra/abstract/10.1103/PhysRevA.101.022305}{Phys Rev A. 2020;101:022305}.

\bibitem{Celeri2021arXiv}Céleri L C, Huerga D, Albarrán-Arriagada F, Solano E, Sanz M. Digital-analog quantum simulation of fermionic models. \href{https://arxiv.org/abs/2103.15689}{arXiv:2103.15689 [quant-ph] (2021)}.

\bibitem{GarciaMolina2021arXiv}García-Molina P, Martin A, Sanz M. Noise in Digital and Digital-Analog Quantum Computation. \href{https://arxiv.org/abs/2107.12969}{arXiv:2107.12969 [quant-ph] (2021)}.

\bibitem{GonzalezRaya2021PRXQuantum}Gonzalez-Raya T, Asensio-Perea R, Martin A, Céleri L C, Sanz M, Lougovski P, Dumitrescu E F. Digital-Analog Quantum Simulations Using the Cross-Resonance Effect. \href{https://journals.aps.org/prxquantum/abstract/10.1103/PRXQuantum.2.020328}{PRX Quantum. 2021;2:020328}.

\bibitem{QiuNPJQI}Qiu X, Zou J, Qi X, Li X. Precise programmable quantum simulations with optical lattices \href{https://www.nature.com/articles/s41534-020-00315-9}{npj Quantum Inf. 2011;333:996}.

\bibitem{Blatt2012NatPhys}Blatt R, Roos CF. Quantum simulations with trapped ions \href{https://www.nature.com/articles/nphys2252/}{Nat Phys. 2012;8:277}.

\bibitem{Paraoanu2014JLTP}Paraoanu GS. Recent Progress in Quantum Simulation Using Superconducting Circuits. \href{https://link.springer.com/article/10.1007/s10909-014-1175-8}{J Low Temp Phys. 2014;175:633}.

\bibitem{Schmidt2013AnnPhys} Schmidt S. Koch J. Circuit QED lattices: Towards quantum simulation with superconducting circuits. \href{https://onlinelibrary.wiley.com/doi/full/10.1002/andp.201200261}{Ann Phys. 2013;525:395}.

\bibitem{Devoret2013Science}Devoret MH, Schoelkopf RJ. Superconducting Circuits for Quantum Information: An Outlook. \href{https://science.sciencemag.org/content/339/6124/1169}{Science. 2013;339:1169}.

\bibitem{Arute2019Nature} Arute F, et al. Quantum supremacy using a programmable superconducting processor. \href{https://www.nature.com/articles/s41586-019-1666-5}{Nature. 2019;574:505}.

\bibitem{Wu2021PRL}Wu Y, et al. Strong Quantum Computational Advantage Using a Superconducting Quantum Processor. \href{https://journals.aps.org/prl/abstract/10.1103/PhysRevLett.127.180501}{Phys Rev Lett. 2021;127:180501}.

\bibitem{Lanyon2010Nat}Lanyon BP, et al. Towards quantum chemistry on a quantum computer. \href{https://www.nature.com/articles/nchem.483}{Nat Chem. 2010;2:106}.

\bibitem{ArgueloLuengo2019Nature}Arg\"uello-Luengo J, Gonz\'alez-Tudela A, Shi T, Zoller P, Cirac JI. Analogue quantum chemistry simulation. \href{https://www.nature.com/articles/s41586-019-1614-4}{Nature. 2019;574:215}.

\bibitem{Babbush2014SciRep}Babbush R, Love P J, Aspuru-Guzik A. Adiabatic Quantum Simulation of Quantum Chemistry. \href{https://www.nature.com/articles/srep06603}{Sci Rep. 2014;4:6603}.

\bibitem{MacDonell2021ChemSci}MacDonell R J, Dickerson C E, Birch C J T, Kumar A, Edmunds C L, Biercuk M J, Hempel C, Kassal I. Analog quantum simulation of chemical dynamics. \href{https://pubs.rsc.org/en/content/articlelanding/2021/SC/D1SC02142G}{Chem Sci. 2021;12:9794}.

\bibitem{Lamata2007PRL}Lamata L, León J, Schätz T, Solano E. Dirac Equation and Quantum Relativistic Effects in a Single Trapped Ion \href{https://journals.aps.org/prl/abstract/10.1103/PhysRevLett.98.253005}{Phys Rev Lett. 2007;98:253005}.

\bibitem{Gerritsma2011PRL}Gerritsma R, et al. Quantum Simulation of the Klein Paradox with Trapped Ions \href{https://journals.aps.org/prl/abstract/10.1103/PhysRevLett.106.060503}{Phys Rev Lett. 2011;106:060503}.

\bibitem{Gerritsma2010Nature}Gerritsma R, Kirchmair G, Z\"ahringer F, Solano E, Blatt R, Roos CF. Quantum simulation of the Dirac equation. \href{https://www.nature.com/articles/nature08688}{Nature. 2010;463:68}.

\bibitem{Bauer2019arXiv}Nachman B, Provasoli D, de Jong WA, Bauer CW. Quantum Algorithm for High Energy Physics Simulations. \href{https://journals.aps.org/prl/abstract/10.1103/PhysRevLett.126.062001}{Phys Rev Lett. 2021;126:062001}.

\bibitem{Casanova2012PRL}Casanova J, Mezzacapo A, Lamata L, Solano E. Quantum Simulation of Interacting Fermion Lattice Models in Trapped Ions. \href{https://journals.aps.org/prl/abstract/10.1103/PhysRevLett.108.190502}{Phys Rev Lett. 2012;108:190502}.

\bibitem{Hensgens2017Nature}Hensgens T, et al. Quantum simulation of a Fermi–Hubbard model using a semiconductor quantum dot array. \href{https://www.nature.com/articles/nature23022}{Nature. 2017;548:70}.

\bibitem{Kim2010Nature}Kim K, Chang M-S, Korenblit S, Islam R, Edwards EE, Freericks JK, Lin G-D, Duan L-M, Monroe C. Quantum simulation of frustrated Ising spins with trapped ions. \href{https://www.nature.com/articles/nature09071}{Nature. 2010;465:590}.

\bibitem{Arrazola2016SciRep}Arrazola I, Pedernales JS, Lamata L, Solano E. Digital-Analog Quantum Simulation of Spin Models in Trapped Ions. \href{https://www.nature.com/articles/srep30534}{Sci Rep. 2016;6:30534}.

\bibitem{AlbarranArriagada2018PRA}Albarr\'an-Arriagada F, Lamata L, Solano E, Romero G, Retamal JC. Spin-1 models in the ultrastrong-coupling regime of circuit QED. \href{https://journals.aps.org/pra/abstract/10.1103/PhysRevA.97.022306}{Phys Rev A. 2018;97:022306}.

\bibitem{Martin2020PRR}Martin A, Lamata L, Solano E, Sanz M. Digital-analog quantum algorithm for the quantum Fourier transform. \href{https://journals.aps.org/prresearch/abstract/10.1103/PhysRevResearch.2.013012}{Phys Rev Res. 2020;2:013012}.

\bibitem{Babukhin2020PRA}Babukhin DV,  Zhukov AA, Pogosov WV. Hybrid digital-analog simulation of many-body dynamics with superconducting qubits. \href{https://journals.aps.org/pra/abstract/10.1103/PhysRevA.101.052337}{Phys Rev A. 2020;101:052337}.

\bibitem{Averin2003PRL}Averin DV, Bruder C. Variable Electrostatic Transformer: Controllable Coupling of Two Charge Qubits. \href{https://journals.aps.org/prl/abstract/10.1103/PhysRevLett.91.057003}{Phys Rev Lett. 2003;91:057003.}

\bibitem{Hutter2006EPL}Hutter C, Shnirman A, Makhlin Y, Sch\"on G. Tunable coupling of qubits: Nonadiabatic corrections. \href{https://iopscience.iop.org/article/10.1209/epl/i2006-10054-4/meta}{EPL. 2006;74:1088.}

\bibitem{Johansson2010PRA}Johansson JR, Johansson G, Wilson CM, Nori F. Dynamical Casimir effect in superconducting microwave circuits. \href{https://journals.aps.org/pra/abstract/10.1103/PhysRevA.82.052509}{Phys Rev A. 2010;82:052509}.

\bibitem{Wilson2011Nature}Wilson CM, Johansson G, Pourkabirian A, Simoen M, Johansson JR, Duty T, Nori F, Delsing P. Observation of the dynamical Casimir effect in a superconducting circuit. \href{https://www.nature.com/articles/nature10561}{Nature. 2011;479:376}.

\bibitem{Vool2017arXiv}Vool U, Devoret M. Introduction to quantum electromagnetic circuits. \href{https://onlinelibrary.wiley.com/doi/abs/10.1002/cta.2359}{Int J Circ Theor App. 2017;45:897}.

\bibitem{Molnar2019ADP}Moln\'ar G, Mikolasek M,  Ridier K, Fahs A, Nicolazzi W, Bousseksou A. Molecular Spin Crossover Materials: Review of the Lattice Dynamical Properties. \href{https://onlinelibrary.wiley.com/doi/abs/10.1002/andp.201900076}{Ann Phys. 2019;531:1900076}.

\bibitem{Rota2017PRB}Rota R, Storme F, Bartolo N, Fazio R, Ciuti C. Critical behavior of dissipative two-dimensional spin lattices. \href{https://journals.aps.org/prb/abstract/10.1103/PhysRevB.95.134431}{Phys Rev B. 2017;95:134431}.

\bibitem{Hegade2021PRApplied}Hegade N N, Paul K, Ding Y, Sanz M, Albarrán-Arriagada F, Solano E, Chen X. Shortcuts to Adiabaticity in Digitized Adiabatic Quantum Computing. \href{https://journals.aps.org/prapplied/abstract/10.1103/PhysRevApplied.15.024038}{Phys Rev Applied. 2021;15:024038}.

\bibitem{Derby2021PRB}Derby C, Klassen J, Baush J, Cubitt T. Compact fermion to qubit mappings. \href{https://journals.aps.org/prb/abstract/10.1103/PhysRevB.104.035118}{Phys Rev B. 2021;104:035118}.

\bibitem{Tasaki1998JPCM}Tasaki H. The Hubbard model - an introduction and selected rigorous results. \href{https://iopscience.iop.org/article/10.1088/0953-8984/10/20/004/meta}{J Phys: Condens Matter. 1998;10:4353}.

\bibitem{Suzuki1990JP}Suzuki M. Fractal decomposition of exponential operators with applications to many-body theories and Monte Carlo simulations. \href{https://www.sciencedirect.com/science/article/abs/pii/037596019090962N}{Phys Lett A. 1990;146:319}. 

\bibitem{Lamata2014EPJ}Lamata L, Mezzacapo A, Casanova J, Solano E. Efficient quantum simulation of fermionic and bosonic models in trapped ions. \href{https://link.springer.com/article/10.1140/epjqt9}{EPJ Quantum Technol. 2014;1:9}.

\bibitem{Place2021}Place A P M, et al. New material platform for superconducting transmon qubits with coherence times exceeding 0.3 milliseconds. \href{https://www.nature.com/articles/s41467-021-22030-5#citeas}{Nat Commun. 2021;12:1779}.

\end{thebibliography}







\end{backmatter}
\end{document}